\documentclass[english,aps,prb,twocolumn,nofootinbib,floatfix,superscriptaddress]{revtex4-2}

\usepackage{amsmath,amssymb,braket,bbold,float,color,mathtools,graphicx,enumitem}
\usepackage[table,xcdraw]{xcolor}
\usepackage{multirow,ragged2e,makecell}
\usepackage{tikz,svg}

\usepackage[unicode=true,pdfusetitle,
 bookmarks=true,bookmarksnumbered=false,bookmarksopen=false,
 breaklinks=false,pdfborder={0 0 0},pdfborderstyle={},backref=false,colorlinks=true]
 {hyperref}

\hypersetup{colorlinks,allcolors=blue}

\usepackage[caption=false]{subfig}
\newcommand{\phantomsubfloat}[1]{
    {
        \captionsetup[subfigure]{labelformat=empty}
        \subfloat[][]{#1}     }%
}

\usepackage{simpler-wick}

\usepackage[mono]{inconsolata}
\usepackage{xspace}
\def\QSpace{\texttt{QSpace}\xspace} 

\def\SigmaY{\Sigma^y}

\begin{document}

\title{Relevance
of Anisotropy in the Kondo Effect: Lessons From the Symplectic Case}

\author{Matan Lotem }
\email{matanlotem@post.bgu.ac.il}
\affiliation{School of Physics and Astronomy, Tel Aviv University, Tel Aviv 6997801, Israel}
\affiliation{Department of Physics, Ben-Gurion University of the Negev, Beer-Sheva 84105, Israel}
\author{Sarath Sankar}
\affiliation{School of Physics and Astronomy, Tel Aviv University, Tel Aviv 6997801, Israel}
\author{Tianhao Ren}
\affiliation{Condensed Matter Physics and Materials Science Division, Brookhaven National Laboratory, Upton, New York 11973, USA}
\author{Moshe Goldstein}
\affiliation{School of Physics and Astronomy, Tel Aviv University, Tel Aviv 6997801, Israel}
\author{Elio. J. König}
\affiliation{Max-Planck-Institut für Festkörperforschung, Heisenbergstraße 1, 70569 Stuttgart, Germany}
\author{Andreas Weichselbaum}
\affiliation{Condensed Matter Physics and Materials Science Division, Brookhaven National Laboratory, Upton, New York 11973, USA}
\author{Eran Sela}
\affiliation{School of Physics and Astronomy, Tel Aviv University, Tel Aviv 6997801, Israel}
\author{Alexei M. Tsvelik}
\email{atsvelik@bnl.gov}
\affiliation{Condensed Matter Physics and Materials Science Division, Brookhaven National Laboratory, Upton, New York 11973, USA}

\begin{abstract}
A Kondo model with symplectic symmetry was recently put forward as the effective low-energy theory of a superconducting-island device coupled to multiple leads. This model, which possesses non-Fermi liquid physics and effective anyons, was argued to belong to the class of topological Kondo effects. Here, we clarify the extent of stability of its exotic fixed point using perturbative and numerical renormalization group in conjunction with bosonization and conformal field theory.
In contrast to previous claims, we show that asymmetry in the coupling to the leads destabilizes the non-Fermi liquid. Other destabilizing perturbations include asymmetry in the superconducting pairing or internal energy of the individual quantum dots in the island. Nevertheless, these perturbations all generate the same relevant operators. Thus, only a small number of couplings need to be tuned individually, and these can be selected according to experimental convenience. Our results highlight a common misconception that anisotropy in single-channel Kondo couplings is always irrelevant. As demonstrated, relevant terms will emerge whenever the group generators do not span the full space of impurity operators. This calls for a more detailed inspection of models that exhibit this property, such as large-spin impurities and $\mathrm{SO}(M)$ Kondo models. 
\end{abstract}

\maketitle

\section{Introduction}
The multichannel Kondo effect is a prime example of quantum criticality that gives rise to non-Fermi-liquid (NFL) physics. It occurs when multiple fermionic channels compete to screen an impurity with internal degrees of freedom. Due to frustration, at low temperatures these degrees of freedom fractionalize. This can be quantified by the residual impurity entropy $\mathrm{S_{imp}}(T{\to}0)=\ln{d}$ with a noninteger $d$.\footnote{$\mathrm{S_{imp}}$ is defined as the difference between the thermodynamic entropy of the impurity and channels together and that of only the channels when decoupled from the impurity -- see Eq.~\eqref{eq:SimpDef}. Thus, it quantifies the effective dimension of the degrees of freedom associated with the impurity.} Interestingly, a fractional dimension is also a hallmark of non-Abelian anyons. These are predicted exotic quasiparticles that can be harnessed for topologically protected quantum computation. There, the dimension of the fusion space---the degenerate Hilbert space generated by $M$ anyons---scales as $d^M$ with $d$  called the anyonic quantum dimension. A mathematical relation between the multichannel Kondo effect and specific anyonic theories has long been known. For the paradigmatic $k$-channel $\mathrm{SU(2)}$ spin-$1/2$ impurity Kondo model [referred to below as $\mathrm{SU}(2)_k$], we have \cite{tsvelickThermodynamicsMultichannelKondo1985}
\begin{equation}\label{eq:Simp}
    \mathrm{S_{imp}}(T{\to}0)=\ln d_{k},\ \ d_{k}=2\cos\left(\frac{\pi}{k+2}\right).
\end{equation}
This matches the quantum dimension of the so-called $\mathrm{SU}(2)_k$ non-Abelian anyons, with $d_2=\sqrt{2}$ and $d_3=(1{+}\sqrt{5})/2$ corresponding to the dimension of Ising (or Majorana) and Fibonacci anyons, respectively. We stress that fractional entropy is in fact a consequence of a deeper relation between Kondo physics and anyons: The low-energy theory of multichannel Kondo systems is governed by so-called $\mathrm{SU}(2)_k$ anyonic fusion rules \cite{affleckExactConformalfieldtheoryResults1993}. 

The two- and three-channel Kondo effects were observed in setups where the localized moment was imitated by a quantum dot and criticality was achieved by fine tuning of the device 
\cite{PotokGoldhaberGordon2007,KellerGoldhaberGordon2015,iftikharTwochannelKondoEffect2015,iftikharTunableQuantumCriticality2018,PouseGoldhaberGordon2021}.  These experiments ignited the hope to observe direct signatures of non-Abelian anyons in Kondo systems, and potentially exploit them for quantum computation \cite{lopesAnyonsMultichannelKondo2020,gabayMultiimpurityChiralKondo2022,lotemChiralNumericalRenormalization2023,lotemManipulatingNonAbelianAnyons2022,komijaniIsolatingKondoAnyons2020,renTopologicalQuantumComputation2024}. 
A major challenge is the quantum critical nature of the NFL: 
The paradigmatic $\mathrm{SU}(2)_k$  model requires equal coupling to all channels. Any asymmetry in the couplings effectively decouples the channels with smaller couplings and destroys the NFL state. Experimentally, this was overcome by fine tuning of multiple gates. However, such an approach might not be feasible when one has to deal with multiple impurities, as required in order to test the anyonic behavior.

Single-channel Kondo models that exhibit NFL physics could help circumvent the  issue of fine tuning. This expectation is based on the fact that, typically, anisotropic single-channel models flow to isotropy \cite{cox_zawadowski_1998,konikInterplayScalingLimit2002,Kurilovich_2017,Kogan_2018,Kogan_2019,Kogan_2021} because of the flow of all Kondo couplings to infinity (we will return to this point). Note that more than two fermionic species are still required to generate NFL physics and the simplest way to get these is by starting from some multichannel model. But, if channel asymmetry in the original model maps to anisotropy in an effective single-channel model (with the same number of fermionic species), one could expect stability. This is exactly what happens in the $\mathrm{SO}(M)$ ``topological'' Kondo effect  \cite{beriTopologicalKondoEffect2012,altlandMultiterminalCoulombMajoranaJunction2013,Beri_2013,altlandMultichannelKondoImpurity2014,altlandBetheAnsatzSolution2014,Galpin_2014,buccheriThermodynamicsTopologicalKondo2015,herviouManyterminalMajoranaIsland2016,Beri_2017,michaeliElectronTeleportationStatistical2017,landauTwochannelKondoPhysics2017,snizhkoParafermionicGeneralizationTopological2018,mitchellNonFermiLiquidCoulomb2021,libermanCriticalPointSpinflavor2021,liMultichannelTopologicalKondo2023,wauters2023topologicalkondomodelequilibrium,BollmannKoenig2024}. Most of these proposals rely on coupling $M$ spinless fermionic channels to $M$ topological Majorana edge modes, and stability to channel asymmetry is indeed demonstrated. From a practical perspective, although, it would be desirable to have such an effect without relying on preexisting Majorana fermions. One example is an $\mathrm{SO}(5)$ Kondo model~\cite{mitchellNonFermiLiquidCoulomb2021,libermanCriticalPointSpinflavor2021} based on the experimental setup in Refs.~\cite{PotokGoldhaberGordon2007,KellerGoldhaberGordon2015}, that was argued to be stable to channel asymmetry. Searching for more examples, one can expand the scope of Kondo models to arbitrary Lie groups \cite{kimuraABCDKondoEffect2021}. Single-channel $\mathrm{SU}(N)$ models (in the defining representation) are not helpful for this cause, as they always lead to a fully-screened Fermi liquid (FL). This leaves only the symplectic group $\mathrm{Sp}(2k)$. Note that, as $\mathrm{Sp}(4)\simeq \mathrm{SO}(5)$,  the detailed analysis in Refs.~\cite{mitchellNonFermiLiquidCoulomb2021,libermanCriticalPointSpinflavor2021} suggests that single-channel symplectic Kondo models could indeed be stable to channel asymmetry.

Recently, a simple physical realization for the $\mathrm{Sp}(2k)$ Kondo effect with arbitrary $k$ was introduced and analyzed by a variety of techniques \cite{liTopologicalSymplecticKondo2023,konigExactSolutionTopological2023,renTopologicalQuantumComputation2024}. It is based on a symplectic Anderson-like model with $k$ spinful quantum dots in proximity to a standard BCS superconducting (SC) island. The dots are coupled to $k$ spinful leads, that together can be regarded as a single $\mathrm{Sp}(2k)$ channel. At $T{\to}0$, the symplectic Kondo model shares the same impurity entropy with the $k$-channel $\mathrm{SU}(2)$ Kondo model [$\mathrm{SU}(2)_k$], either with a spin-$1/2$ or spin-$(k{-}1)/2$ impurity, and the two-channel $\mathrm{SU}(k)$ Kondo model [$\mathrm{SU}(k)_2$]. It stems from the fact that the low-energy  $\mathrm{Sp}(2k)_1$ and $\mathrm{SU}(2)_k$ fixed points are governed by the same anyonic fusion rules.\footnote{As a result, the low-energy fixed points of these models have the same finite-size spectra, up to phase shifts and a redefinition of the quantum numbers. The $\mathrm{SU}(k)_2$ and $\mathrm{SU}(2)_k$ effects are linked by level-rank duality, and thus also share the same finite-size spectrum. However, this does not imply the fixed points are identical. Indeed, they have different leading irrelevant operators and, thus, different power laws when approaching the fixed point (see Sec.~\ref{sec:CFT}).} Note also that the symplectic Kondo model and the spin-$1/2$ impurity $\mathrm{SU}(2)_k$ model have the same strong coupling limit: a spin-$(k{-}1)/2$ impurity  $\mathrm{SU}(2)_k$ model \cite{liTopologicalSymplecticKondo2023}.\footnote{Again, up to a redefinition of the quantum numbers between the two models. In the symplectic case, the $\mathrm{SU}(2)$ symmetry is in Nambu (particle-hole) space, while for the $\mathrm{SU}(2)_k$ model, this symmetry is in spin space.} It was initially argued that, like the $\mathrm{SO}(M)$ models, the single-channel $\mathrm{Sp}(2k)$ Kondo effect is stable with respect to channel asymmetry. As will be shown in this paper, this is a more delicate issue. We stress that the presence of instabilities does not imply that such systems cannot be realized experimentally [as demonstrated for the $\mathrm{SU}(2)_k$ case]. Importantly, the symplectic Kondo model, on one hand, does not require Majorana fermions as the topological Kondo models do, and, on the other hand, may not require so much fine tuning as the multichannel one. 

The goal of this paper is to systemically analyze the stability of the symplectic Kondo effect and corresponding symplectic Anderson model. We base this analysis on the structure of the $\mathrm{Sp}(2k)$ group. Operators acting on the symplectic impurity are separated into two sets (or irreducible representations): The $\mathrm{Sp}(2k)$ generators, and all other complementary operators.  The former are odd under time reversal and can be thought of as ``symplectic dipoles'', while the later can be thought of as ``symplectic quadrupoles'' and are even under time reversal. Only the generators enter the Kondo Hamiltonian in Eq.~\eqref{eq:H-Kondo} and define the NFL physics, whereas the complementary operators are those that destabilize it. This differs from our intuition for $\mathrm{SU}(N)$ in the defining representation, e.g., an $\mathrm{SU}(2)$ spin-$1/2$ impurity. There, the generators span the full space of impurity operators [e.g., the Pauli matrices for $\mathrm{SU}(2)$]. One should recall that for larger $\mathrm{SU}(2)$ spins, the dipoles $S^{x,y,z}$ no longer span the space of impurity operators, and one has to consider the effect of complementary terms, e.g., spin quadrupoles such as $(S^z)^2$. 
Moreover, in the mutlichannel $\mathrm{SU}(2)_k$ case, generators  correspond to relevant operators, i.e., a magnetic field destabilizes the NFL state.

In the single-channel symplectic case, on the other hand, the generators are marginal. Thus, they do not inherently break the NFL state \cite{konigExactSolutionTopological2023}, which is perceived as a source of stability for the symplectic Kondo effect. However, we also have the complementary operators, and these turn out to be relevant. Breaking channel symmetry in the symplectic Anderson model of \citet{liTopologicalSymplecticKondo2023} generates exactly this type of perturbations, and thus destabilizes the NFL. These include channel-asymmetric dot-lead tunneling and dot-dependent chemical potentials, that were previously argued to have no significant effect. Moreover,  combinations of two marginal generator perturbations can effectively generate complementaries, and thus are also relevant. This, however, is a higher-order effect, and so typically is weak. We employ perturbative renormalization group (RG) to link a variety of channel-symmetry-breaking perturbations to the complementaries, and establish their relevance with conformal field theory (CFT). We explicitly back these finding by numerical renormalization group (NRG) \cite{wilsonRenormalizationGroupCritical1975,bullaNumericalRenormalizationGroup2008} calculations for $k{=}2$ and $3$ channels [i.e., $\mathrm{Sp}(4)$ and $\mathrm{Sp}(6)$, respectively], and an exact solution via bosonization in the spirit of \citet{emeryMappingTwochannelKondo1992} (EK) for $k{=}2$.

The remainder of this paper is structured as follows. In Sec.~\ref{sec:Models} we review the key properties of the unperturbed symplectic Kondo model and of the accompanying Anderson-like realization. We then give a crude outline of the various types of perturbations that can affect it. Section \ref{sec:Main} consists the bulk of this paper, and fuses input from multiple theoretical techniques to establish a coherent and rigorous picture of the relevant perturbations. The remaining sections and appendices provide more details for each of the employed methods. In Sec.~\ref{sec:EK} we analyze the $k{=}2$ case via the exact EK solution. This serves as the basis for generalizing to arbitrary $k$ using CFT arguments in Sec.~\ref{sec:CFT}. In Sec.~\ref{sec:SchriefferWolff} we rederive the mapping from the symplectic Anderson model to the effective Kondo model, and systematically analyze how different perturbations go through this mapping. Appendix \ref{sec:Sp2k-properties} reviews important mathematical properties of the symplectic group. In Appendix \ref{sec:poor-mans} we derive the perturbative RG equations for a generic anisotropic Kondo model. In Appendix \ref{sec:NRG} we provide details of the NRG implementation and analysis. Finally, we conclude in Sec.~\ref{sec:Sumamry} and comment on the relevance of the results here to the $\mathrm{SO}(M)$ topological Kondo models.

\section{Models}\label{sec:Models}

Throughout this paper we will move back and forth between the Anderson-like model and the emergent low-energy Kondo model. The latter enables a clean and systematic analysis of all possible perturbations. But the existence of relevant perturbations does not immediately imply instability, and an important question is which of these perturbations can arise in a given physical realization. As an example consider the $\mathrm{SO}(5)\simeq\mathrm{Sp}(4)$ Kondo model discussed by \citet{libermanCriticalPointSpinflavor2021}, in which channel asymmetry maps to irrelevant perturbations. In contrast, channel asymmetry in the symplectic Anderson model considered in this paper maps onto relevant perturbations in the $\mathrm{Sp}(2k)$ Kondo model.

\subsection{\texorpdfstring{$\mathrm{Sp}(2k)$}{Sp(2k)} Kondo Model}

Before defining the symplectic Kondo model, let us first introduce the $\mathrm{Sp}(2k)$ generators $T^A$ with $A{=}1,\dots,k(2k{+}1)$. In the defining representation, these are  $2k{\times}2k$  Hermitian matrices that satisfy the symplectic condition
\begin{equation}
   [T^A]^T = -\SigmaY \, T^A\, \SigmaY
\ \ \big[ {\scriptstyle=-(\sigma_y{\otimes}\mathbb{1}_k) T^A (\sigma_y{\otimes}\mathbb{1}_k)} \big], \label{eq:Sp2k-gen}
\end{equation}
with $\SigmaY {\equiv} \sigma_y{\otimes}\mathbb{1}_k$ where $\sigma_y{=}(\begin{smallmatrix} 0&-i\\i&0 \end{smallmatrix})$ and $\mathbb{1}_k$ is the $k{\times}k$ identity. Note that for the $\mathrm{Sp}(2){\simeq}\mathrm{SU}(2)$ case,  $T^{A=1,2,3}$ are simply the Pauli matrices (up to normalization). The symplectic ``spin'' (in the defining representation) forms a $2k$-dimensional state space, with operators $S^A$. Formally, we relate these to the generators by writing $S^A{\equiv} \textbf{f}^\dagger T^A \textbf{f}$  with auxiliary fermions $\textbf{f}^\dagger \equiv (f_{1\uparrow}^\dagger,\dots,f_{k\uparrow}^\dagger,f_{1\downarrow}^\dagger,\dots,f_{k\downarrow}^\dagger)$ and an implicit projection to the singly-occupied space $\textbf{f}^\dagger\textbf{f}{=}1$. We then understand $S^A$ as the subset of operators acting on the impurity state space that are odd under time reversal.  Explicitly, the symplectic condition \eqref{eq:Sp2k-gen} can be rewritten as $\Theta S^A\Theta^{-1}=-S^A$ with  $\Theta=\SigmaY K$ the time reversal operator and $K$ the complex conjugation operator \cite{flintHeavyElectronsSymplectic2008,flintSymplecticTimeReversal2009}. Note that $S^A$ do not span the full space of bilinears of $\textbf{f}$ for $k{>}1$, and other combinations are possible, including ones that are even under time reversal. This will prove crucial in what follows. Also, to avoid confusion, we stress that $\textbf{f}$ differ from the  $\textbf{d}$ impurity operators of the Anderson model in Eq.~\eqref{eq:H-Anderson}.

The symplectic Kondo Hamiltonian is then given by the sum of a Kondo term and $k$ spinful fermionic channels
\begin{equation}\label{eq:H-Kondo}
    H = \overbrace{\lambda\sum_A S^A J^A(0)}^{\equiv H_K}
    + i v_F\int dx\, \textbf{c}^\dagger(x) \partial_x \textbf{c}(x),
\end{equation}
with the bath fermions represented by creation operators $\textbf{c}^\dagger(x)\equiv(c_{1\uparrow}^\dagger(x),\dots,c_{k\uparrow }^\dagger(x),c_{1\downarrow}^\dagger(x),\dots,c_{k\downarrow }^\dagger(x))$ and a corresponding $\mathrm{Sp}(2k)_1$ current $J^A(x)\equiv \textbf{c}^\dagger (x) T^A \textbf{c}(x)$. In this paper, we consider only a single impurity at the origin $x=0$. Thus, without loss of generality, we can take the channels to be chiral with some Fermi velocity $v_F$. The Kondo coupling $\lambda>0$ is positive and is initially assumed to be isotropic, i.e., independent of $A$. As discussed in the introduction, the low-energy fixed point of this model is a non-Fermi liquid (NFL) with the fractional  impurity entropy specified in Eq.~\eqref{eq:Simp}. The leads have an $\mathrm{O}(4k)$ symmetry, that can be broken down to  $\mathrm{SU}(2){\times}\mathrm{Sp}(2k)$ symmetry. This is similar to the $\mathrm{SU}(2)_k$ case, that has  $\mathrm{Sp}(2k)$ channel-charge (including particle-hole) symmetry, and  $\mathrm{SU}(2)$ spin symmetry, with the impurity coupling to the spin sector. Here, however, the $\mathrm{SU}(2)$ symmetry is in the charge (particle-hole) sector, while the impurity couples to the $\mathrm{Sp}(2k)$ spin-channel symmetry sector.

\subsection{ \texorpdfstring{$\mathrm{Sp}(2k)$}{Sp(2k)} Anderson Model}
\begin{figure}
    \centering
    \includegraphics[width=1\columnwidth]{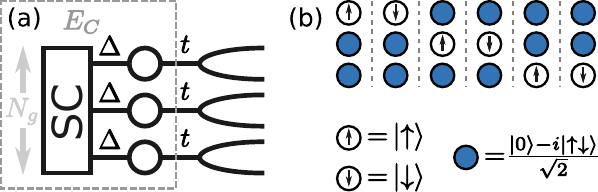}
   \caption{(a) Sketch of the $\mathrm{Sp}(6)$ Anderson model, consisting of $k{=}3$ dots in proximity to a superconducting (SC) island, and each tunnel-coupled to its own lead. (b) Schematic description of the sixfold degenerate ground space, with blue circles corresponding to BCS states.}
    \label{fig:Model}
\end{figure}

We now turn to introduce the symplectic Anderson model of Refs.~\cite{liTopologicalSymplecticKondo2023,konigExactSolutionTopological2023} as depicted in Fig.~\ref{fig:Model}. The impurity consists of $k$ spinful fermionic dots with creation operators $\textbf{d}^\dagger \equiv (d_{1\uparrow}^\dagger,\dots,d_{k\uparrow}^\dagger,d_{1\downarrow}^\dagger,\dots,d_{k\downarrow}^\dagger)$.\footnote{We again stress that despite the resemblance, these $\textbf{d}$ fermions are not the $\textbf{f}$ auxiliary fermions of the Kondo model. See Eq.~\eqref{eq:gs_states} for a definition of  the latter in terms of the former after a Schrieffer-Wolff transformation.} These are placed in proximity to a mesoscopic BCS superconductor and are also tunnel-coupled to the leads. Explicitly, we replace the Kondo term $H_K$ in Eq.~\eqref{eq:H-Kondo} with an Anderson-like term
\begin{equation}\label{eq:H-Anderson}
    \begin{split}
         H_{A} & =E_{C}\left(2\hat{N}_{C}+\hat{n}_{d}-N_{g}\right)^{2}\\
          & +\Delta \sum_{j=1}^k i e^{-i\phi} d_{j\uparrow}^{\dagger} d_{j\downarrow}^{\dagger} + \mathrm{H.c.} \\
          & +t \textbf{d}^\dagger \textbf{c}(0) + \mathrm{H.c.}
    \end{split}
\end{equation}
The first line corresponds to the impurity charging energy $E_C$, with  $\hat{N}_{C} = \rm i\partial/\partial\phi$ the Cooper-pair number operator, $\hat{n}_d=\textbf{d}^\dagger \textbf{d}$ the occupation of all dots together, and $N_g$  controlled externally by a gate voltage. The second line describes the pairing between the superconductor and the dots, with  $e^{-i\phi}$ annihilating a Cooper pair, and $\Delta$ the superconducting order parameter. Each dot is then tunnel-coupled to a corresponding lead at the origin, initially assuming equal tunneling amplitudes $t$, and with the lead fermions $\textbf{c}(x)$ defined below Eq.~\eqref{eq:H-Kondo}. This model has a  $\mathrm{U}(1){\times}\mathrm{Sp}(2k)$ symmetry, with the $\mathrm{U}(1)$ charge symmetry promoted to $\mathrm{SU}(2)$ at the particle-hole symmetric points.

In Sec.~\ref{sec:SchriefferWolff} we will rederive Eq.~\eqref{eq:H-Kondo} as a low-energy limit of this model via a Schrieffer-Wolff transformation \cite{schriefferRelationAndersonKondo1966a,liTopologicalSymplecticKondo2023} and extend the discussion to a variety of perturbations. Let us briefly review this mapping here. Turning off the coupling to the leads (setting $t{=}0$) and fixing an odd total impurity charge $2\hat{N}_{C}+\hat{n}_{d}$ (e.g., by setting $N_g=1$) results in a degenerate ground space of the charging and pairing terms. In each of the ground states, one dot is singly occupied (with either spin up or down, hence $2k$ states). All other dots are in a BCS state, i.e., an equal superposition of the empty and doubly occupied states. This $2k$-dimensional ground space forms the Hilbert space of the symplectic ``spin''. Turning on a small coupling, $t\ll E_C,\Delta$, then gives rise to the  symplectic Kondo model as an effective theory, with isotropic coupling constant
\begin{equation}
    \lambda=\frac{t^2}{E_C-\Delta}+\frac{t^2}{E_C+\Delta}.
\end{equation}
Note that Ref.~\cite{liTopologicalSymplecticKondo2023} contains only the first of these two terms. This is justified in the limit $(E_C{-}\Delta)\ll E_C,\Delta$, whereas here we relax this assumption.
\begin{table}[t]
    \centering
    \begin{tabular}{|p{0.4\columnwidth} p{0.35\columnwidth} p{0.2\columnwidth}|} \hline
        
        \multicolumn{2}{|p{0.75\columnwidth}}{\textbf{Gate detuning}\ \ \ (3)}
        & \multirow{2}{*}[-0.1em]{\includegraphics[width=0.2\columnwidth]{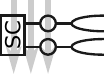}}
        \\ [0.4em]
        $N_g=1{+}\delta N_g$ & $T_\mathrm{FL}\sim\delta N_g^{1+2/k}$ & \\ [0.4em]
        \multicolumn{3}{|p{0.98\columnwidth}|}{Breaks particle-hole symmetry and the degeneracy of excited states, but does not affect the impurity ground space. This generates relevant new coupling terms that result in an anisotropic $\mathrm{SU}(2k)$ model with a FL fixed point below $T_\mathrm{FL}$.} \\ \hline

        \multicolumn{2}{|p{0.75\columnwidth}}{\textbf{Channel-dependent tunneling}\ \ \ (2)}
        & \multirow{2}{*}[-0.3em]{\includegraphics[width=0.2\columnwidth]{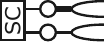}}
        \\ [0.2em]
        $\sum_{j\sigma}\delta t_jd_{j\sigma}^\dagger c_{j\sigma}^{\,}$ & $T^*\sim\delta t_j^{1+k/2}$ & \\ [0.3em]
        \multicolumn{3}{|p{0.98\columnwidth}|}{Effectively decouples the channels with weaker couplings. This is very similar to the effect of channel asymmetry in the  $\mathrm{SU}(2)_k$ Kondo model and breaks the NFL below $T^*$.} \\ \hline

        \multicolumn{2}{|p{0.75\columnwidth}}{\textbf{Channel-dependent SC pairing}\ \ \ (2)}
        & \multirow{2}{*}[-0.3em]{\includegraphics[width=0.2\columnwidth]{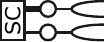}}
        \\ [0.2em]
        $\sum_j \delta\Delta_j ie^{-i\phi}d^\dagger_{j\uparrow}d^\dagger_{j\downarrow}$
        & $T^*\sim\delta \Delta_j^{1+k/2}$ & \\ [0.3em]
        \multicolumn{3}{|p{0.98\columnwidth}|}{Breaks the degeneracy of the ground space and effectively gaps out one (or more) of the channels. Has the same effect as coupling anisotropy and breaks the NFL below $T^*$.} \\ \hline

        \multicolumn{2}{|p{0.75\columnwidth}}{\textbf{Dot-level disorder}\ \ \ (3 or 2)}
        & \multirow{2}{*}[-0.1em]{\includegraphics[width=0.2\columnwidth]{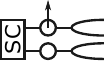}}
        \\ [0.2em]
        $\sum_{j\sigma}\epsilon_j d_{j\sigma}^\dagger d_{j\sigma}^{\,}$ 
        & $\!\!\!\!\!\!\!\!\!\!\!T_\mathrm{FL}\!\sim\bar\epsilon^{1+2/k}$ or   $T^*\!\sim\epsilon_j^{2+k}$ & \\ [0.3em]
        \multicolumn{3}{|p{0.98\columnwidth}|}{Breaks particle-hole symmetry by shifting all levels by an average $\bar\epsilon\equiv\tfrac{1}{k}\sum_j\epsilon_j$. This takes the system to a FL below $T_\mathrm{FL}$, but can be amended by tuning the gate $\delta N_g$. The next-order effect lifts the ground-space degeneracy by generating an effective channel-dependent SC pairing  $\delta\Delta_j\sim\delta\epsilon_j^2/\Delta$ that breaks the NFL below $T^*$.} \\ \hline

        \multicolumn{2}{|p{0.75\columnwidth}}{\textbf{Magnetic field}\ \ \ (1 or 2)}
        & \multirow{2}{*}[-0.1em]{\includegraphics[width=0.2\columnwidth]{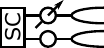}}
        \\ [0.2em]
        $\sum_j B_j(d^\dagger_{j\uparrow}d_{j\uparrow}^{\,}{-}d^\dagger_{j\downarrow}d_{j\downarrow}^{\,})$ 
        & $T^*\sim B_j^{2+k}$ & \\ [0.3em]
        \multicolumn{3}{|p{0.98\columnwidth}|}{A uniform field $B_j{=}B$ is marginal and does not break the NFL. However, it breaks time reversal symmetry and switches to an $\mathrm{SU}(k)_2$ fixed point. Breaking channel symmetry with a dot-dependent magnetic field $B_j{=}B{+}\delta B_j$ is relevant and breaks the NFL below $T^*$.} \\ \hline

        \multicolumn{2}{|p{0.75\columnwidth}}{\textbf{Spin-dependent tunneling}\ \ \ (3)}
        & \multirow{2}{*}[-0.1em]{\includegraphics[width=0.2\columnwidth]{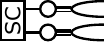}}
        \\ [0.2em]
        $\sum_{j\sigma}\delta t_\sigma d_{j\sigma}^\dagger c_{j\sigma}^{\,}(0)$ &  & \\ [0.3em]
        \multicolumn{3}{|p{0.98\columnwidth}|}{Is found to be irrelevant, both numerically for $k
{=}2,3$, and analytically for $k{=}2$.} \\ \hline
    \end{tabular}
    \caption{List of selected perturbations that could naturally arise in the symplectic Anderson model with the corresponding category indicated in parentheses. For a more comprehensive discussion see Sec.~\ref{sec:SchriefferWolff}. $T^*$ or $T_\mathrm{FL}$ indicate the energy scale at which the NFL breaks as a function of the magnitude of a relevant perturbation, and are plotted for $k{=}2,3$ in Fig.~\ref{fig:NRG-scaling}.}
    \label{tab:Pert-Anderson}
\end{table}

In Table \ref{tab:Pert-Anderson} we list most of the naturally arising perturbations to the Anderson model. We group these perturbations into three categories that will be discussed in detail in the following section:
\begin{enumerate}
    \item \textit{Local generators} break the ground-space degeneracy and violate time-reversal symmetry, but retain the NFL, smoothly transition the system along a line of fixed points to an $\mathrm{SU}(k)_2$ NFL. These include a uniform magnetic field. A combination of generators (e.g., a magnetic field at a single dot), however, can be relevant  \cite{konigExactSolutionTopological2023} by generating complementaries (see below) via RG.
    \item \textit{Local complementaries} break the ground-space degeneracy while retaining time-reversal symmetry and destabilize the NFL  either to a FL or to a different NFL. These encompass all channel-symmetry-breaking perturbations. They couple to the $\Phi_2$ primary field of scaling dimension $h_2{=}k/(k{+}2)$ (see Sec.~\ref{sec:CFT}). Thus, a perturbation of magnitude $\delta$ breaks the NFL below $T^*{\sim}\delta^{1/(1-h_2)}{=}\delta^{1+k/2}$ (or $\delta^{2+k}$ if $\delta^2$ couples to $\Phi_2$).
    \item \textit{Complementary couplings} do not break the  ground-state degeneracy, and include the relevant  particle-hole asymmetry, as well as irrelevant perturbations such as spin-dependent tunneling. By lifting the degeneracy of the excited states, particle-hole asymmetry takes the system to an $\mathrm{SU}(2k)_1$ FL. It couples to an operator with scaling dimension $h'_1{=}2/(k{+}2)$ \cite{liTopologicalSymplecticKondo2023}, thereby giving rise to the crossover scale $T_\mathrm{FL}\sim\delta^{1/(1-h'_1)}=\delta^{1+2/k}$.
\end{enumerate}
While the first and third categories were discussed in previous papers, the significance of the second category, i.e., the complementaries, and in particular of channel-symmetry breaking, was overlooked. In Sec.~\ref{sec:SchriefferWolff} we will also discuss the effect of interdot tunneling, which under a symplectic rotation falls into either the first or second categories.

In the Anderson picture, the effect of channel asymmetry has a clear cartoon description. Lowering the energy of the ground states associated with a given dot gaps out all other dots. As shown in Sec.~\ref{sec:SchriefferWolff}, this can be obtained either by explicitly reducing the potential energy $\epsilon_j$ of dot $j$, or by increasing its SC pairing $\Delta_j$. Similarly, having a stronger tunneling $t_j$ to a specific dot means that it has a lower energy after screening. Thus, this also gaps out the rest of the dots. For $k{=}2$ channels, these will always lead to a Fermi liquid (FL) at low energies, with single-channel Kondo screening ($\pi/2$ phase shift) in one channel and the other decoupling (no phase shift). This is very similar to the effect of channel asymmetry in the couplings of the two-channel $\mathrm{SU}(2)$ Kondo model. For $k{>}2$ we can simultaneously lower the energy or increase the coupling of $m{>}1$ dots. But, from our cartoon picture, the generalization is immediate -- we will get an effective $\mathrm{Sp}(2m)$ Kondo model.

\begin{figure*}[t]
    \centering
    \includegraphics[width=1\textwidth]{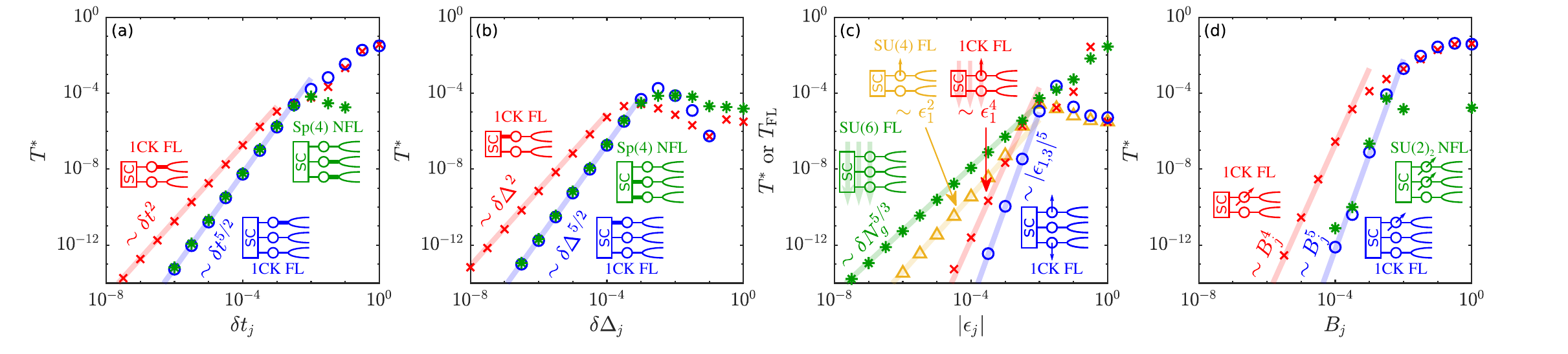}
    \vspace{-2\baselineskip}\phantomsubfloat{\label{fig:NRG-delta-t}}\phantomsubfloat{\label{fig:NRG-delta-Delta}}\phantomsubfloat{\label{fig:NRG-eps}}\phantomsubfloat{\label{fig:NRG-B}}
    \caption{NFL breaking energy scale as a function of the perturbation magnitude for a variety of channel-symmetry-breaking perturbations. These are applied to the isotropic and particle-hole symmetric $\mathrm{Sp}(2k)$ Anderson model with $k{=}2$ (\textcolor{red}{$\boldsymbol{\times}$},\textcolor[rgb]{.929,.694,.125}{$\boldsymbol{\triangle}$}) and $3$ (\textcolor{blue}{$\boldsymbol{\circ}$}, \textcolor[rgb]{0,.6,0}{$\boldsymbol{*}$}) channels. Each perturbation is specified by a cartoon with the resulting low-energy fixed point indicated next to it. The expected scaling for each perturbation (see Table \ref{tab:Pert-Anderson}) is plotted as a guide to the eye, with the corresponding power law appearing beside it. From left to right we have: (a) Tunneling asymmetry with either one or two strongly-coupled leads; (b) SC-pairing asymmetry with either one or two strongly-coupled dots; (c) Dot-dependent shifts in the energy level that for $k{=}2$ break particle-hole symmetry (\textcolor[rgb]{.929,.694,.125}{$\boldsymbol{\triangle}$}) but can be corrected with appropriate gate detuning (\textcolor{red}{$\boldsymbol{\times}$}). For $k{=}3$ the shifts are taken with opposite signs at two dots (\textcolor{blue}{$\boldsymbol{\circ}$}) so as not to break particle-hole symmetry, while the latter is explicitly broken by gate detuning (\textcolor[rgb]{0,.6,0}{$\boldsymbol{*}$}); (d) A magnetic field applied to either one or two dots. All calculations are carried out with isotropic model parameters $E_C/D=0.2, \Delta/D=t/D=0.1$, where $D$ is a sharp high-energy cutoff of the bath density of states, and with NRG parameters $\Lambda=4, z=0, N^*_\mathrm{keep}=2000$ (see Appendix \ref{sec:NRG}).}
    \label{fig:NRG-scaling}
\end{figure*}

The different fixed points are confirmed by comparing the NRG finite-size spectrum with CFT predictions (see Appendix \ref{sec:NRG}). In particular, for $k{=}3$ we get either a FL or an effective $\mathrm{Sp}(4)$ effect below $T^*$, depending on whether the perturbation lowers the energy of one or two dots, respectively. For a detailed discussion of the possible fixed points see Sec.~\ref{sec:CFT}. The energy scale at which the NFL breaks down ($T^*$ or $T_\mathrm{FL}$) as a function of the perturbation magnitude is plotted in Fig.~\ref{fig:NRG-scaling}, along with the resulting fixed point. This scaling will be discussed in detail in the following sections.

\section{Main Results}\label{sec:Main}

The structure of the operator space acting on the symplectic ``spin''
in the defining representation plays a key role in the classification of perturbations.  Let us start with the familiar structure for  $\mathrm{SU}(2)$. There, the space of Hermitian operators acting on some spin-$s$ representation partitions into the irreducible representations of a product of two such spins.\footnote{Generally, the space of $n{\times}n$ Hermitian matrices acting on some $n$-dimensional irreducible representation is the product space of this representation with its dual representation, and is thus a (reducible) representation of the group. For $\mathrm{SU}(2)$, each irreducible representation is isomorphic to its dual.  Thus, the operator space decomposes into irreducible representations like the product space of the representation with itself, i.e., $s{\otimes}s$.} In the defining (spin-$1/2$) representation we have $\frac{1}{2}{\otimes}\frac{1}{2}=0\oplus 1$, corresponding to the identity $\mathbb{1}
$ (singlet) and the generators $S_{x,y,z}$ (triplet). Similarly, for $\mathrm{SU}(N)$ in the  defining representation, one has the identity and $N^2{-}1$ generators. However, for $\mathrm{SU}(2)$ spin-$s>\frac{1}{2}$ representations the operator space is larger, and we will refer to all other operators in it as complementaries. As an example, for a spin-$1$ representation $1\otimes 1=0\oplus 1\oplus 2$, so that besides the identity and the spin operators as before, we get complementary spin-$2$ operators such as $(S^z)^2$.

In contrast to $\mathrm{SU}(N)$, for $\mathrm{Sp}(2k)$, complementaries show up already in the space of operators acting on the $2k$-dimensional defining representation $(10)_k$.\footnote{We use the shorthand $(q_1q_2)_k$ to denote the tuple of $k$ integers $(q_1,q_2,0,\ldots,0)_k$, i.e., with $k{-}2$ trailing zeros, in compact Dynkin notation -- for details see Appendix \ref{sec:qlabels}.} Explicitly, the operator space decomposes into three irreducible representations \cite{georgiLieAlgebrasParticle2019} with the dimension of each indicated below it
\begin{equation}\label{eq:10x10}
\begin{array}{c@{\,}c@{\,}c@{\,}c@{\,}c@{\,}c@{\,}c@{\,}c@{\,}c}
    (10)_k &\otimes& (10)_k  &=& 
        (00)_k &\oplus& (20)_k &\oplus& (01)_k\\
        \scriptstyle
    2k & \scriptstyle\times& \scriptstyle 2k &\scriptstyle =& \scriptstyle \ \ \ \ \  1 \ \ \ \ \ \  &\scriptstyle +& \scriptstyle\ \  k(2k{+}1)\ \  &\scriptstyle+& \scriptstyle (k{-}1)(2k{+}1)
\end{array}.
\end{equation}
As for $\mathrm{SU}(2)$, we identify the scalar representation $(00)_k$ with the identity operator and the adjoint representation $(20)_k$ with the $k(2k{+}1)$ generators $T^A$. This leaves the $(01)_k$ representation that we identify with the $(k{-}1)(2k{+}1)$ complementaries $\mathcal{T^A}$. Together, $T^A$ and $\mathcal{T^A}$ span the full  $[(2k)^2{-}1]$-dimensional space of nontrivial $2k{\times}2k$ Hermitian matrices, and thus form a representation of the generators of $\mathrm{SU}(2k)$. To help visualize, we introduce a specific operator basis for $k{=}2$ (but note that the rest of the discussion is basis independent)
\begin{subequations}
\label{eq:Pauli-basis}
\begin{eqnarray}
    \label{eq:Pauli-basis-gen}
    \!\!\!\!T^A_{k=2}&:&\ 
    \tfrac{1}{2}(\sigma_\mu{\otimes}\mathbb{1}),\ 
    \tfrac{1}{2}(\sigma_\mu{\otimes}\tau_z),\ 
    \tfrac{1}{2}(\sigma_\mu{\otimes}\tau_x),\ 
    \tfrac{1}{2}(\mathbb{1}{\otimes}\tau_y),\  \\
    \label{eq:Pauli-basis-comp}
    \!\!\!\!\mathcal{T}^\mathcal{A}_{k=2}&:&\ 
    \tfrac{1}{2}(\mathbb{1}{\otimes}\tau_z),\  
    \tfrac{1}{2}(\mathbb{1}{\otimes}\tau_x),\ 
    \tfrac{1}{2}(\sigma_\mu{\otimes}\tau_y),
\end{eqnarray}
\end{subequations}
where $\sigma_\mu$ and $\tau_\mu$ are Pauli matrices with $\mu=x,y,z$, and $\mathbb{1}$ is the $2{\times}2$ identity. Thus, for $k{=}2$ we have in total 10 generators and 5 complementaries. One can verify that only $T^A$ satisfy the symplectic defining condition in Eq.~\eqref{eq:Sp2k-gen}, i.e., are odd under time reversal, while $\mathcal{T^A}$ are even. The key point of this paper is that, whereas the generators define the NFL fixed point of the symplectic Kondo effect, it is the complementaries $\mathcal{T^A}$ that destabilize it.

Let us briefly list a few mathematical properties of these two sets of operators, with a more comprehensive discussion in Appendix \ref{sec:Sp2k-properties}. All operators are taken to be traceless, $\mathrm{tr}
\{T^A\}{=}\mathrm{tr}\{\mathcal{T^A}\}{=}0$, and orthonormal, $\mathrm{tr}\{T^A T^B\}{=}\delta_{AB}$, $\mathrm{tr}\{\mathcal{T^A T^B}\}{=}\delta_\mathcal{AB}$, $\mathrm{tr}\{T^A\mathcal{T^B}\}{=}0$. 
By construction, the generators are closed under commutation, but in the symplectic case, under anticommutation they leave the generator  space. Explicitly, we write
\begin{subequations}\label{eq:struct-consts}
\begin{eqnarray}
    [T^A,T^B]\ \!  &=& if_{ABC}T^C, \\
    \{T^A,T^B\} &=& \tfrac{1}{k}\delta_{AB} \mathbb{1}_{2k} + d_{AB\mathcal{C}}\mathcal{T^C},\label{eq:struct-consts-anit-comm}
\end{eqnarray}
\end{subequations}
with $f_{ABC}$ the (basis-dependent) structure constants and $d_{AB\mathcal{C}}$ playing a similar role.  Products of generators can be decomposed as $T^A T^B{=}\frac{1}{2}[T^A,T^B]{+}\frac{1}{2}\{T^A,T^B\}$. Then, the observation that anticommutators yield only elements of the complementaries (and their superposition with the identity) will prove crucial in explaining how complementary terms are generated via RG.

It will also be useful to (partially) define a channel-polarized operator basis  
\begin{equation}\label{eq:chan-basis}
    T^{\mu j} \equiv \tfrac{1}{\sqrt{2}}(\sigma_\mu{\otimes}\Pi_j\mathrm),
    \ \ \mathcal{T}^j \equiv \sqrt{\tfrac{k}{2k-2}}\left(\mathbb{1}{\otimes}\Pi_j-\tfrac{1}{k}\mathbb{1}_{2k}\right),
\end{equation}
where $[\Pi_j]_{mn}\equiv\delta_{mj}\delta_{nj}$ are diagonal $k{\times}k$ matrices with 1 only at the $j$th diagonal element. The rest of the operators in this basis are explicitly specified in Eq.~\eqref{eq:chan-basis-full}. This basis demonstrates two important properties:
\begin{subequations}
\begin{eqnarray}
    \!\!\!\!\![T^{\mu j},T^{\nu j^\prime}] &=& i\delta_{jj^\prime}\varepsilon_{\mu\nu\eta}(\sigma_\eta{\otimes}\Pi_j)=\sqrt{2}i\delta_{jj^\prime}\varepsilon_{\mu\nu\eta}T^{\eta j}, \ \ \ \ \ \\
    \!\!\!\!\!\{T^{\mu j},T^{\nu j^\prime}\} &=& \delta_{jj^\prime}\delta_{\mu\nu}(\mathbb{1}{\otimes}\Pi_j) \propto \delta_{jj^\prime}\delta_{\mu\nu}(\mathcal{T}^j{+}\mathrm{const}),\ \ \ \ \ 
\end{eqnarray}
\end{subequations}
with $\varepsilon_{abc}$ the Levi-Civita symbol. The first property states that the generators form $k$ independent $\mathfrak{su}(2)$ subalgebras; the second states that they square (or anticommute) to channel projectors [in contrast to the generators in Eq.~\eqref{eq:Pauli-basis-gen} that all square to the identity]. As will be discussed later on, these two properties suffice to construct a cartoon picture of channel asymmetry.

In the following subsections, the $T^A$ and $\mathcal{T^A}$ will be promoted to current and impurity operators
\begin{subequations}\label{eq:gen-ops}
\begin{eqnarray}
    J^A(x)&=\textbf{c}^\dagger(x)T^A \textbf{c}(x),\quad &S^A=\textbf{f}^\dagger T^A \textbf{f}, \label{eq:gen-ops-gen} \\
    \mathcal{J^A}(x)&=\textbf{c}^\dagger(x)\mathcal{T^A}\textbf{c}(x),\quad &\mathcal{S^A}=\textbf{f}^\dagger \mathcal{T^A}\textbf{f}, \label{eq:gen-ops-comp}
\end{eqnarray}
\end{subequations}
implicitly assuming that the auxiliary fermions satisfy $\textbf{f}^\dagger\textbf{f}=1$.\footnote{From a symmetry point of view, the set of $\xi=1,\ldots,2k$ operators in ${\bf f}^\dagger \equiv ( f_\xi)$ transform as the Sp($2k$) defining representation $(10)_k$. The set of $(2k)^2$ bilinear operators in $( f_\xi^\dagger f_{\xi'})$ splits into two non-trivial sets of operators, contrary to the well-known case of $\mathrm{SU}(N)$ which only gives the spin operator (generators). To be specific, in the symplectic case for $k{=}2$ were $(10){\otimes}(10)=(00){\oplus}(20){\oplus}(01)$, aside from the identity $(00)$, this contains the ``spin'' operator $S$ in the $10$-dimensional adjoint representation $(20)$, but also the complementary $\mathcal{S}$ in the $5$-dimensional representation $(01)$, such that $4{\times}4=1{+}10{+}5$. The same argument holds for the $\textbf{d}$ operators of the Anderson model.} For conciseness, however,  we will often drop the auxiliary impurity fermionic operators, and simply write $S^A\simeq T^A$ and $\mathcal{S^A}\simeq\mathcal{T^A}$. As will be discussed in Sec.~\ref{sec:CFT}, the complementary operators correspond to the $\Phi_2$  primary field with scaling dimension $k/(k{+}2)$. Thus, they will be related to relevant perturbations. The generator operators, on the other hand, correspond to descendant fields, and will thus be marginal.

\subsection{Local Complementaries}\label{sec:Pert-Complementaries}

The majority of physical perturbations break channel symmetry. This comes in a variety of forms, but as we will see, they all come down to the same principle: They generate complementary terms $\mathcal{S^A}$ that act on the impurity. In this section we focus on perturbations that break channel symmetry but conserve time-reversal symmetry without touching the spin sector. These include a (nonuniform) shift in the potential energy $\epsilon_j$ of the dots, dot-dependent SC pairing $\Delta_j$, and channel-dependent tunneling  $t_j$ (see Table \ref{tab:Pert-Anderson} and Fig.~\ref{fig:NRG-scaling}). As will be shown in Sec.~\ref{sec:SchriefferWolff}, such perturbations map to Kondo perturbations of two types:
\begin{equation}\label{eq:H-pert-comp}
    \delta H_\mathrm{comp}=\sum_\mathcal{C} \mathcal{B_C}\mathcal{S^C} + \sum_{AB}\delta\lambda_{AB} S^A J^B(0),
\end{equation}
with $\mathcal{B^C}$  and $\delta\lambda_{AB}$ real coefficients, and $\mathcal{S^A}$  the complementary terms defined in Eq.~\eqref{eq:gen-ops}. Further assuming symmetric $\delta\lambda_{AB}=\delta\lambda_{BA}$ and noting that the complementaries can be written as anticommutators of the generators $S^A$ as in Eq.~\eqref{eq:struct-consts-anit-comm} highlights that the two terms in Eq.~\eqref{eq:H-pert-comp} transform in the same manner under a symplectic rotation, and are both invariant under time reversal.

Poor man's scaling \cite{andersonPoorManDerivation1970,cox_zawadowski_1998,konikInterplayScalingLimit2002} RG equations are derived in Appendix \ref{sec:poor-mans}, and explicitly demonstrate that these two terms are inherently interconnected. For simplicity, let us assume a uniform local bath density of states $\rho$ in the range $[-D,+D]$ and define dimensionless coefficients $\tilde{\lambda}_{AB}=\rho(\lambda{+}\delta\lambda_{AB})$ and $\tilde{\mathcal{B}}_\mathcal{C}=\mathcal{B_C}/D$. Then, integrating out a small slice of the bandwidth at either end, i.e.,  $D\to e^{-\ell} D$, yields to leading order
\begin{subequations}\label{eq:RGeqMaintext}
\begin{eqnarray}
    \frac{d \tilde \lambda_{AB}}{d \ell} &=& \frac{1}{2} f_{ACE} f_{BDF} \tilde \lambda_{CD} \tilde \lambda_{EF}, \label{eq:RG1} \\
    \frac{d \tilde{\mathcal B}_{\mathcal C}}{d \ell} &=& \tilde{\mathcal B}_{\mathcal C} -  \frac{1}{2} d_{AB\mathcal C }\tilde \lambda_{AD} \tilde \lambda_{BD}, \label{eq:RG2}
\end{eqnarray}
\end{subequations}
with the structure constants defined in Eq.~\eqref{eq:struct-consts} and implicit summation on repeated indices. Thus, we see that, generically, the second term in Eq.~\eqref{eq:H-pert-comp} can generate the first.\footnote{This is already eluded to in Ref.~\cite{konikInterplayScalingLimit2002}. In Appendix \ref{sec:poor-mans} we derive the RG equations to the leading order $\mathcal O(\tilde \lambda_{AB}^2,\mathcal{\tilde B}_{\mathcal A})$, but to higher orders we expect the first term in Eq.~\eqref{eq:H-pert-comp} to affect the flow of the second, i.e., for $\tilde{\mathcal B}_\mathcal{C}$ to enter in Eq.~\eqref{eq:RG1}. This is demonstrated for a specific case in Ref.~\cite{zitkoPropertiesAnisotropicMagnetic2008}.} In contrast to the complementaries, generator terms $B_A S^A$ are not generated by $\delta \lambda_{AB}$ within the present RG scheme.  Starting from $\mathcal{B_C}{=}0$, the complementary coefficients  $\mathcal{B_C}$ will be linear in a weak anisotropy  $\delta\lambda_{AB}\ll\lambda$, and will not be generated in the isotropic case $\delta\lambda_{AB}=0$ (i.e., $\tilde\lambda _{AB}=\delta_{AB}\tilde\lambda$ -- see Appendix \ref{sec:poor-mans-disc}). At the quantum-critical fixed point, the complementary terms $\mathcal{S^C}$ relate to operators with a scaling dimension of $k/(k{+}2)$. Thus, we expect the NFL to break down below $T^*\propto \delta\lambda_{AB}^{1+k/2},\mathcal{B_C}^{1+k/2}$.

We highlight that it is the anticommutators of generators that yield the complementary terms, as evident from the $d_{AB\mathcal C}$ coefficient in Eq.~\eqref{eq:RG2}. These can be understood as bath-mediated interactions of the impurity with itself, in analogy to RKKY \cite{rudermanIndirectExchangeCoupling1954,kasuyaTheoryMetallicFerro1956,yosidaMagneticPropertiesCuMn1957} interimpurity interactions generated in multi-impurity problems. The only difference is that in the latter, $S^B$ and $S^C$  act on different impurities. This behavior is generic, although in many typical cases, such as with an $\mathrm{SU}(2)$ spin-$1/2$ impurity, all generators anticommute trivially, (i.e., to zero or to the identity). However, for a spin-$1$ impurity $\mathrm{SU}(2)$ Kondo model, this effect is important, and uni-axial anisotropy in the couplings $J_z\neq J_\perp$ was shown to generate relevant $(S^z)^2$  impurity terms \cite{konikInterplayScalingLimit2002,schillerPhaseDiagramAnisotropic2008,zitkoPropertiesAnisotropicMagnetic2008}.

Note that the structure constants $f_{ABC}$ are nonzero only for very specific combinations. This imposes strong constraints on the generation of off-diagonal elements when starting from a diagonal $\lambda_{AB}=\delta_{AB}\lambda_A$. Thus, although in the most general case off-diagonal terms might emerge, we develop intuition by focusing on diagonal couplings $\sum_A \delta\lambda_A S^A J^A(0)$. Then, the question of relevance simplifies to: have we perturbed in a direction for which $\{S^A,S^A\}$ is not proportional to the identity? Observe that for $k{=}2$, if one works in the Pauli generator basis of Eq.~\eqref{eq:Pauli-basis-gen}, they all happen to anticommute to the identity. This might be the reason that this effect was previously overlooked. To counter this, consider a crude, but intuitive, cartoon picture based on the channel-polarized generator basis $T^{\mu j}=\tfrac{1}{\sqrt{2}}(\sigma_\mu{\otimes}\Pi_j)$ defined in Eq.~\eqref{eq:chan-basis}. In the very anisotropic limit, where only dot $j$ is coupled to the bath, the Kondo term becomes
\begin{equation}\label{eq:H-channel-1}
    H_K +\delta H_\mathrm{comp} = \lambda_j\!\! \sum_{\mu=x,y,z}\!\! S^{\mu j}J^{\mu j}(0),
\end{equation}
with $S^{\mu j}$ and $J^{\mu j}(x)$ defined according to Eq.~\eqref{eq:gen-ops}. Thus, in the $j$th channel we have a single-channel $\mathrm{SU}(2)$ Kondo effect, while all other channels decouple. We also have decoupled impurity states, but these are gapped out because the Kondo effect lowers the energy of the participating dot $j$ states. From the perturbative RG equations we see that we have generated the local term
\begin{equation}\label{eq:acomm-Smuj-Smuj}
    \{S^{\mu j},S^{\mu j}\} = \mathbb{1}{\otimes}\Pi_j = \sqrt{\tfrac{2k-2}{k}}\mathcal{S}^j + \mathrm{const},
\end{equation} 
with a negative coefficient. Subtracting the constant global shift in the energy, it lies in the complementary subspace $\mathcal{S^A}$ [specifically $\mathcal{S}^1=\tfrac{1}{2}(\mathbb{1}{\otimes}\tau_z)$ for $k{=}2$]. We can then reintroduce similar couplings in the rest of the channels. As long as there is a single maximal coupling, the low-energy fixed point remains the same. Degeneracies between the couplings, however, will yield a variety of fixed points that are unstable to the introduction of the remaining (non-channel-polarized) generators, and are thus not the focus of this paper.

Alternatively, we can start from the channel-symmetric model in Eq.~\eqref{eq:H-Kondo} and perturb in the direction of the $j$th channel according to Eq.~\eqref{eq:H-pert-comp}, either with coupling terms  $\delta\lambda_{ja,ja}$ or local terms $\mathcal{B}_j$. We argue that the cartoon picture still captures the key features of such perturbations. In Sec.~\ref{sec:SchriefferWolff}  we show how such terms are generated from perturbations to the symmetric Anderson model of Eq.~\eqref{eq:H-Anderson}.  As already noted in passing \cite{konigExactSolutionTopological2023},  asymmetric SC pairing generates local terms with  $\mathcal{B}_j\propto\delta\Delta_j$. However, these terms are also generated by asymmetry in the dot levels and in the dot-lead tunneling, with $\mathcal{B}_j\propto\epsilon_j^2$ and  $\mathcal{B}_j\propto\delta t_j$, respectively. Note that asymmetric dot levels can also break particle-hole symmetry due to an average shift of the energy levels, but this can be negated by properly retuning the gate $N_g$ (see Secs.~\ref{sec:Pert-Couplings} and \ref{sec:SW-PH-asym}). In specific limits the local contribution of tunneling asymmetry vanishes \cite{liTopologicalSymplecticKondo2023}, but coupling asymmetry  $\delta\lambda_{ja,ja}\propto\delta t_j$ is always generated. Thus, the local term will emerge via RG to the same order. Applying NRG to the Anderson model with $k{=}2$ and $3$ channels (see details in Appendix \ref{sec:NRG}), we demonstrate the expected scaling behavior in Fig.~\ref{fig:NRG-scaling}.  Plotting the energy scale $T^*$ at which the NFL breaks down as a function of the perturbation magnitude, we indeed observe it to scale as $\delta t_j^{1+k/2}$, $\delta\Delta_j^{1+k/2}$,  and $\epsilon_j^{2+k}$.

The focus on channel-polarized generators and complementaries so far might lead to the wrong impression that these effects are inherently related to channel asymmetry. But they actually only depend on the emergence of complementaries. Section \ref{sec:EK} focuses on the exact EK solution for the $k{=}2$. There, it is convenient to perturb in the direction of specific Cartan generators, chosen to be $S^{z0}\equiv \frac{1}{2}(\sigma_z{\otimes}\mathbb{1})$ and $S^{0y}\equiv\frac{1}{2}(\mathbb{1}{\otimes}\tau_y)$. The main results are summarized in Table \ref{tab:EK}.\footnote{Note that for the derivation of the free-fermionic theory in Sec.~\ref{sec:EK} we temporarily switch notation $S^{z0}\to\tilde{S}^{z0}$  and $S^{0y}\to-\tilde{S}^{0z}$ to make contact with previous studies. But, in Table \ref{tab:EK} we return to the usual notation employed throughout this paper.} We observe that the complementary $\mathcal{S}^{zy}\sim\{S^{z0},S^{0y}\}$ is indeed relevant. We also find that perturbing along $S^{z0}J^{z0}(0)$ or $ S^{0y} J^{0y}(0)$ is irrelevant, while perturbing along the cross terms $S^{0y}J^{z0}(0)$ and $S^{z0} J^{0y}(0)$ directions (which is expected to generate $\mathcal{S}^{zy}$) is relevant. In Sec.~\ref{sec:CFT} we use CFT to generalize this to arbitrary $k$. We argue that the complementaries relate to the $\Phi_2$ primary field that has a scaling dimension $h_2 = k/(k{+}2)$. Thus, it results in a relevant perturbation (of magnitude $\delta$) that breaks the NFL below $T^*\sim\delta^{1/(1-h_2)} = \delta^{1+k/2}$. This is in perfect agreement with the scaling obtained via NRG for $k{=}2$ and $3$, as demonstrated in Fig.~\ref{fig:NRG-scaling}.

\subsection{Local Generators}\label{sec:Pert-Generators}
We now turn to consider local impurity perturbations in the direction of the symplectic ``spin'' operators $S^A$ (i.e., generators)
\begin{equation}\label{eq:H-pert-gen}
    \delta H_\mathrm{gen} = \sum_A B_A S^A.
\end{equation}
These explicitly violate time-reversal symmetry, e.g., due to an external magnetic field. In $\mathrm{SU}(2)$ Kondo models, perturbations of this form (i.e., a magnetic field) break the Kondo effect, and thus destabilize any associated NFL. In the symplectic case, on the other hand, the generators correspond to marginal operators (see Sec.~\ref{sec:CFT}).  Thus, despite the fact that like the complementaries $\mathcal{S^A}$, any generator $S^A$ lifts the $2k$-fold degeneracy of the symplectic ``spin'' states, in some cases  $\delta H_\mathrm{gen}$ does not destabilize the NFL. Specifically, these are the directions in the generator space for which  $(\delta H_\mathrm{gen})^2$ is proportional to the identity operator. Generator perturbations in any other direction are relevant due to the emergence of relevant complementary terms via RG.

Let us start with the effect of a uniform magnetic field, e.g., in the $z$ direction. Such a field couples to the generator $S^z\equiv\frac{1}{\sqrt{2k}}(\sigma_z{\otimes}\mathbb{1}_k)$  and does not break the NFL ($S^z$ squares to the identity). In the Anderson picture, such a field gaps out half the states, leaving a $k$-fold degenerate spin-polarized ground space. Under a Schrieffer-Wolff transformation, these correspond to an $\mathrm{SU}(k)$ degree of freedom. Interestingly, this degree of freedom still couples to both spin species in the lead: a singly occupied (spin-polarized) dot can transition to the BCS state either by the fermion tunneling out to the lead with the same spin, or an opposite-spin fermion tunneling in from the opposite-spin lead. Thus, we get an effective two-channel [$\mathrm{SU}(k)_2$] model \cite{konigExactSolutionTopological2023}, and the resulting NFL has the same fractional impurity entropy as $\mathrm{Sp}(2k)_1$. 

\begin{figure}[t]
    \centering
    \includegraphics[width=1\columnwidth]{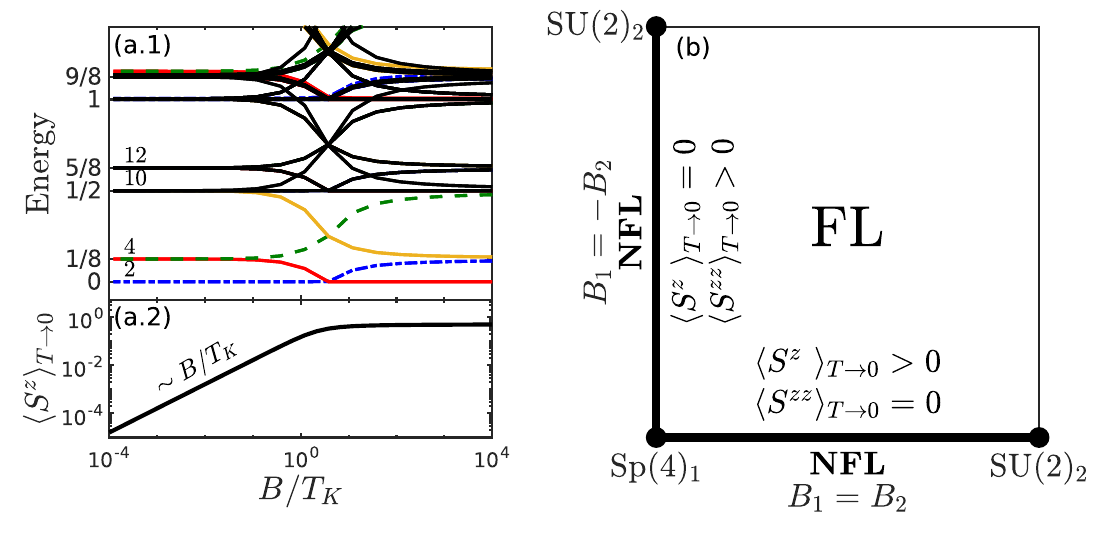}
    \vspace{-3\baselineskip}\phantomsubfloat{\label{fig:SU2-2-FSS}}\phantomsubfloat{\label{fig:Sp4-SU2-2}}
    \caption{(a.1) Low-energy finite-size spectrum along the line of fixed points generated by varying the magnitude of a uniform magnetic field. The energy is given in rescaled units, with the degeneracy indicated next to each level. The lowest lying levels are plotted in color in order to emphasize their flow. Note that each color corresponds to a doubly degenerate level.  (a.2) The impurity magnetization $\braket{S^z}_{T\to 0}$ as a function of a uniform magnetic field. (b) Fixed points and lines of the $\mathrm{Sp}(4)$ Kondo model in the presence of magnetic fields.}
    \label{fig:Generators}
\end{figure}

A clear indication that the low-energy fixed point has changed, however, is that we now have a finite impurity magnetization $\braket{S^z}_{T\to 0}$. Numerically we find a line of NFL fixed points stretching from the $\mathrm{Sp}(2k)_1$ fixed point at one end to the $\mathrm{SU}(k)_2$ fixed point at the other. This is demonstrated in Fig.~\ref{fig:SU2-2-FSS} for $k{=}2$. For small magnetic fields we have an $\mathrm{Sp}(4)$-like spectrum that, around the Kondo temperature,\footnote{Here, the Kondo temperature is defined as  the midpoint between the free (unscreened) and over-screened symplectic ``spin'', i.e., for which $\mathrm{S_{imp}}(T_K)=\frac{1}{2}(2\ln 2{-}\frac{1}{2}\ln 2)$ at $B{=}0$.} crosses over to an $\mathrm{SU}(2)_2$-like spectrum at large fields. While the energy levels at the two ends are the same, the quantum numbers assigned to each level differ. This is what enables the emergence of a finite impurity magnetization.

The same argument holds for any generator arising due to equal magnitude but arbitrary sign of magnetic fields at all of the impurities [for $k{=}2$ think of $S^{zz}\equiv\frac{1}{2}(\sigma_z{\otimes}\tau_z)$, that also squares to the identity]. We will then get other lines of NFL fixed points, culminating in $\mathrm{SU}(k)_2$ fixed points, as depicted in Fig.~\ref{fig:Sp4-SU2-2} for $k{=}2$. While the quantum numbers associated with the levels at the two  $\mathrm{SU}(2)_2$ fixed points are different, they can smoothly be deformed from one to the other, e.g., by rotating the magnetic field at the second impurity. Thus, these two NFL lines are actually part of a large NFL manifold. Crucially, however, such a deformation requires introducing more generators and lifting out of the two-dimensional plane in Fig.~\ref{fig:Sp4-SU2-2}. 

For any perturbation on the plane (but off the lines), $(\delta H_\mathrm{gen})^2$ will not be proportional to the identity (because $S^z$ and $S^{zz}$ do not anticommute), which will destabilize the NFL. Consider for example a magnetic field $B_1$ that couples to the generator $S^{z1}=\frac{1}{\sqrt{2}}(S^{z}{+}S^{zz})$. Via RG it will generate its square
\begin{equation}
    (S^{z1})^2 =\tfrac{1}{2}
    \underbrace{\{S^z,S^{zz}\}}_{= \mathcal{S}^1} + \tfrac{1}{4},
\end{equation}
and thus the relevant complementary $\mathcal{S}^1$ [see Eq.~\eqref{eq:chan-basis}] that takes the system to a FL. Another way to think of the instability of the NFL is in terms of a two-stage gapping out process \cite{konigExactSolutionTopological2023}: First, we perturb in the direction of $S^z$ which takes us from $\mathrm{Sp}(4)_1$ to $\mathrm{SU}(2)_2$ along a specific NFL line. Then we perturb in the direction of $S^{zz}$. But, in contrast to the $\mathrm{Sp}(4)_1$ fixed point, this line is not stable to perturbations in the $S^{zz}$ direction. We stress that this argument holds for any two generators that square to the identity, such as those in Eq.~\eqref{eq:Pauli-basis-gen}. In the EK picture we indeed observe that perturbing in the direction of either of the Cartan generators $S^{z0}\equiv S^z$ and $S^{0y}$ does not break the NFL (see first two rows of Table \ref{tab:EK}). However, perturbing in an intermediate direction, e.g., $S^{z0}{+}S^{0y}$ (that would generate the complementary $\mathcal{S}^{zy}$), is found to be relevant (see third row of Table \ref{tab:EK}).

The generalization to larger $k$ is not complicated. Let us demonstrate this point with an example. A magnetic field $B_j$ at dot $j$ couples to the channel-polarized generator $S^{zj}$ defined in Eq.~\eqref{eq:chan-basis}. It squares according to Eq.~\eqref{eq:acomm-Smuj-Smuj}, yielding the complementary $\mathcal{S}^j$.\footnote{As for $k{=}2$, this can also be described as a two stage process. One can write $S^{zj}=\frac{\sqrt{k}}{2}(S^z{+}\bar S^{z  j})$ with the square of both  $S^z=\frac{1}{\sqrt{k}}\sum_j S^{zj}$ and $\bar S^{zj}\equiv \frac{2}{\sqrt{k}}S^{zj}{-}S^z$ proportional to the identity. Thus, each takes us along an NFL line, but then the other deviates from it. Note that by construction $\{S^z,\bar S ^{zj}\}=\frac{4}{k}(S^{zj})^2{-}\frac{1}{k}\propto \mathcal{S}^j$.} Because of time-reversal symmetry, the prefactor of $\mathcal{S}^j$ must be an even power of $B_j$ (in the simplest case $B_j^2$). Thus, the $\mathrm{SU}(k)_2$ NFL will break down below $T^*\sim B_j^{2+k}$, as demonstrated in Fig.~\ref{fig:NRG-B} for $k{=}2$ and $3$. For a magnetic field at a single dot, the system would always flow to a FL. But if we apply a magnetic field at $1{<}m{\leq}k$ dots, we would get an $\mathrm{SU}(m)_2$ fixed point (or line) \cite{konigExactSolutionTopological2023}. This has the same impurity entropy as the  $\mathrm{Sp}(2m)_1$ fixed point, and emerges for the same reason: We have generated $m$ complementary terms, similar to the effect of simultaneously lowering the energy of $m$ dots. For further discussion see Sec.~\ref{sec:CFT}. Thus, we have established the relevance of any local generator perturbation in Eq.~\eqref{eq:H-pert-gen} that does not square to the identity.

\subsection{Complementary Couplings}\label{sec:Pert-Couplings}
Having covered all possible impurity perturbations in Secs.~\ref{sec:Pert-Complementaries} and \ref{sec:Pert-Generators}, we now turn to discuss coupling perturbations. We have already discussed perturbations involving generator terms at both the impurity and the lead, and now consider the complementaries.

\subsubsection{Particle-Hole Asymmetry}
In Sec.~\ref{sec:SW-PH-asym} we discuss two mechanisms that break particle-hole symmetry: gate detuning, $N_g\to 1{+}\delta N_g$, and an average shift of the dot levels, $\bar\epsilon=\tfrac{1}{k}\sum_j\epsilon_j$. Both do not affect the $2k$-fold degeneracy of the impurity subspace, but break the degeneracy of transitions to excited states with a missing or extra fermion. Importantly, they do not break the symplectic symmetry, and thus can only generate symplectic scalars. Specifically, we find them to generate [see Eq.~\eqref{eq:phasym_heff}]
\begin{equation}\label{eq:H-pert-ph}
    \delta H_\mathrm{ph} = \Lambda\sum_\mathcal{A}\mathcal{S^A}\mathcal{J^A}(0),
\end{equation}
with $\mathcal{S^A}$ and $\mathcal{J^A}(x)$ the complementaries to the generators as defined in Eq.~\eqref{eq:gen-ops}. As $T^A$  and $\mathcal{T^A}$ together form the set of generators of $\mathrm{SU}(2k)$, this is a specific anisotropic case of the single-channel $\mathrm{SU}(2k)$ Kondo model. Thus, the system flows to the isotropic fully-screened $\mathrm{SU}(2k)_1$ Fermi liquid \cite{liTopologicalSymplecticKondo2023}. This fixed point differs from the one to which the system flows as a result of channel asymmetry. In the latter, a single spin-$1/2$ channel screens the impurity degrees of freedom and acquires a $\pi/2$ phase shift. Here, on the other hand, all channels participate, and the total $\pi$ phase shift is equally distributed between the $2k$ spin and channel species.

As $\Lambda$ couples to an operator with scaling dimension $2/(k{+}2)$ [see Sec.~\ref{sec:CFT}] and is linear in a weak perturbation [see Eqs.~\eqref{eq:Lam-dNg} and \eqref{eq:Lam-barEps}], the NFL breaks below $T_\mathrm{FL}\propto \delta N_g^{1+2/k},\bar{\epsilon}^{1+2/k}$.  This scaling is confirmed in the EK picture for the specific term $\mathcal{S}^{zy}\mathcal{J}^{zy}(0)$, and by NRG in Fig.~\ref{fig:NRG-eps} for $k{=}2,3$. In Sec.~\ref{sec:SW-PH-asym} we will explicitly consider the effect of a uniform shift $\epsilon_j{=}\bar\epsilon$ and demonstrate it can be canceled by appropriately tuning  $\delta N_g$. Empirically, we make a stronger statement: Even for a nonuniform shift, tuning $\delta N_g$ can cancel the average shift $\bar\epsilon$, thereby effectively restoring particle-hole symmetry (red crosses). Moreover, choosing  $\epsilon_j$ to cancel each other out also prevents the generation of $\Lambda$  (blue circles). Then, nonuniformity of $\epsilon_j$, or channel asymmetry, enters to the next order and breaks the symplectic symmetry as discussed in Sec.~\ref{sec:Pert-Complementaries}.

\subsubsection{Spin-Dependent Tunneling}
As shown in Sec.~\ref{sec:SW-Spin-Tun}, a spin dependence of the lead-dot tunneling amplitudes generates terms of the form
\begin{equation}\label{eq:H-pert-gen-comp}
    \delta H_\text{gen-comp}=\sum_{A\mathcal{B}} \mathit{\Lambda}_{A\mathcal{B}}S^A \mathcal{J^B}(0),
\end{equation}
as well as generator-generator couplings $\lambda_{AB}$ as in Eq.~\eqref{eq:H-pert-gen}. From the EK analysis one can show that for $k{=}2$, any term of the form of Eq.~\eqref{eq:H-pert-gen-comp} are irrelevant. We further note that tunneling imposes strong constraints on both $\lambda_{AB}$ and $\mathit{\Lambda}_{A\mathcal{B}}$ [see Eq.~\eqref{eq:SW_spin_tun}], and for $k{=}2,3$ observe numerically that such perturbations are irrelevant. The question if generally terms of the form of Eq.~\eqref{eq:H-pert-gen-comp} are irrelevant is left for future work.

\section{Analysis of the \texorpdfstring{$k{=}2$}{k=2} case Using Emery-Kivelson Fermions } \label{sec:EK}

We now turn to bolster our perturbative an numerical arguments with and exact solution. The method developed by Emery and Kivelson in their famous paper on the two-channel Kondo model~\cite{emeryMappingTwochannelKondo1992} can be applied to the $k{=}2$ case~\cite{renTopologicalQuantumComputation2024}. Instead of studying the problem of stability for the ideal $\mathrm{Sp}(4)$ symmetry, we consider a strong anisotropy of the couplings of the Kac-Moody currents $H_K\to\sum_A\lambda_A S^A J^A(0)$, corresponding to Cartan generators in a specific basis. At this point, the Kondo model is found to be equivalent to a model of free fermions, which drastically simplifies the problem of stability.  Since we are interested in exploring a large list of possible perturbations, as was recently done for the equivalent $\mathrm{SO}(5)$ Kondo model~\cite{mitchellNonFermiLiquidCoulomb2021,libermanCriticalPointSpinflavor2021}, we apply the bosonization notations of the latter.

We choose to work with the generator basis $T_{k=2}^A$ of Eq.~\eqref{eq:Pauli-basis-gen}, labeling the spin operators with $\sigma,\tau=0,x,y,z$, and differentiating by font between those corresponding to generators $S^{\sigma\tau}$ (Roman) and complementaries $\mathcal{S}^{\sigma\tau}$ (calligraphic). We select two Cartan (mutually commuting) generators
\begin{subequations}
\begin{eqnarray}
    S^{z0} &\equiv& \tfrac{1}{2}(\sigma_z{\otimes}\mathbb{1}) 
        = +\tilde{S}^{z0} = d^\dagger d-\tfrac{1}{2}= +id_+d_-,\ \  \\
    S^{0y} &\equiv&\tfrac{1}{2}(\mathbb{1}{\otimes}\tau_y) 
        = -\tilde{S}^{0z} =\tfrac{1}{2} -a^\dagger a= -ia_+a_-.\ \ 
\end{eqnarray}
\end{subequations}
The tilde corresponds to a change of basis $\tau_y\to\tau_z$ that will be specified explicitly below, and $a_\pm,d_\pm$ are four Majorana fermions that span the $4$-dimensional impurity space and will be introduced in Eq.~\eqref{eq:EK-Majoranas}. The lead degrees of freedom will be represented by 8 Majorana fields. All but one will effectively decouple from the impurity, resulting in the Hamiltonian $H_{\mathfrak{sp}(4)}$ in Eq.~\eqref{eq:H-EK-sp4}. Importantly, the impurity Majorana $a_+$ also decouples, resulting in the $\ln \sqrt{2}$ residual entropy. Perturbations that leave it decoupled will not break the NFL. We now turn to derive $H_{\mathfrak{sp}(4)}$. A reader interested only in the effect of perturbations can jump ahead to Sec.~\ref{sec:EK-perturbations}.

\subsection{Deriving the Free-Fermion Hamiltonian}
Upon replacing $\tau_y$ and $\tau_z$, the generators coincide with the 10 generators of $\mathrm{SO}(5)$. We implement this replacement by defining $\textbf{c}'(x)=e^{-i \frac{\pi}{4} (\mathbb{1}{\otimes}\tau_x)} \textbf{c}(x)$ for the conduction electrons, and $\tilde{S}^{\sigma\tau}= 
e^{-i \frac{\pi}{2} \mathcal{S}^{0x} } S^{\sigma\tau}  e^{i \frac{\pi}{2} \mathcal{S}^{0x}}$ for the impurity operators (similarly for the complementaries $\mathcal{S}^{\sigma\tau}$). Explicitly, we get, $\tilde{S}^{\sigma 0}=S^{\sigma0}$, $\tilde{S}^{0z}=-S^{0y}$, $\tilde{\mathcal{S}}^{0y}=\mathcal{S}^{0z}$, and $\tilde{\mathcal{S}}^{0x}=\mathcal{S}^{0x}$.

We then bosonize the conduction electrons using the conventions of Refs.~\cite{zarandAnalyticalCalculationFinitesize2000,vondelftFiniteSizeBosonization2Channel1998},
\begin{equation}
c'_{j \sigma}(x)=F_{j \sigma} a^{-1/2} e^{-i \phi_{j \sigma}(x)},
\end{equation}
where $a$ is a short distance cutoff set to unity. The Klein factors $F_{j \sigma}$ retain the fermionic commutation relations, and also act as raising and lowering operators on the number operator for each fermionic species. We write the free part of the Hamiltonian as $H_0=\sum_{j \sigma} v_F \int_{-\infty}^{\infty} \frac{dx}{2\pi}\frac{1}{2} (\partial_x \phi_{j \sigma})^2$ and re-express the bosonic field in the basis of charge, spin, flavor, and spin-flavor degrees of freedom,
\begin{equation}
\begin{split}
\phi_c=\frac{1}{2}(\phi_{1 \uparrow}+\phi_{1 \downarrow} +\phi_{2 \uparrow}+\phi_{2 \downarrow}), \\
\phi_s=\frac{1}{2}(\phi_{1 \uparrow}-\phi_{1 \downarrow} +\phi_{2 \uparrow}-\phi_{2 \downarrow}), \\
\phi_f=\frac{1}{2}(\phi_{1 \uparrow}+\phi_{1 \downarrow} -\phi_{2 \uparrow}-\phi_{2 \downarrow}), \\
\phi_{sf}=\frac{1}{2}(\phi_{1 \uparrow}-\phi_{1 \downarrow} -\phi_{2 \uparrow}+\phi_{2 \downarrow}).
\end{split}
\end{equation}
We define new Klein factors in this new basis satisfying~\cite{zarandAnalyticalCalculationFinitesize2000,vondelftFiniteSizeBosonization2Channel1998}
\begin{equation}
\begin{split}
F_{sf}^\dagger F_s^\dagger = F^\dagger_{1 \uparrow} F_{1 \downarrow},~~~F_{sf} F_s^\dagger = F^\dagger_{2 \uparrow} F_{2 \downarrow}, \\
F_{sf}^\dagger F_f^\dagger = F^\dagger_{1 \uparrow} F_{2 \uparrow},~~~F_{c}^\dagger F_s^\dagger = F^\dagger_{1 \uparrow} F^\dagger_{2 \uparrow}.
\end{split}
\end{equation}
We then construct all 10 Kac-Moody currents, which we denote as $t_{A=1,\dots 10}$, and  also label the corresponding coupling constants,
\begin{eqnarray}
\label{eq:EK-gens}
    J_\perp:~~  t_1=t_2^\dagger 
        &=& {c'}^\dagger (\sigma_+{\otimes}\mathbb{1}) c' \nonumber\\
        &=& (F_{sf}^\dagger e^{i \phi_{sf}}+F_{sf} e^{-i \phi_{sf}}) F_s^\dagger e^{i \phi_s},~~~\nonumber \\
    Q_z: ~~ t_3=t_4^\dagger
        &=& {c'}^\dagger (\sigma_z{\otimes}\tau_+) {c'} \nonumber\\
        &=& (F_{sf}^\dagger e^{i \phi_{sf}}+F_{sf} e^{-i \phi_{sf}}) F_f^\dagger e^{i \phi_f},~~~\nonumber \\
    Q_\perp: ~~ t_5=t_8^\dagger
        &=& {c'}^\dagger (\sigma_+{\otimes}\tau_+) {c'}
         = F_s^\dagger e^{i \phi_s} F_f^\dagger e^{i \phi_f}, \nonumber \\
    ~~~~ ~~ t_6=t_7^\dagger
        &=& {c'}^\dagger (\sigma_-{\otimes}\tau_+) {c'}
         = F_s  e^{-i \phi_s} F_f^\dagger e^{i \phi_f}, \nonumber \\
    J_z: ~~~~~~\! t_9 ~~~ 
        &=&{c'}^\dagger (\sigma_z{\otimes}\mathbb{1}) {c'} \sim \partial_x \phi_s,~~~\nonumber \\
    V_{z}: ~~~~~~\! t_{10} ~~
        &=&{c'}^\dagger (\mathbb{1}{\otimes}\tau_z) {c'} \sim \partial_x \phi_f.
\end{eqnarray}
Here $\sigma_\pm=(\sigma_x \pm \sigma_y)/2$ and  $\tau_\pm=(\tau_x \pm \tau_y)/2$.

Exactly as in the $\mathrm{SO}(5)$ case~\cite{mitchellNonFermiLiquidCoulomb2021,libermanCriticalPointSpinflavor2021}, we proceed with the unitary transformation $U=e^{i \gamma_s \tilde{S}^{z0} \phi_s(0)+ i\gamma_f \tilde{S}^{0z} \phi_f(0)}$. The combination of $H_0$ with the Cartan (last two) generators with anisotropic couplings yields
\begin{equation}
\label{eq:deltaJ_z}
(J_z-\gamma_s v_F) \tilde{S}^{z0} \partial_x \phi_s 
+ (V_z-\gamma_f v_F) \tilde{S}^{0z} \partial_x \phi_f + H_0.
\end{equation}
For $\gamma_s=J_z/v_F$, $\gamma_f=V_z/v_F$ the first two terms cancel. When furthermore $\gamma_s=\gamma_f=1$, all the vertex operators $e^{\pm i\phi_s}$ and $e^{\pm i\phi_f}$ cancel in Eq.~\eqref{eq:EK-gens}, and we obtain a free fermion model.

We refermionize using new fermionic operators
\begin{equation}\begin{split}
\psi_\alpha(x)&=F_\alpha e^{-i \phi_\alpha(x)},~~~(\alpha=c,s,f,sf), \\
\chi_{\alpha+}&=\frac{\psi_\alpha^\dagger + \psi_\alpha}{\sqrt{2}},~~~\chi_{\alpha-}=\frac{\psi_\alpha^\dagger - \psi_\alpha}{i\sqrt{2}},
\end{split}\end{equation}
and also define local fermions
\begin{equation}\label{eq:EK-Majoranas}
\begin{split}
    a&=F_f^\dagger \tilde{\mathcal{S}}^{0-},
    ~~~a_+=\frac{a^\dagger + a}{\sqrt{2}},~~~a_-=\frac{a^\dagger - a}{i\sqrt{2}}, \\
    d&=F_s^\dagger \tilde{S}^{-0},
    ~~~d_+=\frac{d^\dagger + d}{\sqrt{2}},~~~d_-=\frac{d^\dagger - d}{i\sqrt{2}},
\end{split}
\end{equation}
with $\tilde{S}^{\pm 0}{=}\tilde{S}^{x0}{\pm}i \tilde{S}^{y0}$ and $\tilde{\mathcal{S}}^{0\pm}{=}\tilde{\mathcal{S}}^{0x}{\pm}i\tilde{\mathcal{S}}^{0y}$. This gives $a^\dagger a{=}\frac{1}{2}{+}\tilde{S}^{0z}{=}\frac{1}{2}{+}i a_- a_+$ and $d^\dagger d{=}\frac{1}{2}{+} \tilde{S}^{z0}{=}\frac{1}{2}{+}i d_- d_+$ with $a_\pm^2{=}\frac{1}{2}$ and $d_\pm^2{=}\frac{1}{2}$. It leads to the free-fermionic fixed-point Hamiltonian
\begin{equation}\label{eq:H-EK-sp4}
    \begin{split}
        \!\!H_{\mathfrak{sp}(4)}&=iJ_\perp d_- \chi_{sf+}(0) \\
        &-2 Q_z a_- \chi_{sf+}(0) d_- d_+ \\
        &-2i Q_\perp d_+ a_- + H_0
    \end{split}\ \Rightarrow 
    \begin{tikzpicture}[baseline=-1.5em]
    \begin{scope}[every node/.style={circle,draw}]
        \node (dm) at (1.0,0) {\ \ };
        \node (dp) at (1.0,-0.8) {\ \ };
        \node (am) at (1.8,0) {\ \ };
        \node (ap) at (1.8,-0.8) {\ \ };
    \end{scope}
    \draw[] (0.4,0) arc (0:90:0.2);
    \draw[] (0.4,0) arc (360:270:0.2);
    \draw[] (0.2,0.2) -- (-0.3,0.2);
    \draw[] (0.2,-0.2) -- (-0.3,-0.2);
    \begin{scope}[every node/.style={font=\scriptsize}]
        \node (chi0) at (0,0) {$\chi_{sf+}\!$};
        \node at (1.0,0) {$d_-$};
        \node at (1.0,-0.8) {$d_+$};
        \node at (1.8,0) {$a_{-}$};
        \node at (1.8,-0.8) {$a_+$};
    \end{scope}
    \begin{scope}[every edge/.style={draw, thick,font=\scriptsize,color=blue}]
        \path [-] (chi0) edge node [above] {$J_\perp$} (dm);
        \path [-] (chi0) edge node [below] {$Q_z$} (dm);
        \path [-] (dp) edge node [above,sloped,outer sep = -0.1em] {$Q_\perp$} (am);
    \end{scope}
    \end{tikzpicture},
\end{equation}
The $J_\perp$ term is relevant and sets the Kondo temperature $T_K$, below which the Majorana $d_-$ gets absorbed into the Majorana field $\chi_{sf+}$, resulting in the NFL. This yields~\cite{selaNonequilibriumCriticalBehavior2009,selaNonequilibriumTransportDouble2009}
\begin{equation}\label{eq:EK-d--chisf+}
    d_- \cong \frac{\chi_{sf+}(0)}{\sqrt{T_K}}, ~~~T_K \sim J_\perp^2.
\end{equation}
Thus, $d_-$ becomes a field with scaling dimension $1/2$. The $Q_\perp$ term gaps out the $d_+$ and $a_-$ operators, so that $\langle i d_+ a_- \rangle$ has a finite expectation value. As a result, the $Q_z$ term obtains the same form as the $J_\perp$ term, just renormalizing its value. The Majorana fermion $a_+$ remains decoupled at the critical point, as depicted in Eq.~\eqref{eq:H-EK-sp4}.

Let us comment on an artificial lack of symmetry in the roles played by the two Cartan generators in  Eq.~\eqref{eq:H-EK-sp4}, where the free Majorana $a_+$ is associated with one of them. This is because of symmetry breaking definitions of both the impurity fermions and the Klein factors.\footnote{Observe in Eq.~\eqref{eq:EK-Majoranas} that the fermion $d$ is defined by a generator $\tilde{S}^{-0}$, while $a$ is defined by a complementary $\tilde{\mathcal{S}}^{0-}$. A redefinition of $a\to F_f^\dagger \tilde S^{z-}$ with $\tilde{S}^{z\pm}{\equiv}\tilde S^{zx}{\pm}i\tilde S^{zy}$, i.e., in terms of generators restores the symmetry between the $Q_z$ and $J_\perp$ in Eq.~\eqref{eq:H-EK-sp4}. However, the $Q_\perp$ term still breaks symmetry between the two Cartan generators due to an arbitrary choice of the order of Klein factors in Eq.~\eqref{eq:EK-gens}. } For consistency with Ref.~\cite{libermanCriticalPointSpinflavor2021} we retain this choice. Nevertheless, as already observed in the graphics in Eq.~\eqref{eq:H-EK-sp4}, the couplings $J_\perp$ and $Q_z$ to the two Cartans eventually enter symmetrically, and all physical statements will be symmetric.

\subsection{Perturbations to the Free-Fermion Hamiltonian}\label{sec:EK-perturbations}
We now turn to analyze a variety of perturbations, covering the three categories discussed in Sec.~\ref{sec:Main}. It is convenient to consider perturbations aligned with the selected Cartan generators, i.e., the impurity operators $S^{z0},S^{0y}$ and the corresponding current $J^{z0}(x),J^{0y}(x)$. Note that we have returned to the operators without the tildes, i.e., prior to the transformation $\tau_y\to\tau_z$. We summarize the results in Table \ref{tab:EK}.

\newcommand{\EK}{
\begin{scope}[every node/.style={circle,draw}]
    \node (dm) at (0.8,0) {\ \ };
    \node (dp) at (0.8,-0.6) {\ \ };
    \node (am) at (1.4,0) {\ \ };
    \node (ap) at (1.4,-0.6) {\ \ };
\end{scope}
\draw[] (0.4,0) arc (0:90:0.2);
\draw[] (0.4,0) arc (360:270:0.2);
\draw[] (0.2,0.2) -- (-0.3,0.2);
\draw[] (0.2,-0.2) -- (-0.3,-0.2);
\begin{scope}[every node/.style={font=\scriptsize}]
    \node (chi0) at (0,0) {$\chi_{sf+}\!$};
    \node at (0.8,0) {$d_-$};
    \node at (0.8,-0.6) {$d_+$};
    \node at (1.4,0) {$a_{-}$};
    \node at (1.4,-0.6) {$a_+$};
\end{scope}
\begin{scope}[every edge/.style={draw, thick}]
    \path [-] (chi0) edge (dm);
    \path [-] (dp) edge (am);
\end{scope}
}

\newcommand{\EKr}[1]{
\begin{tikzpicture}[baseline=0]
    \EK
    \begin{scope}[every edge/.style={draw,red,line width=0.2 em}]
              #1
    \end{scope}
\end{tikzpicture}
}

\begin{table}[t]
    \centering
\begin{tabular}{|p{0.17\columnwidth}|m{0.26\columnwidth}|p{0.13\columnwidth}|m{0.35\columnwidth}|}\hline
    Physical &Operator          & EK& Graphics \\ \hline
    \multirowcell{2}{$B_1=B_2$ \\ $B_1{=}{-}B_2$}&$S^{z0}{=}\tfrac{1}{2}(\sigma_z{\otimes}\mathbb{1})$& $id_+d_-$ & \EKr{\path (dm) edge (dp);}\\ \cline{2-4}
    &$S^{0y}{=}\tfrac{1}{2}(\mathbb{1}{\otimes}\tau_y)$& $ia_+a_-$& \EKr{\path (am) edge (ap);}\\ \hline$|B_1|{\neq}|B_2|$&$S^{z0}+S^{0y}$& $id_+d_-$ ${\ \ }+$ $ia_+a_-$& \EKr{\path (dm) edge (dp); \path (am) edge (ap);}\\ \hline
    \multirowcell{3}{$t_1\neq t_2$\\$\Delta_1\neq\Delta_2$\\$|\epsilon_1|\neq|\epsilon_2|$}&$\mathcal{S}^{zy}{=}\tfrac{1}{2} (\sigma_z{\otimes}\tau_y)$& \multirow{3}{*}{$i d_-a_+$}& 
    \multirow{3}{*}{\EKr{\path (dm) edge (ap);}}\\ \cline{2-2}&$(S^{z0}{+}S^{0y})J^{z0}$& & \\     \cline{2-2}&$(S^{z0}{+}S^{0y})J^{0y}$& & \\ \hline
    &$S^{z0}J^{z0},S^{0y}J^{0y}$&  \multicolumn{2}{c|}{irrelevant}  \\ \hline
    \multirowcell{1}{\vspace{-0.7em}\\$\delta N_g$ \\ $\epsilon_1=\epsilon_2$}&$\mathcal{S}^{zy}\mathcal{J}^{zy}$& $ia_+\chi_{sf-}$& 
                \begin{tikzpicture}[baseline=0]
                    \EK
                    \draw[] (2,-0.4) arc (90:270:0.2);
                    \draw[] (2,-0.4) -- (2.5,-0.4);
                    \draw[] (2,-0.8) -- (2.5,-0.8);
                    \begin{scope}[every node/.style={}]
                        \node (chi0m) at (2.2,-0.6) {\scriptsize$\chi_{sf-}\!$};
                    \end{scope}
                    \begin{scope}[every edge/.style={draw,red,line width=0.2 em}]
                    \path [-] (chi0m) edge (ap);
                    \end{scope}
                \end{tikzpicture} \\ \hline
    ${\ \ }t_\uparrow\neq t_\downarrow$&$S^{0y}\mathcal{J}^{zy},S^{z0}\mathcal{J}^{zy}$&\multicolumn{2}{c|}{irrelevant} \\\hline
\end{tabular}
    \caption{Classification of perturbations to the $k{=}2$ symplectic Kondo model. The second column lists the perturbations considered in the EK analysis, while the first lists physical perturbations that fall into the same class. The third column gives the EK form for small deviations from the EK  point and the fourth represents this graphically (with the perturbation in red).}
    \label{tab:EK}
\end{table}

\subsubsection{Impurity Perturbations}
A magnetic field $B$ along the $z$ direction on the impurity generates a term $S^{z0}=id_+ d_-$. It does not affect the decoupled Majorana fermion $a_+$, leaving the system in an NFL state. A channel perturbation $S^{0y}$ (generated for example by imaginary interdot hopping) generates a term $ia_+ a_-$. Together with $id_+a_-$ it couples to a linear combination of $a_+$ and $d_+$, leaving the orthogonal-combination Majorana fermion free. This is consistent with the statement that for the $\mathrm{Sp}(4)$ isotropic case, generator impurity perturbations are marginal.

A linear combination of two such marginal perturbations, however, is immediately seen to be relevant and yields a flow to a FL. In the free-fermion theory, this is easily demonstrated for the two perturbations $S^{z0}$ and $S^{0y}$, that, importantly, do not anticommute. The product of these two generators yields a complementary term $\mathcal{S}^{zy}=(id_+d_-)(ia_+a_-)$. As $id_+a_-$ obtains a finite expectation value, this leaves $\mathcal{S}^{zy}\sim id_-a_+$. Thus, it destabilizes the NFL, and has a scaling dimension of $1/2$, consistently with the expected $k/(k{+}2)$ dimension of the complementaries. 

\subsubsection{Coupling Perturbations}
First, we consider a deviation from the EK point $J_z{=}\gamma_s v_F,V_z{=}\gamma_fv_F$, and rewrite the coupling terms in Eq.~\eqref{eq:deltaJ_z} in terms of the EK fermions,
\begin{subequations}
\begin{eqnarray} \label{eq:EK_dev}
    (J_z{-}\gamma_s v_F)S^{z0}J^{z0}(0) &\sim& 
        (J_z{-}\gamma_s v_F) d_-d_+  \chi_{s+} \chi_{s-},\\
    (V_z{-}\gamma_f v_F) S^{0y}J^{0y}(0) &\sim&
        (V_z{-}\gamma_f v_F) a_- a_+  \chi_{f+} \chi_{f-}.\qquad
\end{eqnarray}
\end{subequations}
Each one of these terms, separately to first order, excites the ``molecule" formed between Majorana fermions $d_+$ and $a_-$. However, to second order, these operators generate both FL and NFL corrections to the fixed point, corresponding to the leading irrelevant operator in Eq.~\eqref{eq:Hirr-k2}.  This is consistent with the statement that coupling perturbations of the form $S^A J^A$ for which $\{S^A,S^A\}\propto\mathbb{1}$ are irrelevant. 

On top of these, we add couplings between two different generators, such as $S^{z0} J^{0y}$ (and $S^{0y} J^{z0}$). Terms with this structure are generated by channel asymmetry.  As above, to first order these operators are gapped. But to second order, when combined with the $J_z{-}\gamma_s v_F$ term (or $V_z{-}\gamma_f v_F$ term), they yield the relevant complementary $\{S^{z0} ,S^{0y}\}=\mathcal{S}^{zy}$. The emergence of such local terms is in perfect agreement with the discussion in Sec.~\ref{sec:Pert-Complementaries} and the poor man's RG equations.

Next consider the perturbations generated by breaking of particle-hole symmetry, e.g., due to detuning of the  gate voltage $N_g\to 1{+}\delta N_g$. This inserts coupling terms involving the five additional $\mathrm{SU}(4)$ generators that are not $\mathrm{Sp}(4)$ generators, i.e., the complementaries $\mathcal{T}^\mathcal{A}_{k=2}$ in Eq.~\eqref{eq:Pauli-basis-comp}. Explicitly consider $\mathcal{S}^{zy}\mathcal{J}^{zy}$. In the lead $\mathcal{J}^{zy} \sim \partial_x \phi_{sf} \sim i \chi_{sf+} \chi_{sf-}$, while $\mathcal{S}^{zy} \sim id_-a_+$, as discussed above. Using Eq.~\eqref{eq:EK-d--chisf+}, we get a $\delta N_g i\chi_{sf-} a_+$ perturbation with scaling dimension $1/2$. Thus, it takes the system to a FL below an energy scale $\sim \delta N_g^2$, consistently with the discussion in Sec.~\ref{sec:Pert-Couplings}.

Finally, consider a coupling term consisting of the product of a generator acting on the impurity and a complementary acting on the lead. Such terms arise due to spin-dependent tunneling.  For example, consider $S^{z0} \mathcal{J}^{zy} \sim id_+\chi_{sf-}$, where we have used Eq.~\eqref{eq:EK-d--chisf+}. To second order, this operator only generates irrelevant operators.\footnote{Note a typo in Ref.~\cite{libermanCriticalPointSpinflavor2021}, where this operator was listed as marginal.}

\section{CFT} \label{sec:CFT}
Following the CFT analysis of the symplectic Kondo model \cite{renTopologicalQuantumComputation2024}, we extend it to include the complementary terms. For a full classification of operators at the fixed point we consider the conformal embedding  $\mathrm{Sp}(2k)_1 \otimes \mathrm{SU}(2)_k$. The latter theory corresponds to the isospin $\mathrm{SU}(2)$ charge (and particle-hole) symmetry whose Casimir corresponds to the total charge. Using the results of Ref.~\cite{kimuraABCDKondoEffect2021}, the central charge of the two theories are  $c_{\mathrm{Sp}(2k)_1}=\frac{k(2k+1)}{k+2}$ and $c_{\mathrm{SU}(2)_k}=\frac{3k}{k+2}$, correctly adding up to that of $k$ spinful channels, $c_{\mathrm{Sp}(2k)_1}+c_{\mathrm{SU}(2)_k}=2k$.  We denote the  primary fields of $\mathrm{Sp}(2k)_1$ as $\Phi_a$, with $a=0,1,\dots,k$ and scaling dimension $h_a = \frac{a(2k+2-a)}{4(k+2)}$ \cite{bouwknegtExclusionStatisticsConformal1999}. They transform as $d_a$-dimensional irreducible representations with $d_0=1$ for the vacuum, $d_1=2k$ for the defining representation, and $d_{a\geq 2}=\binom{2k}{a}{-}\binom{2k}{a-2}$. The primary fields of $\mathrm{SU}(2)_k$ have spin $j=0,1/2,\dots k$ and scaling dimension $h'_j=\frac{j(j+1)}{k+2}$.  

The fusion rules of $\mathrm{Sp}(2k)_1$ are identical to those of  $\mathrm{SU}(2)_k$ up to a factor of two ($a{\leftrightarrow}2j$),
\begin{equation}
\label{eq:fusion_rules}
    \!\!\!\!a{\times}a^\prime = |a{-}a^\prime|, |a{-}a^\prime|{+}2, \dots, \min\{a{+}a^\prime,2k{-}a{-}a^\prime\}.
\end{equation}
Let us apply CFT to obtain the relevant operators. We illustrate this for $k{=}2$, but the generalization to any $k$ is immediate. The lowest levels of the free fermion spectrum, with energy $E$ in units of the level spacing, are
\begin{center}
\begin{tabular}{||c c c c||} 
 \hline
 $a$ & $j$ & $E$ & \# \\ [0.5ex] 
 \hline\hline
 $0$ & $0$ & 0 & 1 \\ 
 \hline
 $1$ & $1/2$ & 1/2 & $4 \times 2=8$ \\ 
 \hline
 $2$ & $1$ & $1$ & $5 \times 3=15$ \\ 
 \hline
\end{tabular}
\end{center}
Applying fusion~\cite{affleckConformalFieldTheory1995} with the $a{=}1$ field, corresponding to the $4$-dimensional representation, we obtain the finite-size spectrum of the symplectic fixed point [cf.~Fig.~\ref{fig:SU2-2-FSS}]
\begin{center}
\begin{tabular}{||c c c c||} 
 \hline
 $a$ & $j$ & $E-\frac{3}{16}$  & \# \\ [0.5ex] 
 \hline\hline
 $1$ & $0$ & 1/8 & 4 \\ 
 \hline
 $0$ & $1/2$ & 0 & $2$ \\ 
 \hline
 $2$ & $1/2$ & 1/2 & $10$ \\ 
 \hline
 $1$ & $1$ & $5/8$ & $12$ \\ 
 \hline
\end{tabular}
\end{center}
Then, after double fusion~\cite{affleckConformalFieldTheory1995}, in addition to the identity, we obtain relevant operators of dimension 1/2,
\begin{center}
\begin{tabular}{||c c c c||} 
 \hline
 $a$ & $j$ & scaling~dimension  & \# \\ [0.5ex] 
 \hline\hline
  $0$ & $0$ & 0 & $1$ \\ 
 \hline
  $2$ & $0$ & 1/2 & $5$ \\ 
 \hline
 $0$ & $1$ & $1/2$ & $3$ \\ \hline
 $1$ & $1/2$ & $1/2$ & $16$ \\ 
 \hline
\end{tabular}
\end{center}
We identify the $a{=}2$ field ($\Phi_2$) with the $5$-dimensional representation of EK lead Majorana fermions $\{\chi_{s+}, \chi_{s-}, \chi_{f+}, \chi_{f-}, \chi_{sf+}\}$. These will enter when violating the symplectic symmetry, e.g., due to channel asymmetry.  The $j{=}1$ field is then identified with the remaining fermions $\{\chi_{c+},\chi_{c-},\chi_{sf-}\}$. Note that $\chi_{c\pm}$ are related to the total charge, which is conserved throughout this paper, but $\chi_{sf-}$ will enter when violating particle-hole symmetry. The last line in the table simply corresponds to the non-Abelian bosonization of the original fermions, which are not allowed as perturbations in the critical point. 

Although for $k{=}2$ both fields have scaling dimension $1/2$, generally $\Phi_2$ will have $h_2=k/(k{+}2)$, while the $j{=}1$ field will have $h^\prime_1=2/(k{+}2)$. The latter corresponds to the violation of particle-hole symmetry, e.g., due to gate detuning. Thus, the NFL breaking energy scale $T_\mathrm{FL}\sim\delta N_g^{1/(1-h'_1)}=\delta N_g^{1+2/k}$, as demonstrated by NRG for $k{=}2,3$ in Fig.~\ref{fig:NRG-eps}. The effect of $\Phi_2$ will be discussed in the following sections.

\subsection{Impurity Operators}

The Kac-Moody currents $J^A$ transform according to the $k(2k{+}1)$-dimensional adjoint representation $(20)_k$ and have scaling dimension $1$. They correspond to descendant fields (Kac-Moody descendants of the identity) since no primary field in the above list  transforms according this representation. The complementary currents $\mathcal{J}^{\mathcal{A}}$ are those required to complete $J^A$ to the Kac-Moody currents of $\mathrm{SU}(2k)_1$. The $\mathcal{J}^{\mathcal{A}}$'s transform according to a $(k{-}1)(2k{+}1)$-dimensional representation, just like the primary field $\Phi_2$.  However, while $\Phi_2$ has scaling dimension $h_2{=}k/(k{+}2)$, the currents $\mathcal{J^A}$ have scaling dimension $1$. In terms of EK fermions the currents are given by bilinears
\begin{equation}
\label{eq:EK-currents}
    J^A \in \{ i\chi_\mathcal{A}\chi_\mathcal{B} \},\quad  
    \mathcal{J}^{\mathcal{A}}=i\chi_\mathcal{A}\chi_{sf-},
\end{equation}
with $\mathcal{A}{\neq}\mathcal{B}$ belonging to the five fermions listed before, yielding the $10$ symplectic generators. A sixth fermion $\chi_{sf-}$ (belonging to the $j{=}1$ field) then completes these bilinears to the $15$ generators of $\mathrm{SU}(4)$.

To write down the impurity operators at the symplectic fixed point, we fuse the $a{=}1$ symplectic ``spin'' to the leads. We then we get an operator expansion of the form
\begin{equation}\label{eq:CFT-imp}
    S^A \sim J^A+(J_{-1}\Phi_2)^A \Gamma,\quad \mathcal{S}^{\mathcal{A}}  \sim 
\mathcal{J}^{\mathcal{A}}+\Phi_2^{\mathcal{A}} \Gamma,
\end{equation}
where all fields are implicitly assumed at the location of the impurity. $\Gamma$ is explicitly included to account for the emergent fractional degree of freedom. For multi-impurity cases $\Gamma$ acts on the anyonic fusion space \cite{gabayMultiimpurityChiralKondo2022,renTopologicalQuantumComputation2024}. While the equation for $S^A$ appeared before~\cite{gabayMultiimpurityChiralKondo2022,renTopologicalQuantumComputation2024}, the equation for $\mathcal{S}^{\mathcal{A}}$ is new. It follows from symmetry and from the operator content of the critical point. Note that specific components of this equation were already obtained in our EK analysis above. While the leading term in $S^A$ is $J^A$, which has scaling dimension $1$ (i.e., is marginal), the leading term of the complementaries, i.e., $\Phi_2^{\mathcal{A}} \Gamma$, has scaling dimension $k/(k{+}2)$, meaning they are relevant, in perfect agreement with the NRG results. In terms of the EK fermions, substituting Eq.~\eqref{eq:EK-currents} into Eq.~\eqref{eq:CFT-imp} yields
\begin{equation}
\begin{split}
    S^\mathcal{AB} &\sim i\chi_\mathcal{A}\chi_\mathcal{B} 
    + \chi_\mathcal{C}\chi_\mathcal{D}\chi_\mathcal{E}a_+,\\
    \mathcal{S^A} &\sim i\chi_{\mathcal{A}}\chi_{sf-} + i\chi_{\mathcal{A}} a_+,
\end{split}
\end{equation}
with $\Gamma{=}a_+$ being the decoupled impurity Majorana fermion and an implicit prefactor before the second term in each expression. Observe that each $S^\mathcal{AB}$ involves all five different fermions (hence they are labeled by different letters). Explicitly, we can check that
\begin{equation}
\begin{split}
    S^{z0} &\sim i\chi_{s+}\chi_{s-} + \chi_{f+}\chi_{f-}\chi_{sf+}a_+, \\
    S^{0y} &\sim i\chi_{f+}\chi_{f-} + \chi_{s+}\chi_{s-}\chi_{sf+}a_+, \\
    \mathcal{S}^{zy} &\sim i\chi_{sf+}\chi_{sf-} + i\chi_{sf+}a_+.
\end{split}
\end{equation}

\subsection{Sequence of Fixed Points}

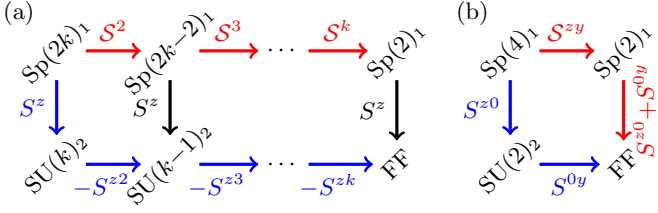
\begin{figure}
    \centering
    \begin{tikzpicture}
\begin{scope}[every node/.style={rotate=45}]
    \node (Sp2k) at (0,0) {$\mathrm{Sp}(2k)_1$};
    \node (Sp2k-2) at (1.5,0) {$\mathrm{Sp}(2k{-}2)_1$};
    \node (Sp2) at (4.5,0) {$\mathrm{Sp}(2)_1$};
    \node (SUk) at (0,-1.5) {$\mathrm{SU}(k)_2$};
    \node (SUk-1) at (1.5,-1.5) {$\mathrm{SU}(k{-}1)_2$};
    \node (FF) at (4.5,-1.5) {$\mathrm{FF}$};
\end{scope}
\node (Sp-dd) at (3,0) {$\cdots$};
\node (SU-dd) at (3,-1.5) {$\cdots$};
\begin{scope}[every edge/.style={draw,very thick,color=red}]
    \path [->] (Sp2k) edge node [above] {$\mathcal{S}^{2}$} (Sp2k-2);
    \path [->] (Sp2k-2) edge node [above] {$\mathcal{S}^{3}$} (Sp-dd);
    \path [->] (Sp-dd) edge node [above] {$\mathcal{S}^{k}$} (Sp2);
\end{scope}
\begin{scope}[every edge/.style={draw, very thick,color=blue}]
    \path [->] (Sp2k) edge node [left] {$S^{z}$} (SUk);
    \path [->] (SUk) edge node [below] {$\!\!\!\!\!\!-S^{z2}$} (SUk-1);
    \path [->] (SUk-1) edge node [below] {$\!\!\!\!\!\!-S^{z3}$} (SU-dd);
    \path [->] (SU-dd) edge node [below] {$\!\!\!\!\!\!-S^{zk}$} (FF);
\end{scope}
\begin{scope}[every edge/.style={draw, very thick}]
    \path [->] (Sp2k-2) edge node [left] {$S^{z}$} (SUk-1);
    \path [->] (Sp2) edge node [left] {$S^{z}$} (FF);
\end{scope}

\begin{scope}[every node/.style={rotate=45}]
    \node (Sp4) at (6,-0) {$\mathrm{Sp}(4)_1$};
    \node (Sp2) at (7.5,0) {$\mathrm{Sp}(2)_1$};
    \node (SU2) at (6,-1.5) {$\mathrm{SU}(2)_2$};
    \node (FF) at (7.5,-1.5) {$\mathrm{FF}$};
\end{scope}
\begin{scope}[every edge/.style={draw, very thick,color=red}]
    \path [->] (Sp4) edge node [above] {$\mathcal{S}^{zy}$}(Sp2);
    \path [->] (Sp2) edge node [right,rotate=90,anchor=north] {$S^{z0}{+}S^{0y}$} (FF);
\end{scope}
\begin{scope}[every edge/.style={draw, very thick,color=blue}]
    \path [->] (Sp4) edge node [left] {$S^{z0}$} (SU2);
    \path [->] (SU2) edge node [below] {$S^{0y}$}(FF);
\end{scope}

\begin{scope} []
    \node (a) at (-0.5,0.5) {(a)};
    \node (b) at (5.5,0.5) {(b)};
\end{scope}
\end{tikzpicture}
     \vspace{-2\baselineskip}\phantomsubfloat{\label{fig:CFT-Sp2m}}\phantomsubfloat{\label{fig:CFT-Szy}}
    \caption{(a) The possible fixed points that can be obtained by gapping out the $\mathrm{Sp}(2k)_1$ theory via symplectic-symmetry breaking all the way down to a free-fermionic (FF) theory. The top row (red path) corresponds to gapping out by complementaries through a sequence of $\mathrm{Sp}(2m)_1$ theories with $1\leq m\leq k$. Applying a generator to each of these leads to a corresponding $\mathrm{SU}(m)_2$ theory (bottom row). One can also go through the latter by applying a sequence of Zeeman fields, i.e., symplectic generators (blue path). A specific choice of channel polarized complementaries and generators that lead to each of these fixed points is indicated next to the arrows. (b) Specific example for $k{=}2$ with the impurity generators and complementaries that are naturally considered in the EK analysis.}
    \label{fig:CFT-sequence}
\end{figure}

Having established the relevance of  the complementaries, let us investigate the effect of impurity generators. Here, we start with the EK formulation for $k{=}2$, and then generalize to arbitrary $k$. Ignoring FL-like irrelevant operators, the leading (nontrivial) irrelevant operator at the symplectic fixed point is given by the product of all five lead fermions and the decoupled impurity fermion  \cite{libermanCriticalPointSpinflavor2021}
\begin{equation}\label{eq:Hirr-k2}
    \delta H_\mathrm{irr} \sim \chi_{s+}\chi_{s-} \chi_{f+}\chi_{f-}\chi_{sf+}a_+.
\end{equation}
This operator has a scaling dimension $5/2$. Applying an impurity Zeeman field corresponding to a single generator, e.g., $S^{z0}$,  gaps out the corresponding pair of lead fermions so that
\begin{equation}
    \delta H_\mathrm{irr} \to \braket{\chi_{s+}\chi_{s-}}\chi_{f+}\chi_{f-}\chi_{sf+}a_+.
\end{equation}
This operator has scaling dimension $3/2$, corresponding to an $\mathrm{SU}(2)_2$ fixed point. Applying a second generator, e.g., $S^{0y}$ then leaves
\begin{equation}
    \delta H_\mathrm{irr} \to \braket{\chi_{s+}\chi_{s-}}\braket{\chi_{f+}\chi_{f-}}\chi_{sf+}a_+,
\end{equation}
with scaling dimension $1/2$, which is relevant and gaps out the decoupled Majorana. This process corresponds to the blue path in Fig.~\ref{fig:CFT-Szy}. One can understand the same process as first gapping out the decoupled impurity Majorana all the way to a FL by a complementary perturbation $\mathcal{S}^{yz}=(S^{z0}{+}S^{0y})^2-1/2$. This results in a single-channel Kondo effect. Then applying $S^{z0}{+}S^{0y}$ corresponds to applying a magnetic field, that breaks the Kondo singlet down to a free-fermionic theory. This procedure is describe by the red path in Fig.~\ref{fig:CFT-Szy}.

Generalizing to larger $k$, the leading irrelevant operator must be an $\mathrm{Sp}(2k)$ scalar and takes the form
\begin{equation}\label{eq:Hirr}
    \delta H_\mathrm{irr} \sim \sum_A J_{-1}^A(J_{-1}\Phi_2)^A\Gamma,
\end{equation}
with scaling dimension of $2{+}h_2^{(k)}$, where $h_2^{(k)}{=}k/(k{+}2)$. The process of gapping out the $\mathrm{Sp}(2k)_1$ fixed point using a sequence of Zeeman fields $\{S^{A_1},\dots,S^{A_k}\}$ has been discussed in the Bethe ansatz exact solution~\cite{konigExactSolutionTopological2023}. It is illustrated by the blue path (bottom row) in Fig.~\ref{fig:CFT-Sp2m}. Introducing the impurity generator $S^{A_1}$ gaps out the current operator $J^{A_1}$ in Eq.~\eqref{eq:Hirr}, resulting in an $\mathrm{SU}(k)_2$ theory with scaling dimension $1{+}h_2^{(k)}$. According to the Bethe ansatz, each additional generator takes the $\mathrm{SU}(m)_2$ theory to an  $\mathrm{SU}(m{-}1)_2$ theory, reducing the scaling dimension of the leading irrelevant operator from $1{+}h_2^{(m)}$ to $1{+}h_2^{(m-1)}$. Reaching $\mathrm{SU}(2)_2$, the last generator $S^{A_k}$ reduces it to the free-fermionic (FF) theory, as explicitly shown above for $k{=}2$.

Note that various choices of generators can short-circuit this path. As a concrete example, consider a magnetic field at one dot, introducing $-S^{1z}$. This singles out one state from the ground-state manifold, and immediately gaps the system to a FF theory. Alternatively, one can consider first applying a uniform magnetic field $S^{A_1}=\sum_j S^{jz}\propto S^z$, and then sequentially canceling this field at all other dots by applying $S^{A_{j>1}}=-S^{jz}$, thus traversing through all the $\mathrm{SU}(m)_2$ fixed points.

Let us now demonstrate that the gapping out by $k$ generators is actually a reflection of the gapping out of the  $\mathrm{Sp}(2k)_1$ fixed point by a sequence of $k{-}1$ complementaries $\{\mathcal{S}^{\mathcal{A}_2},\dots,\mathcal{S}^{\mathcal{A}_k}\}$ (and one generator). Each introduced complementary takes the $\mathrm{Sp}(2m)_1$ theory to an $\mathrm{Sp}(2m{-}2)_1$theory, thereby reducing the scaling dimension of the leading irrelevant operator from $2{+}h_2^{(m)}$ to $2{+}h_2^{(m-1)}$. This is illustrated by the red path (upper row) in Fig.~\ref{fig:CFT-Sp2m}.  The vertical arrows describe symplectic generators that take each of the $\mathrm{Sp}(2m)_1$ theories to its corresponding $\mathrm{SU}(m)_2$ theory. These two theories share the same fractional entropy. Moreover, the scaling dimension of their leading irrelevant operator differs by 1, as can be read of from Eq.~\eqref{eq:Hirr} by gapping out a bosonic current operator. The reduction in the scaling dimension along the horizontal arrows ($2{+}h_2^{(m)}{\to}2{+}h_2^{(m-1)}$ for red and $1{+}h_2^{(m)}{\to}1{+}h_2^{(m-1)}$ for blue) is more complicated due to the nontrivial nature of the operators being gapped out. Nevertheless, the correspondence between the two clearly indicates that they are caused by the same operator. In other words, the blue path obtained via the Bethe ansatz is a reflection of the red path.

To further support this claim, let us return to the example of channel-polarized generators $S^{A_j}=-S^{jz}$, with $j=2,\dots,k$, and choose complementaries $\mathcal{S}^{\mathcal{A}_j} \propto\{S^{A_j},S^{A_j}\}+\mathrm{const}$. Explicitly, these are the channel-polarized complementaries, $\mathcal{S}^{\mathcal{A}_j}=\mathcal{S}^j\propto\mathbb{1}{\otimes}\Pi_j+\mathrm{const}$. Each introduced $\mathcal{S}^j$ gaps out two states from the impurity ground space, thus going from $2m\to 2m{-}2$ . At each step, applying a uniform Zeeman field $S^z$ reduces the impurity ground-space degeneracy $2m\to m$. Therefore, we have established that proceeding along the horizontal direction (for both rows) corresponds to gapping out by complementaries, while only the transition from the top to bottom row is inherently an effect of the generators.

\section{From Symplectic Anderson to Kondo}\label{sec:SchriefferWolff}
Let us now rederive the mapping \cite{liTopologicalSymplecticKondo2023} between the Anderson Hamiltonian of Eq.~\eqref{eq:H-Anderson} to the symplectic Kondo model of Eq.~\eqref{eq:H-Kondo}. Along the way we will introduce a variety of perturbations and derive the effective terms that they generate under this mapping.
\subsection{Ground Space and Local Terms}
We start by finding the degenerate impurity ground space. Setting the tunneling to the leads to zero, we rewrite Eq.~\eqref{eq:H-Anderson} together with local perturbations
\begin{equation}
\begin{split}
        H_\mathrm{imp} &= E_C(2 \hat{N}_C+\hat{n}_d-N_g)^2 \\
        &+ \sum_{j=1}^k  \left( \Delta_j i e^{-i \phi} d^\dagger_{j \uparrow} d^\dagger_{j \downarrow}  +\mathrm{H.c.}\right)\\
&+ \sum_{j=1}^k[\epsilon_j (n_{j \uparrow}+n_{j \downarrow}-1)-B_j (n_{j \uparrow}-n_{j \downarrow})] ,
\end{split}
\end{equation}
where $n_{j\sigma} = d^\dagger_{j\sigma}d_{j\sigma}$ so that $\hat{n}_d=\sum_{j\sigma}n_{j\sigma}$ and $e^{-i\phi}$ annihilates a Cooper pair, i.e., is a lowering operator for $\hat N_C$ . Here, we have allowed for dot-dependent SC pairing $\Delta_j$, a variation of the dot energy levels $\epsilon_j$, and a dot-dependent magnetic field $B_j$. This model conserves total charge. In the absence of a magnetic field ($B_j{=}0$) and for uniform $\varepsilon_j$ and $\Delta_j$, it also displays a symplectic symmetry, namely,  $\textbf{d}^\dagger \equiv (d_{1\uparrow}^\dagger,\dots,d_{k\uparrow}^\dagger,d_{1\downarrow}^\dagger,\dots,d_{k\downarrow}^\dagger)$ transforms as the defining representation of $\mathrm{Sp}(2k)$, so that $H_\mathrm{imp}$ is invariant under the transformation  $\textbf{d}\to e^{i\lambda_AT^A} \textbf{d}$ with $T^A$ the symplectic generators.\footnote{The limitation to symplectic generators comes from the SC-pairing term. In its absence the model displays a full $\mathrm{SU}(2k)$ symmetry between all spin and channel species.}

Ignoring the charging term $E_C$, we can diagonalize the state space of each dot separately to get
\begin{equation}\label{eq:dot-space}
    \begin{array}{ll}
        \mathrm{State} &  \mathrm{Energy}\\ 
        \ket{\uparrow}_j   & \ \ -B_j, \\ 
        \ket{\downarrow}_j & \ \ +B_j, \\
        \cos\theta_{j-}\ket{0}_j+i\sin\theta_{j-}\ket{\uparrow\downarrow}_j & \ \ -\Delta_j^\prime ,\\
        \underbracket{\cos\theta_{j+}}_{-\sin\theta_{j-}}\ket{0}_j + i\underbracket{\sin\theta_{j+}}_{\cos\theta_{j-}}\ket{\uparrow\downarrow}_j & \ \ +\Delta_j^\prime.
    \end{array}
\end{equation}
The effective pairing and angles
\begin{equation} \label{eq:BCS-angles}
    \Delta_j^\prime \equiv \sqrt{\Delta^2_j+\epsilon^2_j},
    \quad \tan \theta_{j\mp}=\frac{\epsilon_j\mp\sqrt{\Delta_j^2+\epsilon_j^2}}{\Delta_j},
\end{equation}
define BCS states to which we will refer as paired ($\theta_{j-}$) and antipaired ($\theta_{j+}$). Note that small $\epsilon_j$ yields $\Delta_j^\prime\approx\Delta_j+\epsilon_j^2/2\Delta_j$ and $\theta_{j\mp}\approx\mp\pi/4+\epsilon_j/2\Delta_j$. Implicitly, the doubly occupied states $\ket{\uparrow\downarrow}_j$ have one less Cooper pair in the SC then the empty states $\ket{0}_j$ as total charge is conserved. After reintroducing $E_C$,  these states will serve as building blocks for constructing the ground space and excited states of all dots together. 

Fixing $N_g=1$ while setting $B_j=0$ and a uniform effective pairing $\Delta_j^\prime=\Delta^\prime$, we find the ground space. Assuming without loss of generality that $\Delta^\prime>0$, we wish to place each dot in the paired ($\theta_{j-}$) state. But, due to the fixing of $N_g=1$ we have an odd number of fermions, so that one dot must remain singly occupied. It can be occupied either by an up or down fermion, yielding $2k$ degenerate states. Labeling these states by the index of the singly occupied dot and its spin, we explicitly define
\begin{equation}\label{eq:gs_states}
\!\!\!\ket{j\sigma} =
  \overbrace{
  d_{j\sigma}^\dagger \prod_{j'\ne j}(\cos \theta_{j'-}\!\!+ i\sin \theta_{j'-}d_{j'\uparrow}^\dagger d_{j'\downarrow}^\dagger e^{-i\phi})}^{\equiv f^\dagger_{j\sigma}}\ket{\emptyset},
\end{equation}
with $\theta_{j'-}$ as in Eq.~\eqref{eq:BCS-angles}. The vacuum $\ket{\emptyset}$ is defined with all dots empty  and some fixed number of Cooper pairs in the SC. It is convenient to introduce new fermionic operators  $f_{j\sigma}$ such that $\ket{j\sigma}=f^\dagger_{j\sigma}\ket{\emptyset}$. The energy of a specific $\ket{j\sigma}$ state is given by $-\sum_{j'\neq j}\Delta_{j'}^\prime$ and so for a uniform effective pairing the ground-space energy is $E_\mathrm{GS}=-(k{-}1)\Delta^\prime$.

\subsubsection{Magnetic Fields and Effective Pairing Asymmetry}
Allowing nonuniform angles in Eq.~\eqref{eq:gs_states} breaks the ground-space degeneracy. However, as long as $B_j$ and $\Delta_j^\prime$ are small perturbations, we remain in the $2k$-dimensional space spanned by $\ket{j\sigma}$, for which $\sum_{j\sigma}f_{j\sigma}^\dagger f_{j\sigma}=1$. Explicitly projecting onto this space yields

\begin{equation}
    H_\mathrm{imp} \to \sqrt{\tfrac{2k-2}{k}}\sum_{j=1}^k \delta\Delta_j^\prime\mathcal{S}^j -\sqrt{2} \sum_{j=1}^k B_j S^{zj},
\end{equation}
$S^{zj}$ and $\mathcal{S}^j$ are, respectively, the generators and complementary terms defined in Eq.~\eqref{eq:chan-basis}. As $\sum_j\mathcal{S}^j=0$, we only keep the variations $\delta\Delta_j'\equiv\Delta_j'-\bar\Delta'$ from the average $\bar\Delta'\equiv\frac{1}{k}\sum_j\Delta_j'$, and omit the average ground-state energy $E_\mathrm{GS}=-(k{-}1)\bar\Delta'$. A small variation in the dot levels, while retaining a uniform SC pairing, yields a quadratic effective pairing variation $\delta\Delta_j'\approx\epsilon_j^2/2\Delta$. Thus, both asymmetry in the dot levels and in the SC pairing couple to the complementary $\mathcal{S}^j$ and are relevant. However, due to the quadratic dependence on $\epsilon_j$, the effect of dot-level asymmetry is weak (comes with a power law $\epsilon_j^{2+k}$). An average shift of the dot levels leads to a more pronounced effect due to the breaking of particle-hole symmetry, and will be discussed in Sec.~\ref{sec:SW-PH-asym}.

The magnetic fields couple to the marginal generators. From Eq.~\eqref{eq:gs_states} it is immediately clear that a magnetic field at dot $j$  lowers the energy of one of the two $\ket{j\sigma}$ states and raises that of the other, while leaving all other states unaffected. Thus, equal magnitude but arbitrary sign fields on all dots select $k$ states out of the ground-space manifold. These form an effective $\mathrm{SU}(k)$ impurity that will be screened by both spin species and does not break the NFL \cite{konigExactSolutionTopological2023}. This is in perfect agreement with the observation that the square of such a choice of fields is proportional to the identity, as discussed in Sec.~\ref{sec:Pert-Generators}. Magnetic fields of varying magnitude will further reduce the dimension of the ground space, also in agreement with Secs.~\ref{sec:Pert-Generators} and \ref{sec:CFT}.

\subsubsection{Interdot Tunneling}
Next consider the effect of interdot tunneling
\begin{equation}
    \Delta H_\mathrm{imp} = \sum_{j<j^\prime}\sum_\sigma t_{jj^\prime}d^\dagger_{j\sigma}d_{j^\prime\sigma}^{\,} + \mathrm{H.c.}, 
\end{equation}
where $t_{jj'}$ can be complex. Projecting to the ground space yields
\begin{equation}
\begin{split}
    \Delta H_\mathrm{imp} \to 2\sum_{j<j^\prime} \ & \cos(\theta_{j-}+\theta_{j'-})\mathrm{Re}\{t_{jj^\prime}\} \mathcal{S}^{jj^\prime} \\ 
    -\  &\cos(\theta_{j-}-\theta_{j'-})\mathrm{Im}\{t_{jj^\prime}\}S^{jj^\prime},
\end{split}
\end{equation}
with the complementaries $\mathcal{S}^{jj'}$ and generators $S^{jj'}$ defined in Eq.~\eqref{eq:chan-basis-full}. While the real-tunneling contribution is relevant, due to the angle dependence it is a weak effect that breaks the NFL at a scale $\sim(\mathrm{Re}\{t_{jj'}\}\epsilon_j)^{1+k/2}$ and vanishes in the absence of a variation in the dot levels (the latter is anyhow relevant in its own right). The imaginary contribution is also relevant (except for $k{=}2$), as the $S^{jj'}$ generators do not square to the identity. However, like nonuniform magnetic fields, this is a weak effect that scales as $\sim\mathrm{Im}\{t_{jj'}\}^{2+k}$. Thus, we do not expect small interdot tunneling to be a practical issue. Note that crossed-Andreev reflection is also expected to generate the complementary terms $\mathcal{S}^{jj'}$ \cite{konigExactSolutionTopological2023}, but its effect has not been analyzed here.

\subsection{Hybridization with the Leads}
\begin{figure*}
\centering
\includegraphics[width=0.99\textwidth]{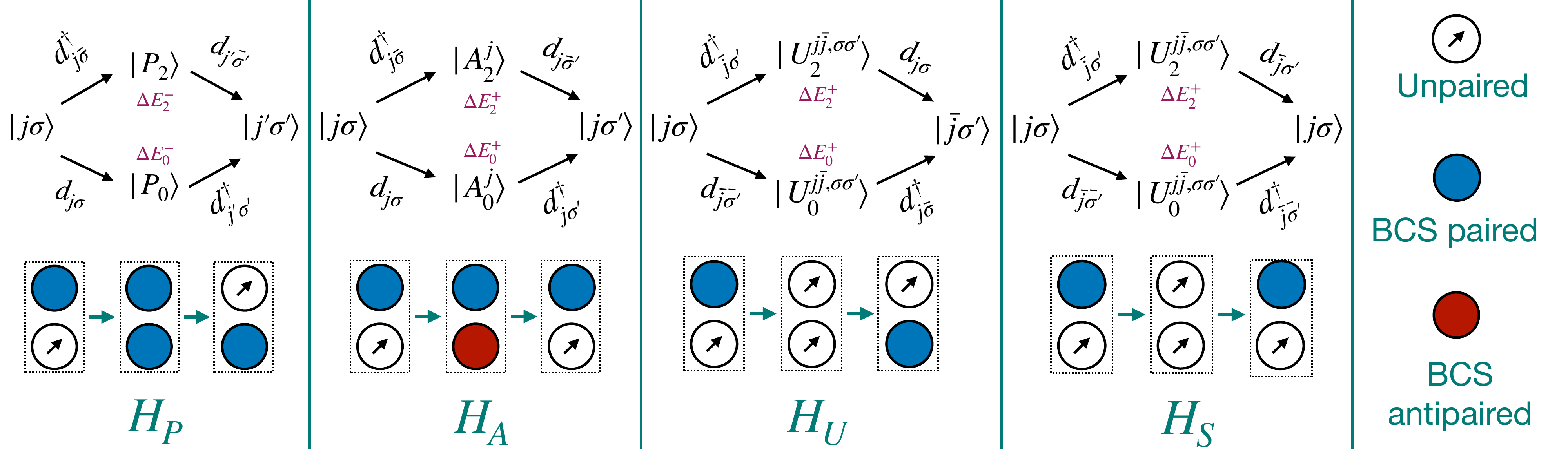}
\caption{All second-order tunneling processes that start from and return to the ground space of the symplectic Anderson model with a single unpaired fermion (i.e., $N_g\approx 1$) -- see Eq.~\eqref{eq:h_eff_form_1}. Each panel corresponds to a different type of process, which is depicted schematically. Blue (red) circles denote BCS paired (antipaired) dots and white circle denote unpaired, i.e., singly occupied, dots. The participating states are stated explicitly, with the involved operators indicated above the arrows. Each transition occurs either via an $m{=}0$ excited state with a fermion first tunneling out and then another tunneling in, or vice versa via an $m{=}2$ excited state. 
\label{fig:SW_wide_fig}}
\end{figure*}
Now turn on weak lead-dot tunneling $t_j\ll E_C,\Delta$,
\begin{eqnarray}
    H_\mathrm{hyb} = \sum_{j\sigma} t_j (d^\dagger_{j\sigma}c_{0j\sigma} + c^\dagger_{0j\sigma}d_{j\sigma}),
\end{eqnarray}
where $c_{0j\sigma}\equiv c_{j\sigma}(0)$ are lead operators at the location of the impurity and assume without loss of generality real and positive $t_j$. The leading contribution of local perturbations has already been accounted for in the previous section, and so here they are set to zero. Deviations from $N_g=1$ that break particle-hole symmetry are allowed, and will be discussed in detail in Sec.~\ref{sec:SW-PH-asym}. There, we will also discuss  the leading effect of an average shift in the dot levels, that here is set to zero.

It is convenient to compactly express the ground states of the impurity Hamiltonian Eq.~\eqref{eq:gs_states} as 
\begin{equation}
    \ket{j\sigma} = f^\dagger_{j\sigma}\ket{\emptyset} = d_{j\sigma}^\dagger \ket{\mathrm{BCS}}_{\{j\}}.
\end{equation}
$\ket{\mathrm{BCS}}_{\{j_1j_2\dots\}}$ denotes the state in which all dots expect $\{j_1j_2\dots\}$ are in their BCS-paired ground state (with $\theta_{j-}=\pi/4$), i.e.,
\begin{equation}
    \ket{\mathrm{BCS}}_{\{j_1j_2\dots\}}=\prod_{\!\!\!\!\!j'\neq j_1j_2\dots\!\!\!\!\!}\tfrac{1}{\sqrt{2}}\left(1- i e^{-i\phi}d^\dagger_{j'\uparrow}d^\dagger_{j'\downarrow}\right)\ket{\emptyset}.
\end{equation}
Tunneling of a fermion in or out of the dots takes us out of the ground space. Thus, second-order processes are considered (with higher orders neglected in the weak-tunneling limit), as depicted in Fig.~\ref{fig:SW_wide_fig}. They involve three types of excited states,
\begin{subequations}\label{eq:ex_states}
\begin{eqnarray}
    &&\ket{P_m} = e^{i\frac{m}{2}\phi}
      \overbrace{
      \tfrac{1}{\sqrt{2}}(1-ie^{-i\phi}d_{j\uparrow}^\dagger d_{j\downarrow}^\dagger) \ket{\mathrm{BCS}}_{\{j\}}}^{= \ket{\mathrm{BCS}}}, \\
    &&\ket{A_m^j} = e^{i\frac{m}{2}\phi}\tfrac{1}{\sqrt{2}}(1+ie^{-i\phi}d_{j\uparrow}^\dagger d_{j\downarrow}^\dagger) \ket{\mathrm{BCS}}_{\{j\}}, \\
    &&\ket{U_m^{jj'\sigma \sigma'}} = e^{i\frac{m-2}{2}\phi} d^\dagger_{j\sigma}d^\dagger_{j'\sigma'}\ket{\mathrm{BCS}}_{\{jj'\}} \ \ (j{>}j'),\ \ 
\end{eqnarray}
\end{subequations}
which we call paired, antipaired, and unpaired, respectively. The index $m{=}0,2$ indicates the total number of fermions at the impurity (dots and SC island together), with $m{=}0$ ($m{=}2$) corresponding to a fermion tunneling out to (in from) the leads. The paired $\ket{P_m}$ and antipaired $\ket{A_m^j}$ states are reached from $\ket{j\sigma}$ by a fermion with spin $\sigma$ (opposite spin $\bar\sigma$) tunneling out from (into) dot $j$. The unpaired state $\ket{U_m^{jj'\sigma\sigma'}}$, on the other hand, is reached from $\ket{j\sigma}$ by a fermion with spin $\sigma'$ ($\bar\sigma'$) tunneling into (out from) a different dot $j'{\neq}j$.

The states $\ket{P_m}$ have an energy of $\Delta E_m^-$ above the ground state, while both  $\ket{A_m^j}$ and  $\ket{U_m^{jj'\sigma\sigma'}}$ have a gap of $\Delta E_m^+=\Delta E_m^-{+}2\Delta$. Explicitly,
\begin{equation}\label{eq:SW_en_denom}
    \begin{split}
        \Delta E_0^\pm &= E_C(1+2\delta N_g) \pm \Delta,\\
        \Delta E_2^\pm &= E_C(1-2\delta N_g) \pm \Delta,
    \end{split}
\end{equation}
with $\delta N_g=N_g{-}1$. Thus, at particle-hole symmetry we can drop the $m$ index in the energies, but importantly not in the states. Note that Ref.~\cite{liTopologicalSymplecticKondo2023} only considers the paired states $\ket{P_m}$. This is valid in the limit $E_C-\Delta\ll\Delta,E_C$, and suffices to capture the symmetry structure of the symplectic Kondo model. Here, we relax this assumption and consider all states involved in second-order processes. 
Note also that there are excited states with a smaller gap from the ground space, but they cannot be reached by a single tunneling event.

The effective Hamiltonian is given to second order by
\begin{equation}
H_\mathrm{eff}=\hat{P}_G H_\mathrm{hyb} \frac{1}{E_\mathrm{GS}-H_\mathrm{imp}}H_\mathrm{hyb} \hat{P}_G,
\end{equation}
where $\hat{P}_G=\sum_{j\sigma}\ket{j\sigma}\bra{j\sigma}$ projects to the ground-state manifold. Inserting complete sets of states $\sum_X\ket{X}\bra{X}=1$ at both sides of the energy denominator we obtain
\begin{equation}
\label{eq:h_eff_form_1}
   \!\! H_\mathrm{eff} = -\!\sum_{\substack{jj'\\\sigma \sigma'}} \!
        f^{\dagger}_{j'\sigma'}f_{j\sigma}\!
        \sum_X \!\frac{\braket{j'\sigma'| H_\mathrm{hyb} |X}\!\!\braket{X| H_\mathrm{hyb}|j\sigma}}{\Delta E_X}.
\end{equation}
The only states with nonzero matrix element are the excited states of Eq.~\eqref{eq:ex_states}, so that the sum can be restricted to  $X\in\{P_m,A_m^j,U_m^{jj'\sigma\sigma'}\}$. As these are eigenstates of $H_\mathrm{imp}$, the denominator is replaced by the corresponding energy gaps. Note that as the total impurity charge is fixed in the ground space, if one matrix elements corresponds to tunneling out, the other corresponds to tunneling in.

We split $H_\mathrm{eff}=H_P+H_A+H_U+H_S$ into four types of process that can be read off from the different panels of Fig.~\ref{fig:SW_wide_fig}. All nonzero matrix elements are of amplitude $\frac{1}{\sqrt{2}}$ with all phases (except the sign $\sigma=\{\uparrow,\downarrow\}=\{+,-\}$) canceling out in Eq.~\eqref{eq:h_eff_form_1}. As an example consider $H_P$. In the transition $\ket{j\sigma}\to\ket{P_0}\to\ket{j'\sigma'}$ first a fermion tunnels out of dot $j$, involving  $d_{j\sigma}$ that yields the matrix element $\frac{1}{\sqrt{2}}t_j c^\dagger_{0j\sigma}$. Then a fermion tunnels into dot $j'$  (can be $j'{=}j$), involving $d^\dagger_{j'\sigma'}$ that yields the matrix element $\frac{1}{\sqrt{2}}t_{j'} c_{0j'\sigma'}$. Similarly, in the transition  $\ket{j\sigma}\to\ket{P_2}\to\ket{j'\sigma'}$ first a fermion of opposite spin $\bar\sigma$ tunnels into dot $j$, involving  $d^\dagger_{j\bar\sigma}$ that yields $\frac{\sigma}{\sqrt{2}}t_j c_{0j\bar\sigma}$. Then a fermion tunnels out yielding $\frac{\sigma'}{\sqrt{2}}t_{j'} c^\dagger_{0j'\bar\sigma'}$. Repeating for all four panels and rearranging, we get
\begin{widetext}
\begin{subequations}\label{eq:H_PAUS}
\begin{eqnarray}
     H_P &=& \sum_{\substack{jj'\\ \sigma \sigma'}} \left[-\frac{t_jt_{j'}}{2\Delta E_0^-} (f^\dagger_{j'\sigma'}  c_{0j'\sigma'})(c^\dagger_{0 j \sigma}f_{j\sigma})  + \frac{t_jt_{j'}}{2\Delta E_2^-}(f^\dagger_{j'\sigma'} \sigma'  c^\dagger_{0j'\bar{\sigma}'})( c_{0j\sigma}\sigma f_{j\bar{\sigma}}) \right], \label{eq:H_P} \\
    H_A &=& \sum_{\substack{j \\ \sigma \sigma'}}\left[-\frac{t_j^2}{2\Delta E_0^+} (f^\dagger_{j\sigma'}  c_{0j\sigma'})(c^\dagger_{0 j \sigma}f_{j\sigma})  + \frac{t_j^2}{2\Delta E_2^+}(f^\dagger_{j\sigma'} \sigma'  c^\dagger_{0j\bar{\sigma}'})( c_{0j\sigma}\sigma f_{j\bar{\sigma}}) \right], \label{eq:H_A} \\
    H_U&=&\sum_{\substack{j\ne j'\\ \sigma \sigma'}}\left[-\frac{t_jt_{j'}}{2\Delta E_2^+} (f^\dagger_{j'\sigma'}  c_{0j'\sigma'})(c^\dagger_{0 j \sigma}f_{j\sigma}) + \frac{t_jt_{j'}}{2\Delta E_0^+}(f^\dagger_{j'\sigma'} \sigma'  c^\dagger_{0j'\bar{\sigma}'})( c_{0j\sigma}\sigma f_{j\bar{\sigma}}) \right], \label{eq:H_U}\\
    H_S &=& \sum_{j \sigma} f^\dagger_{j \sigma}f_{j\sigma}\sum_{\substack{j'\ne j \\ \sigma'}}\left[-\frac{t_{j'}^2}{2\Delta E_2^+} c^\dagger_{0j'\sigma'}c_{0j'\sigma'}   - \frac{t_{j'}^2}{2\Delta E_0^+}c_{0j'\sigma'}c^\dagger_{0j'\sigma'} \right]. \label{eq:H_S}
\end{eqnarray}
\end{subequations}
\end{widetext}
Note the single sum on $j$ in $H_A$. This is because the ground states contain only paired (but not antipaired) dots. Thus, the transitions involving $\ket{A_m^j}$ cannot take $\ket{j\sigma}$ to a different $j$, but can flip its spin. This is in contrast to transitions through the paired excited states in $H_P$, that can reach any state in the ground space. The processes involving the unpaired excited states are split between $H_U$ and $H_S$. In these processes, first $\ket{j\sigma}\to\ket{U_m^{jj'\sigma\sigma'}}$ due to a fermion tunneling out of (into) a different dot $j'{\neq}j$ (note the difference in energy denominators with respect to $H_A$). Then $H_S$ accounts for that same fermion tunneling back into (out of) dot $j'$, thereby returning to $\ket{j\sigma}$, while $H_U$ accounts for a fermion tunneling into (out of) dot $j$, thereby reaching $\ket{j'\sigma'}$. Note that these processes cannot flip the spin of the original unpaired dot, i.e., reach $\ket{j\bar\sigma}$.

It is convenient to add the $j'{=}j$ terms to the sums in $H_U$ and $H_S$ and subtract them from $H_A$, thereby defining $\tilde H_U$, $\tilde H_S$, an $\tilde H_A$, respectively. Then, $\tilde H_U$
has an identical structure to $H_P$ (up to a substitution $\Delta E_m^-\to\Delta E_{2-m}^+$).  Using  $\sum_{j\sigma} f^\dagger_{j \sigma}f_{j\sigma}=1$ and discarding constant contributions to the energy yields
\begin{subequations}
\begin{eqnarray}
\label{eq:H_A_clean}
    \tilde{H}_A &=& \sum_{j\sigma}
        \left(\frac{t_j^2}{2\Delta E_0^+} + \frac{t_j^2}{2\Delta E_2^+}\right)
        f^\dagger_{j\sigma}f_{j\sigma}^{\,}, \\
    \tilde{H}_S &=&  \sum_{j \sigma}
        \left(\frac{t_{j}^2}{2\Delta E_0^+} - \frac{t_{j}^2}{2\Delta E_2^+}\right)
        c^\dagger_{0j\sigma}c_{0j\sigma}^{\,}.
\end{eqnarray}
\end{subequations}
Observe that $\tilde{H}_S$ is just a potential scattering term in the leads that can be discarded. For  $t_j{=}t$, $\tilde{H}_A$  is just a constant contribution to the energy and can also be ignored.

Thus, in the fully symmetric case we are left with $H_\mathrm{eff}=H_P+\tilde H_U$. Then, grouping the fermionic operators into vectors $\textbf{f}^\dagger \equiv (f_{1\uparrow}^\dagger,\dots,f_{k\uparrow}^\dagger,f_{1\downarrow}^\dagger,\dots,f_{k\downarrow}^\dagger)$ and
$\textbf{c}^\dagger_0\equiv(c_{01\uparrow}^\dagger,\dots,c_{0k\uparrow }^\dagger,c_{01\downarrow}^\dagger,\dots,c_{0k\downarrow }^\dagger)$ we can write 
\begin{equation}\label{eq:H-eff-fin}
\begin{split}
      H_\mathrm{eff} = -\frac{\lambda}{2}&\left[(\textbf{f}^\dagger \textbf{c}_0)(\textbf{c}_0^\dagger \textbf{f}) 
    + (\textbf{f}^\dagger\SigmaY \textbf{c}_0^*) 
      (\textbf{c}_0^T\SigmaY \textbf{f}) 
    \right],\\
   &\lambda=\frac{t^2}{E_C-\Delta}+\frac{t^2}{E_C+\Delta},
\end{split}
\end{equation}
with $\SigmaY \equiv \sigma_y{\otimes}\mathbb{1}_k$. 
Using the Fierz identity in Eq.~\eqref{eq:fierz_symplectic} for the $\mathrm{Sp}(2k)$ generators $T^A$, 
we arrive at the symplectic Kondo Hamiltonian of Eq.~\eqref{eq:H-Kondo}
\begin{equation}\label{eq:SW_iso_kondo}
    H_\mathrm{eff}=\lambda\sum_A(\textbf{f}^\dagger T^A \textbf{f})(\textbf{c}_0^\dagger T^A \textbf{c}_0)=\lambda \sum_A S^A J^A(0).
\end{equation}
Having laid the detailed groundwork, we can now consider how different perturbations affect this mapping.

\subsubsection{Channel-Dependent Tunneling}\label{sec:SW-Chan-Tun}
Channel asymmetry of the tunneling amplitudes $t_j=t+\delta t_j$ affects $H_\mathrm{eff}$ in two ways. First, consider its effect on $H_P+\tilde H_U$. These can be treated as in the supplemental of Ref.~\cite{liTopologicalSymplecticKondo2023} by absorbing channel dependence into non-normalized lead fermions $\tilde{c}_{0j\sigma} = \frac{t_j}{t} c_{0j\sigma}$. Introducing the diagonal matrix $\eta=\mathbb{1}{\otimes}\mathrm{diag}(t_1,\dots,t_k)/t$ we write\footnote{In Ref.~\cite{liTopologicalSymplecticKondo2023} $\eta$ as defined as the square root of the $\eta$ defined here in order to split its contribution between the impurity and lead terms. Then note a typo in Eq.~(S18) therein, that should read instead $\lambda_{AB}=\lambda\sum_C\kappa_{AC}\kappa_{CB}$.} 
\begin{equation}
    \textbf{c}_0^\dagger (\eta T^A\eta)\textbf{c}_0 = \tilde{\textbf{c}}_0^\dagger T^A \tilde{\textbf{c}}_0 = \tilde{J}^A(0).
\end{equation}
One can verify that $\eta T^A \eta$ satisfy the defining relation of the symplectic generators in Eq.~\eqref{eq:def-rel-gen}. Thus, although they are not orthogonal with respect to each other, they can be spanned by an orthogonal basis of generators
\begin{equation}
    \eta T^A \eta = \sum_B \frac{\lambda_{AB}}{\lambda} T^B,\quad \lambda_{AB}\equiv \lambda\mathrm{Tr}\{\eta T^A \eta T^B\}.
\end{equation}
Substituting this into Eq.~\eqref{eq:SW_iso_kondo} yields 
\begin{equation}\label{eq:SW_noniso_kondo}
   H_P+\tilde H_U=\lambda\sum_A S^A\tilde{J}^A(0)=\sum_{AB}\lambda_{AB} S^A J^B(0).
\end{equation}
The deviations of $\lambda_{AB}$ from the isotropic coupling $\lambda\delta_{AB}$ are linear in the tunneling perturbations $\delta t_j$.

However, we also have a local contribution from $\tilde{H}_A$ of Eq.~\eqref{eq:H_A_clean}. Using the channel polarized complementaries of Eq.~\eqref{eq:chan-basis}, to leading order in $\delta t_j$ it takes the form
\begin{equation}
    \tilde{H}_A=\sqrt{\tfrac{2k-2}{k}}\sum_j\frac{2\delta t_j}{E_C+\Delta}\mathcal{S}^j +     \mathrm{const}.
\end{equation}
Thus, we immediately see that tunneling asymmetry is relevant. While in specific limits, e.g., $E_C-\Delta\ll E_C,\Delta$, this contribution is negligible, it teaches us that such terms are allowed by symmetry. Hence, they will be generated from Eq.~\eqref{eq:SW_noniso_kondo} via RG according to Eq.~\eqref{eq:RG2} with a coefficient linear in $\delta t_j$. This can be demonstrated for specific examples, and is verified numerically.

\subsubsection{Spin-Dependent Tunneling}\label{sec:SW-Spin-Tun}
Channel symmetric but spin-dependent tunneling terms $t_{j\sigma}=t_\sigma=t{+}\sigma\delta t$ could arise, e.g., due to Zeeman splitting when using integer quantum Hall edge states as leads. They can immediately be incorporated into Eqs.~\eqref{eq:H_PAUS}. In this case, $\tilde{H}_A$ becomes
\begin{equation}
    \tilde{H}_A = \sum_{j\sigma}\frac{t_{\sigma}^2+t_{\bar{\sigma}}^2}{2\Delta E^+}f^\dagger_{j\sigma}f_{j\sigma}^{\,},
\end{equation}
which is a constant, as the prefactor can be pulled out of the sum. For $H_P+\tilde H_U$ we proceed as before, defining $\tilde{c}_{0j\sigma}\equiv \frac{t_\sigma}{t}c_{0j\sigma}$, that results in the diagonal matrix $\eta = \mathbb{1}_{2k} +\frac{\delta t}{t} (\sigma_z{\otimes}\mathbb{1}_k)$. Observe that $(\sigma_z{\otimes}\mathbb{1}_k)\equiv\tilde{T}^z$ is a (non-normalized) symplectic generator that squares to the identity. Then,
\begin{equation}
    \eta T^A \eta = T^A +\left(\frac{\delta t}{t}\right)^2 \tilde{T}^z T^A \tilde{T} ^z + \frac{\delta t}{t}\{\tilde{T}^z,T^A\}.
\end{equation}
The last term yields a complementary, while the first two terms remain in the generator space. Moreover, the first two terms are proportional to  $T^A$ if one works with a basis of generators that are given by a product of a Pauli matrix (or the identity) and a $k{\otimes}k$ matrix (as are all generator bases considered in this paper). Thus, we obtain
\begin{equation}\label{eq:SW_spin_tun}
    H_\mathrm{eff} = \sum_A \lambda_A S^A J^A(0) + \sum_A \mathit{\Lambda}_A S^A \mathcal{J}^A(0),
\end{equation}
where $\mathcal{J}^A$ is defined as the complementary obtained by the anticommutation of $T^A$ with $\tilde{T}^z$ (hence, the Roman index). Note that the deviation of the couplings $\lambda_A$ from isotropy is quadratic in $\delta t$, while the couplings $\mathit{\Lambda}_A$ are linear in $\delta t$. Numerically we find these terms to be irrelevant. We expect that carefully applying the RG equations of Appendix \ref{sec:poor-mans} to the set of $\mathrm{SU}(2k)$ generators (to also account for the $S^A\mathcal{J}^A$ term) will indeed demonstrate that no local terms emerge.

\subsubsection{Particle-Hole Asymmetry}\label{sec:SW-PH-asym}
Both gate detuning, $N_g=1{+}\delta N_g$, and an average (uniform) shift of the dot levels, $\bar\epsilon=\frac{1}{k}\sum_j\epsilon_j$, break the degeneracy of the couplings to the excited states with different $m{=}0,2$. The local contribution  $\tilde{H}_A$ is a constant due to channel symmetry, and can be ignored. However, the two terms in Eq.~\eqref{eq:H-eff-fin} now come with different coefficients
\begin{equation}
        H_\mathrm{eff}=-\frac{\lambda{+}\Lambda}{2} (\textbf{f}^\dagger \textbf{c}_0)(\textbf{c}_0^\dagger \textbf{f}) -\frac{\lambda{-}\Lambda}{2}(\textbf{f}^\dagger\SigmaY\textbf{c}_0^*)(\textbf{c}_0^T \SigmaY \textbf{f}),
\end{equation}
with $\lambda$ the average coupling and $\Lambda$ given by differences [see Eqs.~\eqref{eq:Lam-dNg} and \eqref{eq:Lam-barEps}].  Using the Fierz identities in Eqs.~\eqref{eq:fierz_symplectic} and \eqref{eq:fierz_complementary} for the $\lambda$ and $\Lambda$ terms, respectively, yields
\begin{equation} \label{eq:phasym_heff}
\begin{split}
    H_\mathrm{eff} &= \lambda \sum_A S^A J^A(0) +\Lambda \sum_\mathcal{A} \mathcal{S^A} \mathcal{J^A}(0) \\
    & -\frac{\Lambda}{2k}
    \underbrace{
    \sum_{j\sigma} f^\dagger_{j\sigma} f_{j\sigma}}_{= 1} \sum_{j'\sigma'}c_{0j'\sigma'}^\dagger c_{0j'\sigma'}.
\end{split}
\end{equation}
The second line is just a potential scattering term for the lead fermions, and can be discarded. Observe that both terms in $H_\mathrm{eff}$ are $\mathrm{Sp}(2k)$ scalars, which is expected, as the symplectic symmetry has not been broken. $S^A$ and $\mathcal{S^A}$ together form the generators of $\mathrm{SU}(2k)$, and, thus, the system flows to the isotopic fully-screened FL fixed point \cite{liTopologicalSymplecticKondo2023}.

All that remains is to explicitly express the couplings $\lambda$ and $\Lambda$. In the case of gate detuning $\delta N_g$, we have different energy denominators according to Eq.~\eqref{eq:SW_en_denom}, resulting in
\begin{eqnarray}
     \lambda &=& \frac{t^2}{2\Delta E^-_0} 
                        {+} \frac{t^2}{2\Delta E^-_2} 
                        {+} \frac{t^2}{2\Delta E^+_2} 
                        {+} \frac{t^2}{2\Delta E^+_0 }
                        = \lambda_{N_g=1} 
                        {+} \mathcal{O}(\delta N_g^2), \nonumber\\
     \Lambda&=& \frac{t^2}{2\Delta E^-_0} 
                        {-} \frac{t^2}{2\Delta E^-_2} 
                        {+} \frac{t^2}{2\Delta E^+_2} 
                        {-} \frac{t^2}{2\Delta E^+_0 }
                        \propto \delta N_g^2.\label{eq:Lam-dNg}
\end{eqnarray}
An average shift of the dot levels $\bar\epsilon$, on the other hand, modifies the denominators $\Delta E^\pm  = E_C \pm \sqrt{\Delta^2+\bar{\epsilon}^2}$ but does not break their degeneracy (thus no index $m$). However, the BCS paired and antipaired states [see Eq.~\eqref{eq:dot-space}] no longer consist of equal superpositions of the empty and doubly occupied states. This modifies the matrix elements by factors of $\cos^2\theta_{\mp}$ or $\sin^2\theta_{\mp}$ depending on the type of excited state, resulting in 
\begin{equation}\label{eq:Lam-barEps}
  \begin{split}
     \lambda &= \frac{t^2}{\Delta E^- } + \frac{t^2}{\Delta E^+} = \lambda_{\bar\epsilon=0} + \mathcal{O}(\bar\epsilon^2),
     \\
     \Lambda&=  \left(\frac{t^2}{\Delta E^-}-\frac{t^2}{\Delta E^+}\right) \cos2\theta_- \propto \frac{\bar{\epsilon}}{\Delta}.
 \end{split}
 \end{equation}
Observe that in both cases $\Lambda$ is linear in a small perturbation and can take either sign. Thus, these two contributions can be tuned to cancel each other out.

\section{Summary} \label{sec:Sumamry}
In this paper we systemically analyzed the effect of perturbations to the symplectic $\mathrm{Sp}(2k)$ Kondo model. This model has a non-Fermi Liquid (NFL) low-energy theory with a fractional entropy, similar to the multichannel $\mathrm{SU}(2)_k$ Kondo effect. Previous studies argued that, in contrast to multichannel Kondo, the symplectic model is stable with respect to channel asymmetry. The present  analytical and numerical analysis refutes this claim. We started by partitioning the space of operators that act on a symplectic ``spin'' into two irreducible representations of the symplectic group: The $\mathrm{Sp}(2k)$ generators, and all other complementary operators (excluding the identity). While the generators appear in the symplectic Kondo Hamiltonian, and define its NFL behavior, it is the complementaries that destabilize it. In the language of CFT, the generators correspond to descendant fields and are found to be marginal, while the complementaries correspond to a primary field and are thus relevant.

We then applied these observations to the symplectic Anderson-like model of \citet{liTopologicalSymplecticKondo2023}, that consists of $k$  quantum dots, each coupled to a separate lead, and in proximity to a superconducting island. In this model, channel symmetry can be broken by a variety of perturbations, e.g., asymmetry in the dot-lead tunneling, asymmetry in the SC pairing, dot-dependent energy levels, and dot-dependent magnetic fields. We found that all of these perturbations destabilize the NFL via the same effective complementary terms. This implies that tunable parameters, such as the dot-lead tunneling, can compensate for effects that are harder to control, such as the proximity-induced SC pairing. As a result, ${\sim }k$ physical parameters need to be tuned to achieve the NFL, just as in the $\mathrm{SU}(2)_k$ Kondo model. We also found that dot-dependent energy levels and magnetic fields enter only to second order, strongly suppressing their effect. We further note that global spin asymmetry does not destabilize the NFL. This is of practical advantage if one wishes to link multiple impurities via chiral edge states to implement anyonic braiding \cite{lotemManipulatingNonAbelianAnyons2022,renTopologicalQuantumComputation2024}. Such a setup assumes a uniform magnetic field, that is found to be marginal, as well as irrelevant spin-dependent tunneling due to Zeeman splitting of the edge states.

We now turn to discuss the implications of our analysis to generic Kondo problems. The key message is that when the group generators do not span the full space of impurity operators, we must carefully consider the effect of all complementary operators in this space. In the single-channel symplectic case it is of particular importance as the generators are marginal, but the complementaries are relevant. Then, in the presence of anisotropic Kondo couplings these complementary operators can effectively emerge via RG. Constructively, if the anisotropy involves generators that do not anticommute to zero or the identity, complementary terms will be generated. A known example is an $\mathrm{SU}(2)$ large spin-$s$ impurity coupled to a spin-$1/2$ channel. There are $s^2{-}1$ nontrivial Hermitian operators that can act on the impurity, but only three spin generators $S^{\mu=x,y,z}$. Hence, we have $s^2{-}4$ complementary terms, e.g.,  $(S^\mu)^2=\frac{1}{2}\{S^\mu,S^\mu\}\neq\mathbb{1}$. As a result, anisotropic Kondo couplings, with the easy or strong axis defined as the $\hat{z}$ direction, will generate the relevant impurity term $(S^z)^2$ \cite{konikInterplayScalingLimit2002,schillerPhaseDiagramAnisotropic2008,zitkoPropertiesAnisotropicMagnetic2008} -- see Appendix \ref{sec:poor-mans-SU2}.  A second example is the $\mathrm{SO}(5)\simeq\mathrm{Sp}(4)$ Kondo model of Refs.~\cite{mitchellNonFermiLiquidCoulomb2021,libermanCriticalPointSpinflavor2021}. This model was shown to be stable with respect to both channel asymmetry in the lead-dot tunneling $\delta t$ and an applied magnetic field $B$, as both map to symplectic generator perturbations. However, as shown here, a combination of these two marginal perturbations anticommutes to a relevant complementary operator. Specifically, this is a weak second-order effect, with the NFL-breaking energy scale strongly suppressed as $B^2\delta t^2$.

Thus, the flow to isotropy, or symmetry restoration \cite{konikInterplayScalingLimit2002}, that occurs in many single-channel problems, offers only a partial source of stability. We further note that, while symmetry restoration occurs in the $\mathrm{SO}(M)$ topological Kondo models, it is not the source of their stability, or topological protection. These models consist of $M$ Majorana fermions  $\gamma_{j=1\dots M}$ coupled to $M$ spinless leads. The dimension of the impurity space is $2^{(M-1)/2}$ for odd $M$ and   $2^{(M-2)/2}$ for even $M$.\footnote{This corresponds to the dimension of the fusion space of $M$ Majorana fermions in a specific fusion channel.} The $M(M{-}1)/2$ generators of $\mathrm{SO}(M)$ are then given by Majorana bilinears $i\gamma_j \gamma_{j^\prime}$. The $M{=}3$ case is somewhat trivial -- it maps to an anisotropic $\mathrm{SU}(2)$ spin-$1$ lead and spin-$1/2$ impurity, and so the generators span the impurity space. However, for $M{>}4$ we necessarily have complementary operators, given by the product of more than two Majoranas (for $M{=}5$ we have 10 bilinears, and 5 products of four Majoranas). Nevertheless, as the generators in this specific basis all square to the identity, no complementaries can be generated from asymmetric couplings. This highlights the topological nature of these Kondo models: Stability relies on the nonlocal nature of the impurity degrees of freedom, and not only on the flow to isotropy. 

Let us conclude by stressing that the question of stability cannot be addressed only at the level of the Kondo model. It also depends on the underlying realization and which perturbations arise naturally. The NFL of the symplectic Kondo model is unstable to complementary perturbations, but survives in the presence of the marginal generators. Thus, in the $\mathrm{Sp}(2k)$ Anderson model of \citet{liTopologicalSymplecticKondo2023} channel asymmetry destabilizes the NFL, while in the $\mathrm{Sp}(4)\simeq \mathrm{SO}(5)$ model of Refs.~\cite{mitchellNonFermiLiquidCoulomb2021,libermanCriticalPointSpinflavor2021} it does not. We hope that this understanding would help guide the search for robust Kondo effects, their experimental realization, and the demonstration of fractionalized degrees of freedom.

\begin{acknowledgments}
ML, SS, and ES acknowledge support from the European Research Council (ERC) Synergy funding for Project No.~951541. ML was also supported by the Israel Science Foundation (ISF), Grant No.~154/19.
MG has been supported by the ISF and the Directorate for Defense Research and Development (DDR\&D) Grant No.~3427/21, the ISF grant No.~1113/23, and the US-Israel Binational Science Foundation (BSF) Grant No.~2020072.
AMT and AW where supported by Office of Basic Energy Sciences, Material Sciences and Engineering Division, U.S. Department of Energy (DOE) under Contract No.~DE-SC0012704.  
\end{acknowledgments}

\begin{appendix}

\section{Group Theoretical Details}
\label{sec:Sp2k-properties}

In this appendix we review useful mathematical properties of the space of operators acting on a symplectic ``spin'' in the  $2k$-dimensional defining
representation. In Eq.~\eqref{eq:chan-basis-full} below we introduce an explicit basis for this space of  $2k{\times}2k$ Hermitian matrices, and one can verify all listed properties on it. As the operator space is isomorphic to the product space of two symplectic spins, it decomposes accordingly into three $\mathrm{Sp}(2k)$ irreducible representations of dimension $1$, $k(2k{+}1)$, and $(k{-}1)(2k{+}1)$ \cite{georgiLieAlgebrasParticle2019}, as in Eq.~\eqref{eq:10x10}, which is repeated here for clarity (see Appendix \ref{sec:qlabels} for notation)
\begin{equation}\label{eq:10x10-app}
    (10)_k \otimes (10)_k =  (00)_k \oplus (20)_k \oplus (01)_k.
\end{equation}
We identify the trivial $1$-dimensional representation $(00)_k$ with the identity operator $\mathbb{1}_{2k}$. The $k(2k{+}1)$-dimensional representation corresponds to the adjoint representation $(20)_k$ which we identify with the generators $T^A$ with $A=1,\dots,k(2k{+}1)$. These satisfy the defining relation of the symplectic group
\begin{equation}\label{eq:def-rel-gen}
    [T^A]^T  =- \SigmaY\, T^A\, \SigmaY,
\end{equation}
with $\SigmaY \equiv \sigma_y{\otimes}\mathbb{1}_k$. We then identify the remaining (complementary) operators with the  $(01)_k$ representation of dimension $(k{-}1)(2k{+}1)$. We denote them in calligraphic script $\mathcal{T^A}$ with $\mathcal{A}=1,\dots,(k{-}1)(2k{+}1)$. Aside from the defining representation $(10)_k$, this is another fundamental representation of $\mathrm{Sp}(2k)$ (in the Dynkin labeling convention, these have all-zero labels except for a single entry of~1). As the super-operator that takes an operator $O$  to $[\SigmaY O\, \SigmaY]^T$ squares to the identity, the complementaries necessarily satisfy
\begin{equation}\label{eq:def-rel-comp}
        [\mathcal{T^A}]^T = + \SigmaY\, \mathcal{T^A}\, \SigmaY.
\end{equation}
Note that the defining representation of $\mathrm{SU}(2k)$ is also $2k$-dimensional, and together $T^A$ and $\mathcal{T^A}$ form the $(4k^2{-}1)$-dimensional adjoint representation of its  generators.

The generators and complementaries are by construction traceless, and chosen to be mutually orthogonal and normalized
\begin{equation}\label{eq:gen-norm}
    \mathrm{tr}\{T^A T^B\}=\delta_{AB},\ \ \mathrm{tr}\{\mathcal{T^A T^B}\}=\delta_\mathcal{AB},\ \ \mathrm{tr}\{T^A \mathcal{T^B}\}=0.
\end{equation}
By definition the generators are closed under commutation
\begin{equation}
    [T^A,T^B]=if_{ABC}T^C,
\end{equation}
with $f_{ABC}$ the basis dependent structure constants. The anticommutators of generators, on the other hand, yield
\begin{equation}\label{eq:comp-anticomm}
    \{T^A,T^B\}=\tfrac{1}{k}\delta_{AB}\mathbb{1}_{2k} + d_{AB\mathcal{C}}\mathcal{T^C}, 
\end{equation}
where we have assumed the normalization condition of Eq.~\eqref{eq:gen-norm}. This is a specific case for a similar-looking relation for the generators in the defining representation of $\mathrm{SU}(N{=}2k)$. There, complementaries $\mathcal{T^C}$ are part of the generator space, but in the symplectic case, the generators are excluded from the output of the anticommutators. This can be verified by plugging them into Eq.~\eqref{eq:def-rel-comp}. If inclined, one can write similar equations for commutators and anticommutators of the complementaries or of complementaries with the generators. Using the explicit basis in Eq.~\eqref{eq:chan-basis-full}, we can verify that anticommutators of generators yield the full complementary space. Note that the $d$ ``structure constants'' depend not only on the basis, but also on the representation (as does the space of complementaries).

The generators and the complementaries satisfy  the following Fierz identities:
\begin{subequations}
\begin{eqnarray}
    \sum_A T^A_{ab} T^A_{cd} &=& \tfrac{1}{2}\left[\delta_{ad}\delta_{bc}-\SigmaY_{ca} \SigmaY_{bd}\right] \label{eq:fierz_symplectic},\\
    \sum_\mathcal{A} \mathcal{T}^\mathcal{A}_{ab} \mathcal{T}^\mathcal{A}_{cd} &=& \tfrac{1}{2}\left[\delta_{ad}\delta_{bc}+\SigmaY_{ca} \SigmaY_{bd}\right]-\tfrac{1}{2k} \delta_{ab}\delta_{cd}.\ \  \label{eq:fierz_complementary}
\end{eqnarray}
\end{subequations}
Equation \eqref{eq:fierz_complementary} follows from the fact that together $T^A$ and $\mathcal{T^A}$ form the basis of generators of $\mathrm{SU}(2k)$, and these satisfy the Fierz identity
\begin{equation}
    \sum_A T^A_{ab} T^A_{cd}
    + \sum_\mathcal{A} \mathcal{T}^\mathcal{A}_{ab} \mathcal{T}^\mathcal{A}_{cd} 
    = \delta_{ad}\delta_{bc}-\tfrac{1}{2k}\delta_{ab}\delta_{cd}.
\end{equation}

\subsection{Explicit Bases}

Let us complete the channel polarized basis introduced in Eq.~\eqref{eq:chan-basis}. The space of $2k{\times}2k$ Hermitian matrices can be spanned by tensor products of a Pauli matrix $\sigma_\mu$ or the identity $\mathbb{1}$ and a $k{\times}k$ Hermitian matrix that is either symmetric ($S$) or anti-symmetric ($iA$). One can show that products of the form $\sigma_\mu{\otimes}S$ and $\mathbb{1}{\otimes}iA$ satisfy the defining condition of the generators Eq.~\eqref{eq:def-rel-gen}, while all products of the form $\sigma_\mu{\otimes}iA$ and $\mathbb{1}{\otimes}S$ satisfy Eq.~\eqref{eq:def-rel-comp}. For $k{=}2$ one can choose $iA{=}\tau_y$ and $S{\in}\{\mathbb{1},\tau_z,\tau_x\}$ to get the Pauli basis of Eq.~\eqref{eq:Pauli-basis}. For larger $k$ we can construct $S$ and $iA$ by embedding the Pauli matrices into specific cells of $k{\times}k$ zero matrices. Defining $k$-dimensional basis vectors $\ket{j}$ with all zero elements except the $j$th element which is 1, we explicitly write
\begin{equation}
\begin{split}
    Z_{jj'}&\equiv \Pi_j{-}\Pi_{j^\prime},  \ \ \ \ 
    \Pi_j \equiv \ket{j}\bra{j},\\
    X_{jj'}&\equiv \ket{j}\bra{j'}+\ket{j'}\bra{j},\\
    Y_{jj'}&\equiv- i\left(\ket{j}\bra{j'}-\ket{j'}\bra{j}
    \right)\,.
\end{split}
\label{eq:XYZ}
\end{equation}
With these we can define a complete basis of generators (left column) and complementaries (right column)
\begin{alignat}{3} 
    &T^{\mu j} {=} \tfrac{1}{\sqrt{2}}(\sigma_\mu{\otimes}\Pi_j),
    &&\mathcal{T}^j {=} \sqrt{\tfrac{k}{2k-2}}\left(\mathbb{1}{\otimes}\Pi_j{-}\tfrac{1}{k}\mathbb{1}_{2k}\right),\notag\\
    &T^{\mu jj^\prime} {=} \tfrac{1}{2}(\sigma_\mu{\otimes}X_{jj^\prime}),\ \  
    &&\mathcal{T}^{\mu jj^\prime} {=} \tfrac{1}{2}(\sigma_\mu{\otimes}Y_{jj^\prime}),
    \label{eq:chan-basis-full}\\
    &T^{jj^\prime} {=} \tfrac{1}{2}(\mathbb{1}{\otimes}Y_{jj^\prime}),
    &&\mathcal{T}^{jj^\prime} {=} \tfrac{1}{2}(\mathbb{1}{\otimes}X_{jj^\prime}), \notag 
\end{alignat}
where $\mu=x,y,z$ and we take $j<j^\prime$  to avoid double counting and sign ambiguities. The first line coincides with Eq.~\eqref{eq:chan-basis}. One can readily check that we indeed have $k(2k{+}1)$ generators, corresponding to the adjoint representation of $\mathrm{Sp}(2k)$. Counting the complementaries and subtracting one (as the $\mathcal{T}^j$ terms are linearly dependent) we get $(k{-}1)(2k{+}1)$ elements. Thus, in total we have $4k^2{-}1$ operators, corresponding to the generators of $\mathrm{SU}(2k)$ and spanning the full impurity operator space (excluding the identity). For $k{=}2$, we can relate these operators to the Pauli basis of Eq.~\eqref{eq:Pauli-basis}
\begin{alignat}{3}
    &\begin{array}{@{}l}
         T^{\mu 0} \equiv\tfrac{1}{\sqrt{2}}(T^{\mu 1}{+}T^{\mu 2}),\\
         T^{\mu z} \equiv\tfrac{1}{\sqrt{2}}(T^{\mu 1}{-}T^{\mu 2}),
    \end{array} \ \ 
    && \mathcal{T}^{0z}\equiv\mathcal{T}^1, \notag\\
    &T^{\mu x}\equiv T^{\mu 12}, 
    && \mathcal{T}^{\mu y}\equiv\mathcal{T}^{\mu 12}, \label{eq:Pauli-basis-app}\\
    & T^{0y}\equiv T^{12},
    && \mathcal{T}^{0x}\equiv\mathcal{T}^{12}. \notag
\end{alignat}

\subsection{Root Space}
\label{sec:roots}
\begin{figure}[tb]
\begin{center}
\includegraphics[width=1\linewidth]{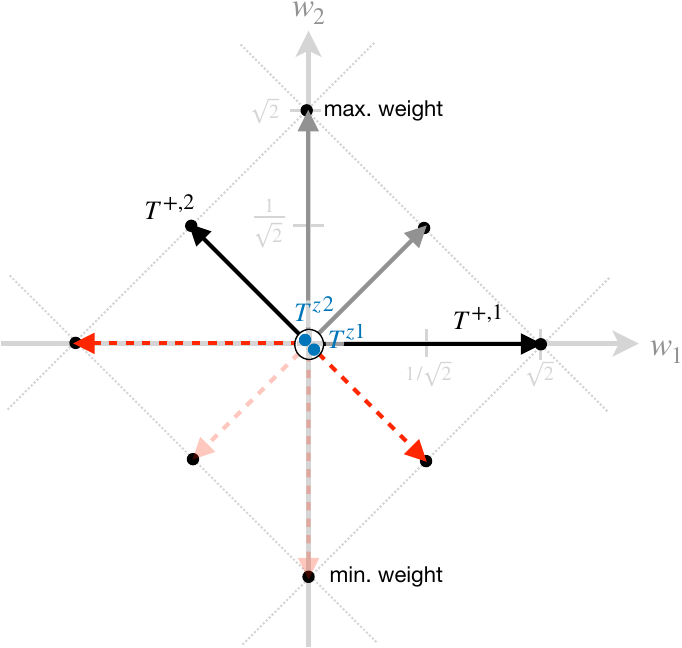}
\end{center}
\caption{Root system of the Lie algebra $\mathfrak{sp}(4)$.  Since $\mathrm{Sp}(2k)$ is a symmetry of rank $k$, the root diagram for  $\mathrm{Sp}(4)$ becomes ($k{=}2$)-dimensional, having weights $(w_1,w_2)$. Roots correspond to the generators $\tilde{T}^A$, and are depicted as points as well as arrows. $\mathrm{Sp}(4)$ has a total of $10$ generators, with the Cartan generators $\tilde{T}^{zj}$  located at the origin (blue points). Roots with opposite locations with respect to the origin are Hermitian conjugates of each other (also the transpose of each other, since all generators $\tilde{T}^A$ are real matrices). Hence, only the Cartan generators $\tilde{T}^{zj}$ are Hermitian (diagonal actually). The black-solid (red-dashed) vectors describe the {\it simple positive (negative) roots}, or the elementary set of raising (lowering) operators, respectively.}
\label{fig:Sp4-roots}
\end{figure}

The generators in Eq.~\eqref{eq:chan-basis-full} can be reorganized into a root system, as depicted in Fig.~\ref{fig:Sp4-roots}. First, $k$ mutually commuting (i.e., Cartan) generators are selected and placed at the center of the diagram. As these mutually commute, here we choose them all to be diagonal
\begin{eqnarray}
  \tilde{T}^{zj} &\equiv&   T^{zj} 
  = \tfrac{1}{\sqrt{2}}(\sigma_z {\otimes}  \Pi_j) \,,
  \quad (j=1,\ldots,k). \qquad
\label{eq:Troot:z}
\end{eqnarray}
Note that the choice of Cartan generators is arbitrary, and indeed for the EK analysis in Sec.~\ref{sec:EK} we make a different choice. To the $k$ Cartan generators we add an elementary set of $k$ raising operators, i.e., the simple positive roots
\begin{eqnarray}\label{eq:Troot:+}
    \tilde{T}^{+,1} &=& \tfrac{1}{\sqrt{2}}(T^{x1} + i T^{y1}) 
        = \sigma_+ {\otimes} \Pi_1,
    \qquad\, (j=1)\notag  \\
    \tilde{T}^{+,j} &=& \tfrac{1}{\sqrt{2}}( T^{z,j-1,j} - i T^{j-1,j} ) \\
   &=& \tfrac{1}{2\sqrt{2}}(
    \sigma_z{\otimes}X_{j-1,j}
    -i\mathbb{1}{\otimes}Y_{j-1,j} ),
    \quad (j=2,\ldots,k) \notag
\end{eqnarray}
with $\sigma_+ \equiv \frac{1}{2}(\sigma_x {+} i\sigma_y)$. These are plotted in Fig.~\ref{fig:Sp4-roots} (solid-black arrows), together with their respective $k$ lowering operators $\tilde{T}^{-,j} \equiv (\tilde{T}^{+,j})^\dagger$, i.e., the simple negative roots (dashed-red arrows).

The complete set of generators $\tilde{T}^A$, including the remainder of the non-diagonal generators, can be obtained by applying the above elementary set of raising and lowering operators to any existing generator in the set $\tilde{T}^A$ as in Eqs.~\eqref{eq:Troot:z} or \eqref{eq:Troot:+}. These are plotted as the remaining arrows in Fig.~\ref{fig:Sp4-roots}.  By applying raising operators,  which in operator space translates to taking commutators, this reaches a ``maximum weight'' operator or state. Repeated application of lowering operators then sequences the full irreducible operator $\tilde{T}$, which thus transforms completely analogous to a multiplet in a particular irreducible representation. As apparent from the above construction, in the strictly $\mathrm{Sp}(2k)$ symmetric case only the generators in Eqs.~\eqref{eq:Troot:z} and \eqref{eq:Troot:+} are required \cite{weichselbaumNonabelianSymmetriesTensor2012,weichselbaumQSpaceOpensourceTensor2024pub}.

The irreducible operator $\tilde{T}$ transforms in the adjoint representation. Each of its generators acquires weight vectors $w^A$ that are obtained by taking the commutators with the Cartan subalgebra $[\tilde{T}^{zj}, \tilde{T}^A] = w^A_j \tilde{T}^A$. By construction, the weights for the generators in the Cartan subalgebra $A \in \{zj'\}$ are trivially all zero. This basis, thus, reflects a particularly simple canonical format of the structure constants. For comparison, e.g., for $\mathrm{SU}(2)$ with $\tilde{T} = \{ -\tfrac{1}{\sqrt{2}} S^+, S^z,
\tfrac{1}{\sqrt{2}} S^- \}$ one has $[S^z,S^+] = +1\, S^+$. Hence, $w^+ = +1$, which corresponds to the state $S_z=+1$ in the adjoint multiplet, i.e., $S=1$ that also contains three elements. Now, because the Cartan subalgebra contains $k$ diagonal generators, there are $j=1,\ldots,k$ weights $w^A_j$ for each generator $\tilde{T}^A$. Hence, $\mathrm{Sp}(2k)$ is referred to as a symmetry of rank $k$. With this, the weight space becomes $k$-dimensional. From a practical point of view, this implies that in numerical simulations, irreducible representations, and thus state multiplets, have typical dimensions up to $\sim 10^k$ \cite{weichselbaumQSpaceOpensourceTensor2024pub} that grows {\it exponentially} with the symmetry rank $k$ since this needs to explore a $k$-dimensional {\it volume} of weights.

Plotting the weights for the generators $\tilde{T}^A$ above provides the root diagram. For $k=2$, this is two-dimensional, and is presented in Fig.~\ref{fig:Sp4-roots}. The maximum weight state shown at the top is destroyed by any of the raising operators (cf.~caption of Fig.~\ref{fig:Sp4-roots}), whereas the minimum weight state at the bottom is destroyed by any lowering operator. The commutator of two generators $\tilde{T}^A$ in root space results in {\it vector addition} of their respective weight vectors (arrows in Fig.~\ref{fig:Sp4-roots}). If the resulting weight is outside the weight diagram (outer gray-dotted lines), that commutator necessarily results in zero.

Observe that for $\mathrm{Sp}(2k)$ we have roots of two lengths. For the Cartan generators as chosen in Eq.~\eqref{eq:Troot:z}, the long roots align along the weight axes, i.e., are $\sigma_\pm{\otimes}\Pi_j$, corresponding to $w^{zj}=\pm \sqrt{2}$ and $w^{zj'}=0$ for $j'\neq j$. The short roots align along the intermediate directions between every two pairs of weight axes, i.e, have $|w^{zj_1}|=|w^{zj_2}|=1/\sqrt{2}$ and $w^{zj'}=0$ for all other $j'\neq j_1,j_2$. It is also instructive to compare the root diagram of $\mathrm{Sp}(4)$ with that of $\mathrm{SU}(4)$. While the generators of the former are a subset of those of the latter, the root diagram changes entirely. To start with, all roots of $\mathrm{SU}(4)$ have the same length. Moreover, while $\mathrm{Sp}(2k{=}4)$ is a rank-$(k{=}2)$ symmetry, i.e., has a two-dimensional root diagram, $\mathrm{SU}(N{=}4)$ is a rank-$(N{-}1{=}3)$ symmetry, i.e., has a $3$-dimensional root diagram with 3 Cartan generators at the center. All of its 12 non-zero roots are uniformly spread. The $\mathrm{SU}(3)$ subgroup of $\mathrm{SU}(4)$, which like $\mathrm{Sp}(4)$ is also two-dimensional, describes a hexagon [6 non-zero roots plus 2 Cartans, giving the expected total of 8 generators for $\mathrm{SU}(3)$], in stark contrast to the root diagram of $\mathrm{Sp}(4)$ in Fig.~\ref{fig:Sp4-roots}. It is the hexagon in the root space of $\mathrm{SU}(3)$ that is inherited by the root diagram of $\mathrm{SU}(4)$.

\subsection{Multiplet Labels}\label{sec:qlabels}
The maximum weight state, e.g., as indicated in Fig.~\ref{fig:Sp4-roots}, is unique to any multiplet. Hence, its weights $w^{\rm max}$ can be used to identify a multiplet. In practice, this includes a linear map $w^{\rm max} \to q = M w^{\rm max}$ with some matrix $M$ \cite{weichselbaumNonabelianSymmetriesTensor2012} which brings the (likely non-standard) maximum weight labels $w^{\rm max}$ to standard Dynkin format $q \equiv (q_1,q_2,\ldots,q_k)$ where any $k$-tuple with $q_i \in \mathbb{N}_0$ represents a valid multiplet. For the case of $\mathrm{SU}(N)$ which is a symmetry of rank $k=N-1$, the Dynkin labels directly identify the respective Young tableau in terms of offsets of the number of boxes having $k$ rows \cite{weichselbaumQSpaceOpensourceTensor2024pub}. For compactness, we adopt the \QSpace notation
\begin{eqnarray}
   q \equiv (q_1 q_2 \ldots q_k)
\end{eqnarray}
without any space or separators, where $q_i$ is extended alphabetically for $q_i \ge 10$ \cite{weichselbaumQSpaceOpensourceTensor2024pub}. Since $\mathrm{Sp}(2k)$ is self-dual, it follows that $\bar{q} = q$.

For example, for $\mathrm{Sp}(4)$ the $4$-dimensional defining representation is given by $q_{\rm def} = (1,0) \equiv (10)$. The $10$-dimensional adjoint is given by $q_{\rm adj} = (2,0) \equiv (20)$. The tensor product of the defining with its dual $\bar{q}_{\rm def} = q_{\rm def}$ must also fuse to the adjoint. Indeed, $(10) \otimes (10) = (00) \oplus (20) \oplus (01)$ with $(00) \equiv (0,0)$ the scalar (singleton) representation, and $(01) \equiv (0,1)$ a representation of dimension $5$. The standard Dynkin labels generalize in a simple way to $\mathrm{Sp}(2k)$ when $k$ is increased: all multiplet labels above simply get zeros appended up to a total
of $k$ integers and we adopt the shorthand notation $(q_1 q_2)_k\equiv(q_1 q_2 0\ldots 0)$ to  indicate trailing zeros in a tuple of integers of length $k$. We then denote the defining representation as $(10)_k$ and the adjoint as $(20)_k$. The fusion of the defining representation with its dual (i.e., with itself) then yields Eqs.~\eqref{eq:10x10} and \eqref{eq:10x10-app}.

\section{Poor Man's Scaling}
\label{sec:poor-mans}

In this appendix we will derive the (poor man's) perturbative RG equations for an arbitrary Lie group and demonstrate how anisotropies in the Kondo couplings generate local complementary terms. The starting point for the poor man's scaling approach \cite{andersonPoorManDerivation1970} is a generic Kondo Hamiltonian
[as in Eq.~\eqref{eq:H-Kondo}]
\begin{eqnarray}\label{eq:PMS:Kondo:H}
   H &=&  \underbrace{
   \sum_a \lambda_{ab} S^{a} J^b(0)}_{\equiv H_K}
   \ +\ \underbrace{
   \sum_{k\xi} \varepsilon_k \,
     c^\dagger_{k\xi} c^{\,}_{k\xi} }_{\equiv H_\mathrm{bath}}.
\end{eqnarray}
$\xi=1,\dots,M$ is a combined index for all fermionic species [e.g., $\xi\equiv(j,\sigma)$ for the channel and spin degrees of freedom in the symplectic case, with $M=2k$]. The $N$-dimensional impurity ``spin'' operators $S^{a}$ couple to the bath ``spin'' current at the location of the impurity
\begin{equation}\label{eq:PMS:Kondo:s0}
   J^a(0) = 
   \frac{1}{L}\sum_{\substack{k k'\\\xi\xi'}}
   c^\dagger_{k\xi}
      T^a_{\xi\xi'}
   c^{\,}_{k'\xi'} \equiv \textbf{c}(0)^\dagger T^a \textbf{c}(0),
\end{equation}
with $T^a$ the group generators in an $M$-dimensional representation, $L$ the size of the bath, and $\mathbf c(0)
\equiv  \frac{1}{\sqrt{L}}\sum_{k} \mathbf c_{k}$ the Fourier transform for the $L$ bath modes $k$ present in the system up to bandwidth $D$ and vectorized in $\xi$. The generator basis in Eq.~\eqref{eq:PMS:Kondo:H} is assumed to be in a standard Hermitian form with the normalization convention ${\rm tr}\{T^a T^b\} = \delta_{ab}$. The structure constants thus take the form $[T^a,T^b]= i f_{abc} T^c$. We use lowercase roman indices to stress that the discussion is not restricted to the specific case of the symplectic operators. Importantly, we do not restrict the ``spin'' operators $S^a$ to be in the same representation as the currents $J^a$, but they obey the same normalization convention as the $T^a$.

\subsection{Derivation of RG Equations}

The following assumes, for simplicity, a uniform density of states $\rho$ and hybridization of the bath levels within the bandwidth $\varepsilon_k \in [-D,D]$. Then, with $D$ much larger than any local energy scale within $H_K$, we split $c_{k\xi}$ into fast modes with $\vert \varepsilon_k \vert \in [D', D]$ and slow modes with $\vert \varepsilon_k \vert \in [0,D']$. The Hamiltonian can be schematically written as 
\begin{equation}
    H = \left (\begin{array}{cc}
       H_{ff}  & H_{fs} \\
       H_{sf}  & H_{ss} 
    \end{array} \right),
\end{equation}
where $H_{ff}$ ($H_{ss}$) only acts in the Hilbert space of fast (slow) modes and $H_{sf},H_{fs}$ mix the sectors. Following a textbook approach~\cite{ColemanBook} we integrate out the fast modes, leading to the following correction of the slow Hamiltonian 
\begin{equation}
    \delta H_{ss}  = - H_{sf} H_{ff}^{-1} H_{fs}.
\end{equation}
In the present case $H_{fs} =  \lambda_{ab} S^a (J^b_{sf} {+} J^b_{fs})$, where the subscript on the $J$ indicates whether the $c$ or $c^\dagger$ is the fast or slow field. In particular, we will drop the Kondo term from $H_{ff}$  to get the leading order RG equations.
Denoting slow momenta by lower case and fast momenta by upper case letters we find (projectors on the slow sector are implied here)
\begin{align}
    &J^{b}_{fs} H_{ff}^{-1} J^{b'}_{sf} + J^{b}_{sf} H_{ff}^{-1} J^{b'}_{fs}  \notag\\
    & = \frac{1}{L^2}  \sum_{\substack{k,p,\\K,P}} \Big( \mathbf{c}^\dagger_{K} T^{b} \mathbf{c}_{k} H_{ff}^{-1} \mathbf{c}^\dagger_{p} T^{b'} \mathbf{c}_{P} + \mathbf{c}^\dagger_{k} T^{b} \mathbf{c}_{K} H_{ff}^{-1} \mathbf{c}^\dagger_{P} T^{b'} \mathbf{c}_{p} \Big ) \notag\\
    & = \frac{1}{L^2} \sum_{k,p,K} \text{tr}\{T^{b} \mathbf{c}_{k} \mathbf{c}^\dagger_p T^{b'}\} \frac{\theta(-K)}{\vert \varepsilon_K \vert} + \mathbf{c}^\dagger_{k} T^{b} T^{b'} \mathbf{c}_p \frac{\theta(K)}{\vert \varepsilon_K \vert} \notag \\
    & = \frac{\rho \ln(D/D')}{L} \sum_{k,p} (\mathbf{c}^\dagger _k [T^{b},T^{b'}] \mathbf{c}_p + \delta_{k,p} \text{tr}\{T^{b} T^{b'}\}) \notag \\
    & = \rho \ln(D/D') [i f_{bb' c} J^{c}_{ss} + \delta_{bb'}].
\end{align}
We will drop the subscript $ss$ in what follows and we used
\begin{equation}
    \frac{1}{L} \sum_{K \text{fast}} \frac{\theta(K)}{\varepsilon_K} \simeq \rho \ln (D/D') \equiv \rho \ell.
\end{equation}
Substituting back into $\delta H_{ss}$ yields 
\begin{equation}
    \delta H_{ss} = -\rho \ell \lambda_{ab} \lambda_{a'b'} S^a S^{a'} (i f_{bb'c} J^c + \delta_{bb'}).
\end{equation}
Then, using $S^a S^{a'}=(\{S^a,S^{a'}\}{+}[S^a,S^a])/2$ and 
\begin{equation}
\{S^a, S^{b}\} = \frac{2}{N}\delta_{ab} \mathbb 1_N + d_{abc} S^c,
\end{equation}
with $N$ the dimension of the impurity ``spin'' space, we arrive at the corrected Hamiltonian
\begin{equation}
    \delta H_{ss} =  \tfrac{1}{2}\rho \ell \lambda_{ab} \lambda_{a'b'} (f_{aa'd}f_{bb'c} S^d J^c -\delta_{bb'} d_{aa'c} S^{c} ).
\end{equation}
Note that we dropped the term proportional to unity in the anticommutator as it provides a mere shift in energy. Resumming these corrections following standard RG concepts yields
\begin{subequations}
\label{eq:RGApp}
\begin{eqnarray}
    \frac{d \tilde \lambda_{ab}}{d \ell} &=& \frac{1}{2} f_{ace} f_{bdf} \tilde \lambda_{cd} \tilde \lambda_{ef},  
\label{eq:RGAppa} \\
      \frac{d \tilde B_c}{d \ell} &=& \tilde B_c -  \frac{1}{2} d_{abc} 
    \tilde \lambda_{ad} \tilde \lambda_{bd}.
\label{eq:RGAppb}
\end{eqnarray}   
\end{subequations}
The tilde denotes dimensionless $\tilde  \lambda_{ab} = \rho \lambda_{ab}$ and $\tilde B_c =  B_c/D$. Here, we have included the trivial engineering dimension of the magnetic fields, that may be attributed to the inclusion of $D$ in $\tilde B_c$.

\subsection{Discussion}\label{sec:poor-mans-disc}
Equations \eqref{eq:RGApp} are the main results of this appendix and the basis for Eqs.~\eqref{eq:RGeqMaintext} of the main text. As mentioned, the derivation here is for arbitrary generators, i.e., any subgroup of $\mathrm{SU}(N)$ and any impurity representation (note that different representations yield different $d_{abc}$). The key observation of this calculation is that the poor man's RG scaling also generates purely local impurity terms, namely the last term in Eq.~\eqref{eq:RGAppb}.

It is insightful to consider these equations for an anisotropic but diagonal model, i.e., $\lambda_{ab} \rightarrow \lambda_a \delta_{ab}$.\footnote{Any real $\lambda_{ab}=U_{ac}\lambda_c V_{cb}^T$ can be brought to diagonal form by a singular value decomposition (SVD) with $U$ and $V$ real orthogonal matrices. These can be absorbed as a change of the generator basis. In the symplectic Kondo model considered in this paper $U=V$ as $\lambda_{ab}$ is symmetric.} Then Eqs.~\eqref{eq:RGApp} take the form
\begin{eqnarray}
    \frac{d\tilde\lambda_{ab}}{d\ell} &=& \frac{1}{2} f_{acd}f_{bcd}\tilde \lambda_c\tilde\lambda_d,\\
    \frac{d\tilde{\mathcal{B}}_c}{d\ell} &=& \tilde{\mathcal{B}}_c-\frac{1}{2}d_{aac}\tilde\lambda_a\tilde\lambda_a,
\end{eqnarray}
with implicit summation on $c$ and $d$ in the first line and on $a$ in the second. Observe that in the fully isotropic case, $\lambda_a=\lambda$, no local terms are generated, as $\sum_a d_{aac} = 2 \sum_{a} \mathrm{tr} \{ (S^a)^2 S^c\}=0$ since the Casimir operator $\sum_{a}(S^a)^2 \propto \mathbb 1_N$. Moreover, the isotropic case remains isotropic since $\sum_{cd} f_{acd} f_{bcd} =\mathrm{tr}(T^a T^b)
\propto \delta_{ab}$. Once we deviate from isotropy, nondiagonal $\lambda_{ab}$ (coupling to $S^aJ^b$) can be generated. Note that, if we cast the model into canonical form, i.e., express the Hamiltonian in terms of roots (see Appendix \ref{sec:roots}), the only nondiagonal $\lambda_{ab}$ that can be generated are those for which both $a$ and $b$ are in the selected Cartan subalgebra.

While the discussion here is generic, in the main text we keep $\lambda_{ab}\neq 0$ for $a,b \rightarrow A,B$ in the symplectic Lie algebra, only. By closure of the Lie algebra, only $\lambda_{AB}$ are generated under RG. However, due to the anticommutator $\{S^A,S^B\}$, local complementary fields $\mathcal{B_C}$ appear as well [cf.~Eq.~\eqref{eq:comp-anticomm}] for sufficiently anisotropic couplings.

\subsection{Test Case: \texorpdfstring{$\mathrm{SU}(2)$}{SU(2)}}\label{sec:poor-mans-SU2}
It is instructive to test the result for $k{=}1$, i.e., $\mathrm{Sp}(2){\simeq}\mathrm{SU}(2)$. In our convention $T^{\mu=x,y,z} = \sigma_\mu/\sqrt{2}$ to ensure $\mathrm{tr}\{T^\mu T^\nu\} = \delta_{\mu\nu}$, and thus
\begin{equation}
    f_{\mu\nu\eta} = -i\text{tr}\{[\sigma_\mu,\sigma_\nu]\sigma_\eta\}/\sqrt{8}  = \sqrt{2}\epsilon_{\mu\nu\eta}.
\end{equation}
Choosing a diagonal basis $\lambda_{\mu\nu} = \lambda_\mu \delta_{\mu\nu}$, the RG equations are $ d \tilde \lambda_x/d \ell  = 2\tilde \lambda_y \tilde \lambda_z$ and cyclic permutations thereof (cf.~Ref.~\cite{ColemanBook}). Note that, even in the presence of anisotropy, no local terms can be generated, as for a spin-$1/2$ we have no complementary terms.

As a simple nontrivial example consider a spin-$1$ Kondo impurity coupled to a spin-$1/2$ lead with anisotropic diagonal couplings as above. To leading order, the renormalization of the $\tilde \lambda_\mu$'s is independent of the representation, i.e., is the same as for the spin-$1/2$ case. However, now we have a larger impurity space and thus complementaries. It is convenient to take the 8 Gell-Mann matrices (which are also denoted by $\lambda$)
\begin{alignat}{3}
    &\lambda_1=& \Big(\begin{smallmatrix} 0&1&0\\1&0&0\\0&0&0 \end{smallmatrix}\Big),\ \  
    &\lambda_2=& \Big(\begin{smallmatrix} 0&-i&0\\i&0&0\\0&0&0 \end{smallmatrix}\Big),\ \  
    &\lambda_3= \Big(\begin{smallmatrix} 1&0&0\\0&-1&0\\0&0&0 \end{smallmatrix}\Big), \notag\\
     &\lambda_4=& \Big(\begin{smallmatrix} 0&0&1\\0&0&0\\1&0&0 \end{smallmatrix}\Big),\ \  
    &\lambda_5=& \Big(\begin{smallmatrix} 0&0&-i\\0&0&0\\i&0&0 \end{smallmatrix}\Big),\ \  
    &\qquad\qquad\\
     &\lambda_6=& \Big(\begin{smallmatrix} 0&0&0\\0&0&1\\0&1&0 \end{smallmatrix}\Big),\ \  
    &\lambda_7=& \Big(\begin{smallmatrix} 0&0&0\\0&0&-i\\0&i&0 \end{smallmatrix}\Big),\ \  
    &\lambda_8= \tfrac{1}{\sqrt{3}}\Big(\begin{smallmatrix} 1&0&0\\0&1&0\\0&0&-2 \end{smallmatrix}\Big),\notag
\end{alignat}
as a basis for the $3{\times}3$ Hermitian matrices acting on the spin-$1$ impurity. Writing the spin-$1$ operators as $(S^\mu)_{\nu\eta}=\frac{1}{\sqrt{2}}i\epsilon_{\mu\nu\eta}$ with $\mu,\nu,\eta \in\{x,y,z\} =\{1,2,3\}$, we identify them with the three asymmetric Gell-Mann matrices, i.e., $S^\mu=\frac{1}{\sqrt{2}} (-\lambda_7,\lambda_5,-\lambda_2)^
\mu$. The symmetric matrices are spin-nematics and form the complementaries $\mathcal {S^A} = \frac{1}{\sqrt{2}}(\lambda_1,\lambda_4,\lambda_6,\lambda_3,\lambda_8)^{\mathcal A}$. Observe that 
\begin{equation}
\{S^\mu,S^\nu\}  = \delta_{\mu\nu} \mathbb{1}_3  - \tfrac{1}{2}(\ket{\mu}\bra{\nu} +\ket{\nu}\bra{\mu}),
\end{equation}
with $\ket{\mu}$ unit vectors in $\mathbb{R}^3$ and so $d_{\mu\nu\mathcal A}=-\tfrac{1}{\sqrt{2}}(\mathcal{S^A})_{\mu\nu}$. Thus, we explicitly rewrite Eq.~\eqref{eq:RGAppb} as
\begin{equation}
    \frac{d \mathcal{B_A}}{d \ell} = \mathcal{B_A} + \frac{1}{\sqrt{2}} (\mathcal{S^A})_{\mu\nu} \tilde \lambda_\mu \tilde \lambda_\nu.
 \end{equation}
For example, for $\mathcal A = 5$, which we relate to the complementary $(S^z)^2$ up to rescaling and a shift by the identity, we find consistently with \cite{zitkoPropertiesAnisotropicMagnetic2008,schillerPhaseDiagramAnisotropic2008}
\begin{equation}    
    \frac{d \mathcal{B}_{\mathcal A = 5}}{d \ell} =  \mathcal{B_A} - \frac{2\tilde \lambda_z^2 - \tilde \lambda_x^2- \tilde \lambda_y^2}{2\sqrt{3}}.
\end{equation}
Observe that multiple Kondo couplings contribute to the same complementary. In the isotropic case these contributions cancel out, but otherwise they can add up to either a positive or a negative correction. Starting from $\mathcal{B^A}=0$, a negative correction will leave the impurity underscreened, but a positive correction will fully self-screen the impurity.

\section{NRG Details}\label{sec:NRG}

All NRG calculations in this paper were carried out using the open-source \QSpace tensor library \cite{weichselbaumNonabelianSymmetriesTensor2012,weichselbaumXsymbolsNonAbelianSymmetries2020,weichselbaumQSpaceOpensourceTensor2024pub} that exploits both Abelian and general non-Abelian symmetries, together with a proprietary \QSpace-based NRG code \cite{leeAdaptiveBroadeningImprove2016}.\footnote{Note that \QSpace also contains equivalent open-source NRG code. The proprietary NRG implementation can be obtained by contacting Seung-Sup Lee.} This enables exploiting the high-symmetry structure of multichannel impurity models: By keeping only a single representative state from each symmetry multiplet, and inferring the rest from generalized Clebsch-Gordan coefficients that are tabulated, the effective Hilbert space of all participating tensors is significantly reduced. Unless stated otherwise, calculations were carried out with logarithmic discretization $\Lambda{=}4$, no $z$-averaging, and $N^*_\mathrm{keep}=2000$ kept multiplets (which corresponds to a larger number of kept states $N_\mathrm{keep}$, depending on the symmetries exploited -- see below). Note that we avoid interleaving the different channels \cite{mitchellGeneralizedWilsonChain2014,stadlerInterleavedNumericalRenormalization2016} along the Wilson chain, as this artificially breaks channel symmetry. While the latter can be restored manually, it complicates exploring symmetry breaking perturbations. Thus, we rely solely on symmetry exploitation to reduce the computational cost. 

\subsection{Symmetry Structure}
As mentioned in Sec.~\ref{sec:Models}, the unperturbed Anderson model has an $\mathrm{Sp}(2k)$ spin-channel symmetry together with a $\mathrm{U}(1)$ charge symmetry. The latter is promoted to $\mathrm{SU}(2)$ at the particle-hole symmetric point $N_g{=}1$.  The unperturbed Kondo model is already at the particle-hole symmetric point, and so has the full $\mathrm{SU}(2){\times}\mathrm{Sp}(2k)$ symmetry. These symmetry setups are now available as part of \QSpace. However, in order to explore symmetry-breaking effects, we naturally need to break some of these symmetries. Working with the Anderson model, we retain the global $\mathrm{U}(1)$ charge symmetry (of all channels together), and break the spin-channel symmetry $\mathrm{Sp}(2k)\to\mathrm{SU}(2)_{\otimes k}$, i.e., we retain an $\mathrm{SU}(2)$ spin symmetry in each channel (thus  $N^*_\mathrm{keep}=2000$ kept multiplets corresponds to an average effective $N_\mathrm{keep}\approx 8000$ and $9000$ kept states for $k{=}2$ and $3$, respectively). This enables exploring all channel and particle-hole symmetry breaking effects, expect for site-dependent magnetic fields. For the latter, we further break the $\mathrm{SU}(2)$ spin symmetry in each channel to a $\mathrm{U}(1)$ symmetry.

We now briefly explain how we set up the Hamiltonian while exploiting non-Abelian symmetries, and refer the interested reader to more details in Refs.~\cite{weichselbaumNonabelianSymmetriesTensor2012,weichselbaumXsymbolsNonAbelianSymmetries2020,weichselbaumQSpaceOpensourceTensor2024pub}. The way \QSpace proceeds internally is that it starts with the local state space of a single fermionic site in Fock space. For example, for $k$ spinful channels this results in a $2^{2k}$-dimensional Hilbert space. This space is organized into symmetry sectors based on elementary symmetry operations such as raising and lowering operators for all non-Abelian symmetries as specified, and subsequently compressed by switching from state-space based description to symmetry multiplets (see examples in Table.~\ref{tab:symmetries}). Once the basis of this symmetry transformation is determined, \QSpace proceeds to the representation of elementary local operators by extracting reduced matrix elements from their Fock-space based representation using the Wigner-Eckart theorem. From the user perspective, however, much of this is hidden, and one can directly start with the a set of representative local operators cast into \QSpace tensors for the specified symmetries. In the fermionic case, this includes a full set of fermionic annihilation operators as well as the fermionic parity. With these one may proceed to construct any other local operator required for a particular model Hamiltonian.

\begin{table}[t]
    \begin{tabular}{c c c}
    $\quad{\rm U}(1) {\otimes} {\rm SU}(2)\quad$&  
    ${\rm U}(1) {\otimes} {\rm SU}(2)_{\otimes 2}$& 
    $\qquad{\rm U}(1) {\otimes} {\rm Sp}(4)\qquad$\\
    $(k=1)$ & $(k=2)$ & $(k=2)$ \\
    \begin{tabular}[t]{|cc|cc|}\hline
        $n$&$j$& $d$ & $\#$ \\\hline
        $0$&$0$&$1$ & \\
        $1$&$\tfrac{1}{2}$&$2$ & \\
        $2$&$0$&$1$ & \\\hline
    \end{tabular}
    &
    \begin{tabular}[t]{|ccc|cc|}\hline
        $n$&$j_1$&$j_2$& $d$ & $\#$\\\hline
        $0$&$0$&$0$&$1$ & \\
        $1$&$\tfrac{1}{2}$&$0$&$2$ & \\
        $1$&$0$&$\tfrac{1}{2}$&$2$ & \\
        $2$&$0$&$0$&$1$ & 2 \\ 
        $2$&$\tfrac{1}{2}$&$\tfrac{1}{2}$&$4$ & \\
        $3$&$\tfrac{1}{2}$&$0$&$2$ & \\
        $3$&$0$&$\tfrac{1}{2}$&$2$ & \\
        $4$&$0$&$0$&$1$ & \\\hline
    \end{tabular}
    & 
    \begin{tabular}[t]{|cc|cc|}\hline
        $n$&$(q_1 q_2)$ & $d$ & $\#$ \\\hline
        $0$&$(00)$&$1$ & \\
        $1$&$(10)$&$4$ & \\
        $2$&$(00)$&$1$ & \\
        $2$&$(01)$&$5$ & \\
        $3$&$(10)$&$4$ & \\
        $4$&$(00)$&$1$ & \\\hline
    \end{tabular}
    \end{tabular}
    \caption{Symmetry structure of the Hilbert space of a single fermionic site having $k$ symmetric spinful channels. Here, the symmetry labels are denoted by $n$ for $\mathrm{U}(1)$ particle number (charge), $j_{(i)}\geq 0$ for $\mathrm{SU}(2)$ multiplets, and $(q_1 q_2) \equiv (q_1, q_2)$ with $q_i\in \mathbb{N}_0$ for $\mathrm{Sp}(4)$ multiplets using the compact symmetry-label notation of Appendix \ref{sec:qlabels}.  The number of degenerate states in a single multiplet is denoted by $d$ and \# specifies how many copies of the multiplet are present ($1$ where not explicitly indicated). From left to right: (i) single channel ($k{=}1$) with $\mathrm{SU}(2)$ symmetry; (ii) two channels ($k{=}2$) with $\mathrm{SU}(2)$ symmetry in each channel, i.e., two copies of the case in (i); (iii) two channels with $\mathrm{Sp}(4)$ symmetry. Since the symmetry in (iii) is larger than in (ii), this gives rise to fewer multiplets. In all cases, the total number of states adds up to the Hilbert space dimension $2^{2k}$, i.e., $4$ in (i) and $16$ in (ii) and (iii). For the shown settings, the multiplets are all unique except for the case $(n,j_1,j_2)=(2,0,0)$ in (ii) which contains $\#{=}2$ multiplets. These represent the doubly occupied state in either of the two channels, with the other channel being empty. 
    \label{tab:symmetries}}
\end{table}

Our basic building block is the set of  $2k$ fermionic operators $\textbf{f} \equiv( f_{1\uparrow  },\dots,f_{k\uparrow  },  f_{1\downarrow},\dots,f_{k\downarrow})^T$ acting on a single site. \QSpace already groups these into so-called irreducible operators (irops), i.e., operators that transform as irreducible representations with respect to the chosen symmetry setting. For example, for the largest symmetry setting (iii) in Table \ref{tab:symmetries}, all $2k$ operators in ${\bf f}$ are represented as a single entity by a single \QSpace tensor that has the operator index as an additional third index of range $2k$. This third index carries the symmetry labels that describe the action of the operator: it reduces particle number by one (hence, $n{=}-1$ in the notation of Table \ref{tab:symmetries}) and transforms in the (dual to the) defining representation of $\mathrm{Sp}(2k)$. Anisotropies reduce the symmetry, such that ${\bf f}$ will be returned as multiple irops consistent with the chosen symmetries. For example, for the setting (ii) in Table \ref{tab:symmetries}, the operator ${\bf f}$ will consist of two irops consisting of $k$ operators each that transform as $(n,j_1,j_2) = (-1,\frac{1}{2},0)$ or $(-1,0,\frac{1}{2})$, respectively.

By taking linear combinations of products $f^\dagger_{j\sigma}f_{j'\sigma'}$ and renaming $f_{j\sigma}\to c_{j\sigma}(0)$ we can construct the current operators $J^A(0)=\textbf{c}^\dagger(0) T^A \textbf{c}(0)$ and $\mathcal{J^A}(0)=\textbf{c}^\dagger(0)\mathcal{T^A}\textbf{c}(0)$ that act in the full $2^{2k}$-dimensional fermionic Hilbert space of a site. After projecting onto the singly-occupied sector ($n{=}1$), we obtain the symplectic ``spin'' operators $S^A=\textbf{f}^\dagger T^A \textbf{f}$  and complementaries $\mathcal{S^A}=\textbf{f}^\dagger\mathcal{T^A}\textbf{f}$ . In the context of \QSpace, the above procedure translates to contracting ${\bf f}^\dagger \ast {\bf f}$ like a matrix product, followed by fusing the pair of open operator indices (e.g., see \textit{complete operator basis} in Ref.~\cite{weichselbaumQSpaceOpensourceTensor2024pub}). In the symplectic case, based on Eq.~\eqref{eq:10x10}, this results in three irops corresponding to the identity $\mathbb{1}$, the generators $S^A$, and the complementaries $\mathcal{S^A}$.

The Hamiltonian is a symmetry scalar, i.e., a trivial irop with symmetry labels of zero for all non-Abelian symmetries. Thus, only scalar perturbations can be introduced to it. With the full $\mathrm{Sp}(2k)$ symmetry irops we can introduce the particle-hole-symmetry breaking perturbation $\sum_\mathcal{A}\mathcal{S^A
 J^A}(0)$. However, in order to introduce anisotropy in the Kondo couplings or local impurity terms, we switch to $\mathrm{U}(1)\otimes\mathrm{SU}(2)_{\otimes k}$ symmetry as in setting (ii) in Table \ref{tab:symmetries}. We can then construct irops from linear combinations of  $f^\dagger_{j\sigma}f_{j'\sigma'}$ that correspond to elements of either $S^A$ or $\mathcal{S^A}$, with some of them being symmetry scalars.

\subsection{Analysis}

\begin{figure}
    \centering
    \includegraphics[width=\columnwidth]{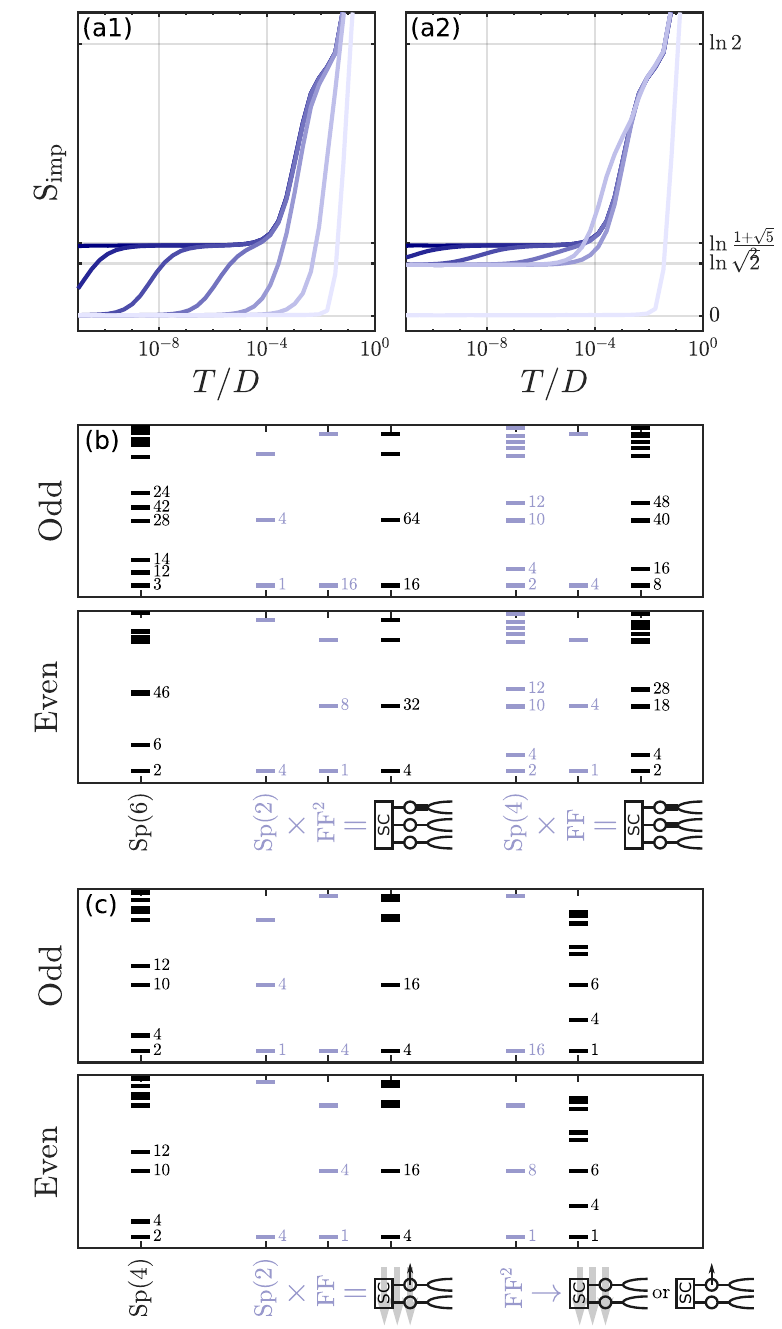}
      \vspace{-2\baselineskip}\phantomsubfloat{\label{fig:NRG-Simp}}\phantomsubfloat{\label{fig:NRG-FSS}}\phantomsubfloat{\label{fig:NRG-FSS-Sp4-SU4}}
    \caption{(a) Impurity entropy for $k{=}3$ channels and a tunneling asymmetry of varying magnitude, with either (a1) a single strongly-coupled lead; or (a2) two strongly-coupled leads (implemented as one weakly coupled lead). Light (dark) colors correspond to a large (small) perturbation. (b) Low-energy finite-size spectra with the degeneracy indicated next to the low-lying levels. The upper (lower) panel corresponds to an odd (even) number of single-particle levels (or Wilson-chain sites). From left to right, in black, the spectra corresponding to: The unperturbed $\mathrm{Sp}(6)$ model; a single strongly-coupled lead; and two strongly-coupled leads. For comparison, in light we have the finite-size spectra that can be multiplied to yield the spectra of the perturbed model. (c)  Low-energy finite-size spectra for $k{=}2$ and a shift in the dot levels. From left to right, in black, the spectra corresponding to: The unperturbed $\mathrm{Sp}(4)$ model; a potential applied to a single dot and rectified by gate detuning to reinstate particle-hole symmetry; the $\mathrm{SU}(4)$ fixed point due to the breaking of particle-hole symmetry either by gate detuning or by a potential at a single dot. All results are obtained with the same parameters as used for Fig.~\ref{fig:NRG-scaling}, i.e., $E_C/D=0.2, \Delta/D=t/D=0.1$ and $\Lambda=4,N^*_\mathrm{keep}=2000$.}
    \label{fig:NRG-Analysis}
\end{figure}

The bulk of this paper is based on analyzing the impurity entropy  $\mathrm{S_{imp}}$ as a function of temperature $T$. It is defined as
\begin{equation}\label{eq:SimpDef}
    \mathrm{S_{imp}}(T)\equiv\mathrm{S_{full}}(T)-\mathrm{S_{bath}}(T),
\end{equation}
i.e., the difference between the thermodynamic entropy of the full system, and that of the leads only. These are obtained from two separate NRG runs in the conventional way \cite{bullaNumericalRenormalizationGroup2008}. For each type of perturbation we plot $\mathrm{S_{imp}}(T)$ as a function of the perturbation magnitude, and extract the energy scales at which it changes.

This is demonstrated in Fig.~\ref{fig:NRG-Simp} for $k{=}3$ and tunneling asymmetry. Once the temperature drops below $E_C$ and $\Delta$ we (barely) observe an $\mathrm{S_{imp}}{=}\ln k$ plateau, corresponding to the $k$-fold degenerate decoupled-dots ground space. Then, below the Kondo temperature $T_K$ of the unperturbed system, the entropy drops to the fractional $\ln d_k$ according to Eq.~\eqref{eq:Simp} and corresponding to the $\mathrm{Sp}(2k)$ NFL. But in the presence of a relevant perturbation, the entropy drops further at $T^*$ (or $T_\mathrm{FL}$) to its final $T\to0$ value, with $T^*$  defined as the midpoint
\begin{equation}
    \mathrm{S_{imp}}(T^*)\equiv \tfrac{1}{2}[\ln d_k + \mathrm{S_{imp}}(T{\to}0)].
\end{equation}
Note that $\mathrm{S_{imp}}(T{\to}0)$ depends on the specific perturbation, as can be seen in the two panels of  Fig.~\ref{fig:NRG-Simp}. We then plot the extracted  $T^*$ as a function of the perturbation magnitude and extract its scaling, as plotted in Fig.~\ref{fig:NRG-scaling}.

The different fixed points are identified according to the final impurity entropy and the NRG finite-size spectrum (from which  $\mathrm{S_{imp}}$ was extracted). In Fig.~\ref{fig:NRG-FSS} we plot (in black) the $T{\to}0$ spectra for the $k{=}3$ unperturbed NFL fixed point, and the two resulting perturbed fixed points:
\begin{itemize}
    \item Unperturbed $\mathrm{Sp}(6)$ NFL with $\mathrm{S_{imp}}{=}\ln\frac{1+\sqrt{5}}{2}$. Has the known spectrum of the $\mathrm{SU}(2)_3$ low-energy fixed point (due to the same fusion rules).
    \item One strongly-coupled lead  $\mathrm{Sp}(2)\simeq\mathrm{SU}(2)$ FL with $\mathrm{S_{imp}}{=}0$ together with two decoupled leads. Has the known spectrum of the single-channel Kondo effect, i.e., a FL with a $\pi/2$ phase shift, multiplied by the spectrum of a two free-fermionic (FF) channels (i.e., FLs with no phase shift).
    \item Two strongly-coupled leads $\mathrm{Sp}(4)$ NFL with $\mathrm{S_{imp}}{=}\ln\sqrt{2}$  together with a single decoupled lead. Has the known spectrum of the $\mathrm{SU}(2)_2$ fixed point, multiplied by the FF spectrum.
\end{itemize}
The level spacing and degeneracies are in excellent agreement with our expectations from the CFT fusion ansatz (up to controlled discretization corrections due to the moderate $\Lambda{=}4$). An elegant way to verify this is to separately calculate the factor spectra (plotted in light color). These display exactly the same discretization finite-size corrections. One can then readily verify that they multiply to the perturbed fixed-point spectra, as expected.

In Fig.~\ref{fig:NRG-FSS-Sp4-SU4} we plot the finite-size spectra for a $k{=}2$ and demonstrate the two different fixed points associated with a shift of the energy level of a single dot $\epsilon_1$. As discussed in Sec.~\ref{sec:SchriefferWolff}, this has two effects: breaking of the channel, or $\mathrm{Sp}(4)$ symmetry, and breaking of particle-hole symmetry. The latter is the leading order effect, and takes the system to a single-channel $\mathrm{SU}(4)$ FL fixed point (rightmost spectra) with a $\pi/4$ phase shift in each of the original spin-$1/2$ channels. This is the same fixed point obtained by gate detuning $N_g{=}1{+}\delta N_g$. Tuning these to cancel out effectively reinstates particle-hole-symmetry. Then channel asymmetry still destabilizes the $\mathrm{Sp}(4)$ NFL (leftmost spectra) by gapping out one of the dots. This results in a single-channel $\mathrm{Sp}(2)\simeq\mathrm{SU}(2)$ Kondo effect in one channel, and a decoupled lead in the other. The resulting finite-size spectra (central black spectra) indeed corresponds to the product of the spectra of a FL with a $\pi/2$ phase shift and FF (with no phases shift).

The procedure of extracting the NFL-breaking energy scale and analyzing the finite-size spectra is repeated for all perturbations discussed in the various sections. The outcome is plotted in Fig.~\ref{fig:NRG-scaling} and demonstrates prefect agreement with the coherent picture established in this paper.

\end{appendix}
\clearpage

\bibliography{bib}

\end{document}